\pdfoutput=1

\documentclass[runningheads,a4paper]{llncs}

\usepackage{amssymb}
\usepackage{graphicx}
\usepackage{amsmath}
\usepackage{algorithmic}
\usepackage{algorithm}
\usepackage{url}
\usepackage{alltt}
\urldef{\mailsa}\path|pat@dcs.gla.ac.uk|

\newlength{\halftextwidth}
\usepackage{a4wide}
\usepackage{multicol}
\usepackage{listings}

\begin{document}

\mainmatter  

\title{Exact Algorithms for Maximum Clique \\
A Computational Study \\
TR-2012-333 }

\titlerunning{Exact Algorithms for Maximum Clique}

\author{Patrick Prosser}

\authorrunning{Exact Algorithms for Maximum Clique}

\institute{Computing Science,\\
Glasgow University, Glasgow, Scotland\\
pat@dcs.gla.ac.uk\\}

\maketitle

\lstset{ %
language=Java,              
basicstyle=\scriptsize,     
numbers=left,               
numberstyle=\scriptsize,    
numbersep=10pt,             
frame=trBL,                 
captionpos=b,               
breaklines=true,            
breakatwhitespace=false,    
showstringspaces=false,     
frameround=fttt
}

\section{Introduction}
\label{sec:intro}
\vspace{-1.5mm}
The purpose of this paper is to investigate a number of recently reported exact algorithms for the
maximum clique problem. The actual program code used is presented and critiqued. 
The computational study aims to show how implementation details, problem features and hardware platforms influence
algorithmic behaviour in those algorithms.\\

\noindent
{\bf The Maximum Clique Problem (MCP):}
A simple undirected graph $G$ is a pair $(V,E)$ where $V$ is a set of vertices and $E$ a set of edges. An edge
$\{u,v\}$ is in $E$ if and only if $\{u,v\} \subseteq V$ and vertex $u$ is adjacent to vertex $v$. A \emph{clique} is a 
set of vertices $C \subseteq V$ such that every pair of vertices in $C$ is adjacent in $G$.
Clique is one of the six basic NP-complete problems given in \cite{gareyJohnson}. It is posed as a decision
problem [GT19]: given a simple undirected graph $G = (V,E)$ and a positive integer $k \leq |V|$ does $G$ contain a clique of size 
$k$ or more? The optimization problems is then to find the \emph{maximum clique}, where $\omega(G)$ is the size of a maximum clique. 

A graph can be coloured, by that we mean that any pair of adjacent vertices must be given different colours. We do not use colours, 
we use integers to label the vertices. The minimum number of different colours required is then the
\emph{chromatic number} of the graph $\chi(G)$, and $\omega(G) \leq \chi(G)$. Finding the chromatic number is NP-complete.\\ 

\noindent
{\bf Exact Algorithms for MCP:}
We can address the decision and optimization problems with an exact algorithm, 
such as a backtracking search 
\cite{fahle,regin2003,wood97,carraghanPardalos90,pardalosRodgers92,prjo2002,segundo2011,segundo2011b,Konc_Janezic_2007,tomita2003,tomita2010,aaai2010,carmoZuge}.
Backtracking search incrementally constructs the set $C$ (initially empty) by choosing a \emph{candidate vertex}
from  the \emph{candidate set} $P$ (initially all of the vertices in $V$) and then adding it to $C$. 
Having chosen a vertex the candidate set is then updated, removing vertices
that cannot participate in the evolving clique. If the candidate set is empty then $C$ is maximal (if it is a maximum we save it)
and we then backtrack. Otherwise $P$ is not empty and we continue our search, selecting from $P$ and adding to $C$.

There are other scenarios where we can cut off search, i.e. if what is in $P$ is insufficient
to unseat the champion (the largest clique found so far) search can be abandoned. That is, an upper bound can be computed.
Graph colouring can be used to compute an upper
bound during search, i.e. if the candidate set can be coloured with $k$ colours then it can contain a clique no larger than $k$
\cite{wood97,fahle,segundo2011,Konc_Janezic_2007,tomita2003,tomita2010}. There are also heuristics that can be used when selecting
the candidate vertex, different styles of search, different algorithms to colour the graph and different
orders in which to do this.\\

\noindent
{\bf Structure of the Paper:}
In the next section we present in Java the following algorithms: Fahle's Algorithm 1 \cite{fahle}, Tomita's MCQ
\cite{tomita2003}, MCR \cite{tomita2007} and MCS \cite{tomita2010} and San Segundo's BBMC \cite{segundo2011}. 
By using Java and its inheritance mechanism algorithms are presented as modifications of previous algorithms.
Three vertex orderings are presented, one being new to these algorithms. Starting with the basic algorithm MC
we show how minor coding details can significantly impact on performance. Section 3 presents a chronological review
of exact algorithms, starting at 1990. Section 4 is the computational study. The study investigates MCS and determines
where its speed advantage comes from, measures the benefits resulting from the bit encoding of BBMC, the
effectiveness of three vertex orderings and the potential to be had from tie-breaking within an ordering. New benchmark
problems are then investigated. Finally an established technique for calibrating and scaling results is put to the test
and is shown to be unsafe. Finally we conclude.

\section{The Algorithms: MC, MCQ, MCS and BBMC}
\label{sec:algorithms}
We start by presenting the simplest algorithm \cite{fahle} which I will call MC. This sets the scene.
It is presented as a Java class, as are all the algorithms,
with instance variables and methods. And we might say that I am abusing Java by using the class merely as a place holder
for global variables and methods for the algorithms, and inheritance merely to over-ride methods so
that I can show a program-delta, i.e. to present the differences between algorithms. 

Each algorithm is first described textually and then the actual implementation is given in Java. Sometimes a program trace is given to better expose
the workings of the algorithm. It is possible to read this section skipping the Java descriptions, however the Java code
makes it explicit how one algorithm differs from another and shows the details that can severely
affect the performance of the algorithm.

\begin{figure}
\centering
\includegraphics[height=9.2cm,width=10.2cm]{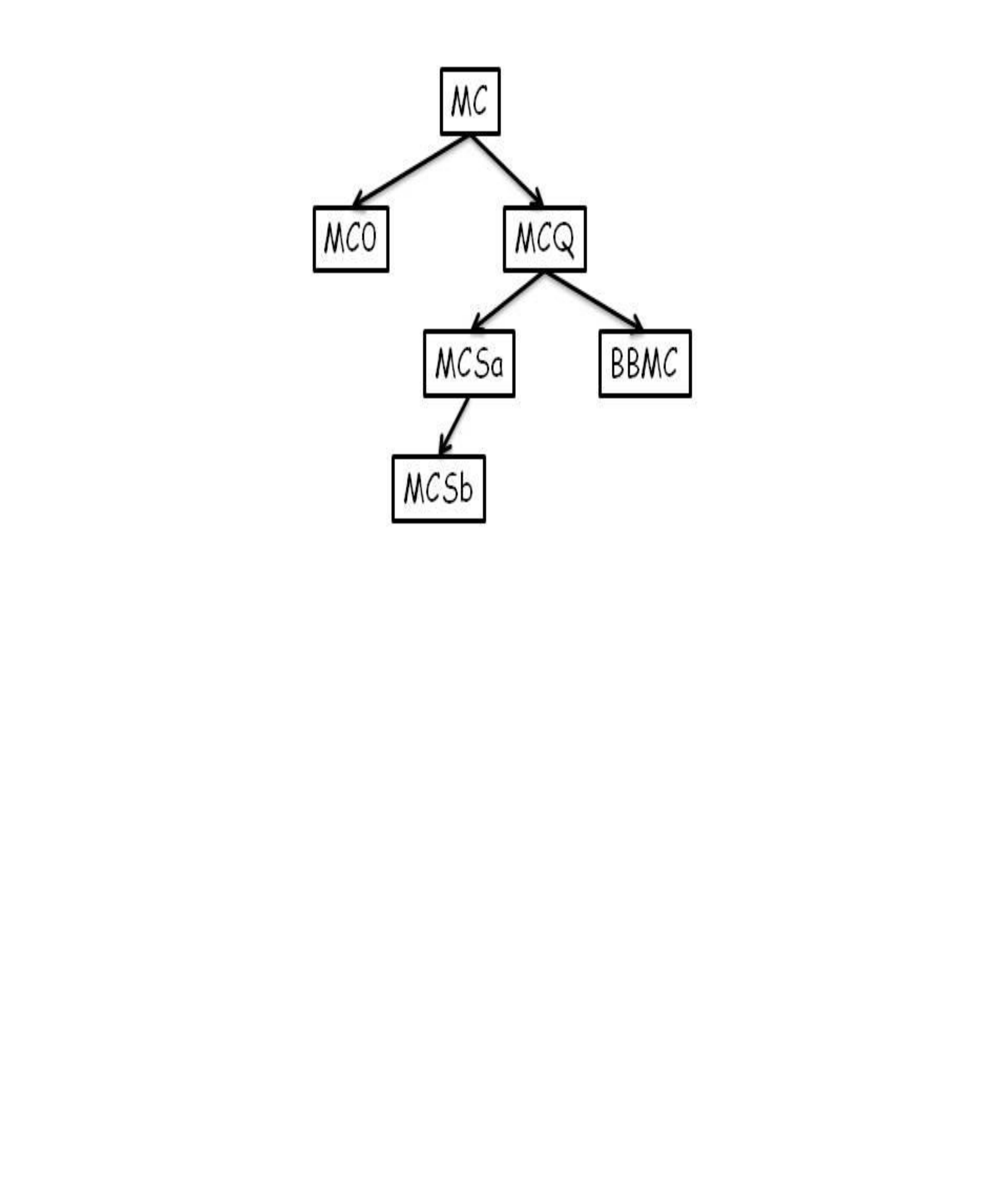}
\vspace{-50mm}
\caption{The hierarchy of algorithms.}
\label{inheritance}
\end{figure}

We start with MC. MC is essentially a \emph{straw man}: it is elegant but too simple to be of any practical worth.
Nevertheless, it has some interesting features. MCQ is then presented as an extension to MC, our first algorithm that uses a tight integration of 
search algorithm, search order and upper bound cut off. Our implementation of MCQ allows three different vertex orderings 
to be used, and one of these corresponds to MCR. The presentation of MCQ is somewhat laborious but this pays off when we present
two variants of MCS as minor changes to MCQ. BBMC is presented as an extension of MCQ, but is essentially MCSa with sets implemented using
bit strings. Figure \ref{inheritance} shows the hierarchical structure for the algorithms presented.
\vspace{-5mm}

\noindent
Appendix 1 gives a description of the execution environment (the MaxClique class) that allows us to 
read clique problems and solve them with a variety of algorithms and colour orderings. It also shows how to actually run our programs
from the command line.

\subsection{MC}
\label{sec:mcCode}
We start with  a simple algorithm, similar to Algorithm 1 in \cite{fahle}. Fahle's Algorithm 1 uses two sets: $C$ the growing clique
(initially empty) and $P$ the candidate set (initially all vertices in the graph). $C$ is \emph{maximal} when $P$ is empty and if
$|C|$ is a \emph{maxima} it is saved, i.e. $C$ becomes the \emph{champion}. If $|C| + |P|$ is too small to unseat the champion
search can be terminated. Otherwise search iterates over the vertices in $P$ in turn selecting a vertex $v$,
creating a new growing clique $C'$ where $C' = C \cup \{v\}$ and a new candidate
set $P'$ as the set of vertices in $P$ that are adjacent to $v$ (i.e. $P' = P \cap neighbours(v)$), 
and recursing. We will call this MC.

\subsubsection{MC in Java}
Listing \ref{codeMC} can be compared to Algorithm 1 in \cite{fahle}.
The constructor, lines 14 to 22, takes three arguments: $n$ the number of vertices in the graph, $A$ the adjacency 
matrix where $A[i][j]$ equals 1 if and only if vertex $i$ is adjacent to vertex $j$, and $degree$ where $degree[i]$ is
the number of vertices adjacent to vertex $i$ (and is the sum of $A[i]$). The variables $nodes$ and $cpuTime$ are used
as measures of search performance, $timeLimit$ is a bound on the run-time, $maxSize$ is the size of the largest clique found so far,
$style$ is used as a flag to customise the algorithm with respect to ordering of vertices and the array $solution$ is the largest clique found
such that $solution[i]$ is equal to 1 if and only if vertex $i$ is in the largest clique found.

The method $search()$ finds a largest clique\footnote{There may be more than one largest clique so we say we find ``a largest
clique''} or terminates having exceeding the allocated $timeLimit$. Two sets are produced: the \emph{candidate set} $P$
and the \emph{current clique} $C$\footnote{At least two of the published algorithms name the candidate set $P$, maybe 
for ``potential'' vertices, for example \cite{fahle} and more recently \cite{eppstein2011}.}.  
Vertices from $P$ may be selected and added to the growing clique $C$. Initially all 
vertices are added to $P$ and $C$ is empty (lines 27 and 28).
The sets $P$ and $C$ are represented using Java's ArrayList, a re-sizable-array implementation of the List interface. Adding
an item is an $O(1)$ operation but removing an arbitrary item is of $O(n)$ cost. This might appear to be a damning 
indictment of this simple data structure but as we will see it is the cost we pay if we want to maintain order in $P$ and
in many cases we can work around this to enjoy $O(1)$ performance.

The search is performed in method $expand$\footnote{We use the same name for this method as Tomita \cite{tomita2003,tomita2010}.}.
In line 34 a test is performed to determine if the cpu time limit has been exceeded, and if so search terminates. Otherwise we
increment the number of nodes, i.e. a count of the size of the backtrack search tree explored. The method then iterates
over the vertices in $P$ (line 36), starting with the last vertex in $P$ down to the first vertex in $P$. This form
of iteration over the ArrayList, getting entries with a specific index, is necessary when entries are deleted (line 45)
as part of that iteration. A vertex $v$ is selected from $P$ (line 38), 
added to $C$ (line 39), and a new candidate set is then created (line 40)
$newP$ where $newP$ is the set of vertices in $P$ that are adjacent to vertex $v$ (line 41). Consequently all vertices in $newP$
are adjacent to all vertices in $C$ and all pairs of vertices in $C$ are adjacent (i.e. $C$ is a clique). If $newP$ is empty $C$ is 
maximal and if it is the largest clique found it is saved (line 42). If $newP$ is not
empty then $C$ is not maximal and search can proceed via a recursive call to $expand$ (line 43). 
On returning from the recursive call $v$ is removed from $P$ and from $C$ (lines 44 and 45).

There is one ``trick'' in $expand$ and that is at line 37: if the combined size of the current clique and the candidate set
cannot unseat the best clique found so far this branch of the backtrack tree can be abandoned. This is the simplest 
upper bound cut-off and corresponds to line 3 from Algorithm 1 in \cite{fahle}.
The method $saveSolution$ does as it says: it saves off the current maximal clique and records its size.

\begin{figure}
\lstset{caption={The basic clique solver},label=codeMC}
\begin{lstlisting}
import java.util.*;

public class MC {
    int[] degree;   // degree of vertices
    int[][] A;      // 0/1 adjacency matrix
    int n;          // n vertices
    long nodes;     // number of decisions
    long timeLimit; // milliseconds
    long cpuTime;   // milliseconds
    int maxSize;    // size of max clique
    int style;      // used to flavor algorithm
    int[] solution; // as it says

    MC (int n,int[][]A,int[] degree) {
	this.n = n;
	this.A = A;
	this.degree = degree;
	nodes = maxSize = 0;
	cpuTime = timeLimit = -1;
	style = 1;
	solution = new int[n];
    }

    void search(){
	cpuTime              = System.currentTimeMillis();
	nodes                = 0;
	ArrayList<Integer> C = new ArrayList<Integer>();
	ArrayList<Integer> P = new ArrayList<Integer>(n);
	for (int i=0;i<n;i++) P.add(i);
	expand(C,P);
    }

    void expand(ArrayList<Integer> C,ArrayList<Integer> P){
	if (timeLimit > 0 && System.currentTimeMillis() - cpuTime >= timeLimit) return;
	nodes++;
	for (int i=P.size()-1;i>=0;i--){
	    if (C.size() + P.size() <= maxSize) return;
	    int v = P.get(i);
	    C.add(v);
	    ArrayList<Integer> newP = new ArrayList<Integer>();
	    for (int w : P) if (A[v][w] == 1) newP.add(w);
	    if (newP.isEmpty() && C.size() > maxSize) saveSolution(C);
	    if (!newP.isEmpty()) expand(C,newP);
	    C.remove((Integer)v);
	    P.remove((Integer)v);
	}
    }

    void saveSolution(ArrayList<Integer> C){
	Arrays.fill(solution,0);
	for (int i : C) solution[i] = 1;
	maxSize = C.size();
    }
}
\end{lstlisting}
\end{figure}

\subsubsection{Observations on MC}
\begin{figure}
\centering
\includegraphics[height=9.2cm,width=10.2cm]{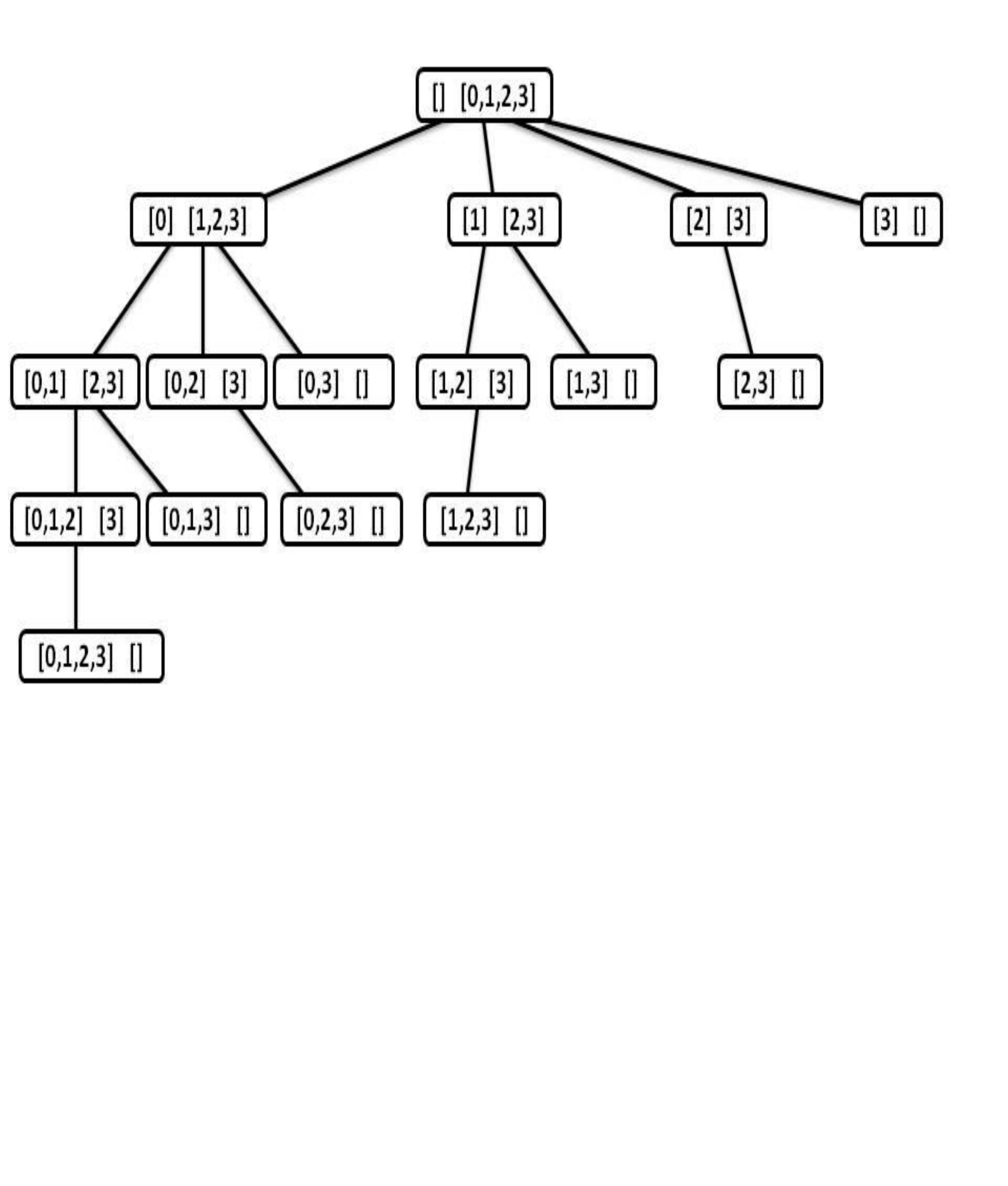}
\vspace{-35mm}
\caption{A binomial search tree producing the power set of $\{0,1,2,3\}$.}
\label{binomialTree}
\end{figure}

There are several points of interest. This first is the search process itself. If we commented out lines 37 and changed
line 41 to add to $newP$ all vertices in $P$ other than $v$, method $expand$ would produce the power set of $P$ and
at each depth $k$ in the backtrack tree we would have $n \choose k$ calls to $expand$. That is, $expand$ produces a 
\emph{binomial backtrack search tree} of size $O(2^{n})$ (see page 6 and 7 of \cite{dek}). This can be compared to a bifurcating
search process, where on one side we take an element and make a recursive call, and on the other side reject it and
make a recursive call, terminating when $P$ is empty. This generates the power set on the
leaf nodes of the backtrack tree and explores $2^{n+1}-1$ nodes. This is also $O(2^{n})$ but in practice is 
often twice as slow as the binomial search. In Figure \ref{binomialTree} we see a binomial search produced
by a simplification of MC, generating the power set of $\{0,1,2,3\}$. Each node in the tree contains two sets: the set
that will be added to the power set and the set that can be selected from at the next level. We see 16 nodes and at each
depth $k$ we have $n \choose k$ nodes. The corresponding tree for the bifurcating search (not shown) has 31 nodes with the
power set appearing on the 16 leaf nodes at depth 4.

The second point of interest is the actual Java implementation. Java gives us
an elegant construct for iterating over collections, the for-each loop, used in line 41 of Listing \ref{codeMC}. This is
rewritten in class MC0 (extending MC, overwriting the $expand$ method) Listing \ref{codeMC0} lines 15 to 18: 
one line of code is replaced with 4 lines. MC0 gets the
$j^{th}$ element of $P$, calls it $w$ (line 16 of Listing \ref{codeMC0}) and if it is adjacent to $v$ it is added to
$newP$ (line 17 of Listing \ref{codeMC0}). In MC (line 41 of Listing \ref{codeMC}) the for-each statement implicitly creates an
iterator object and uses that for selecting elements. This typically results in a 10\% reduction in runtime for MC0.

Our third point is how we create our sets. In MC0 line 15 the new candidate set is created with a capacity of $i$. 
Why do that when we can just create 
$newP$ with no size and let Java work it out dynamically? And why size $i$? In the loop of line 10 $i$ counts down from the
size of the candidate set, less one, to zero. Therefore at line 14 $P$ is of size $i+1$ and we can
set the maximum size of $newP$ accordingly. If we do not set the
size Java will give $newP$ an initial size of 10 and when additions exceed this $newP$ will be re-sized. By grabbing this space
we avoid that. This results in yet another measurable reduction in run-time.  

Our fourth point is how we remove elements from our sets. In MC we remove the current vertex $v$ from $C$ and $P$ 
(lines 44 and 45) whereas in MC0 we remove the \emph{last element} in $C$ and $P$ (lines 21 and 22). Clearly $v$ will always
be the last element in $C$ and $P$. The code in MC results in a sequential scan to find and then delete the last
element, i.e. $O(n)$, whereas in MC0 it is a simple $O(1)$ task. And this raises another question: $P$ and $C$ are really stacks
so why not use a Java Stack? The Stack class is represented using an ArrayList and cannot be initialised with a size,
but has a default initial size of 10. When the stack grows and exceeds its current capacity the capacity is doubled
and the contents are copied across. Experiments showed that using a Stack increased run time by a few percentage points.

Typically MC0 is 50\% faster than MC. In many cases a 50\% improvement in run time would be considered 
a significant gain, usually brought about by changes in the algorithm. Here, such a gain can be achieved by
moderately careful coding. And this is our first lesson: when comparing published results we need to be cautious
as we may be comparing programmer ability as much as differences in algorithms.

The fifth point is that MC makes more recursive calls that it needs to. At line 37 $|C| + |P|$ is sufficient
to proceed but at line 43 it is possible that $|C| + |newP|$ is actually too small and will generate a failure at
line 37 in the next recursive call. We should have a richer condition at line 43 but as we will soon see the
algorithms that follow do not need this. 

The sixth point is a question of space: why is the adjacency matrix an array of integers when we could have used booleans, 
surely that would have been more space efficient? In Java a boolean is represented as an integer with 1 being true,
everything else false. Therefore there is no saving in space and only a minuscule saving in time (more code is generated to 
test if $A[i][j]$ equals 1 than to test if a boolean is true). Furthermore by representing the adjacency matrix
as integers we can sum a row to get the degree of a vertex.

And finally, Listing \ref{codeMC} shows \emph{exactly} what is measured. Our run time starts at line 25, at the start of search.
This will include the times to set up the data structures peculiar to an algorithm, and any reordering of vertices. It does
not include the time to read in the problem or the time to write out a solution. There is also no doubt about what we mean by a 
\emph{node}: a call to $expand$ counts as one more node.

\lstset{caption={Inelegant but 50\% faster, MC0 extends MC},label=codeMC0}
\begin{lstlisting}
import java.util.*;

public class MC0 extends MC {

    MC0 (int n,int[][]A,int[] degree) {super(n,A,degree);}

    void expand(ArrayList<Integer> C,ArrayList<Integer> P){
	if (timeLimit > 0 && System.currentTimeMillis() - cpuTime >= timeLimit) return;
	nodes++;
	for (int i=P.size()-1;i>=0;i--){
	    if (C.size() + P.size() <= maxSize) return;
	    int v = P.get(i);
	    C.add(v);
	    ArrayList<Integer> newP = new ArrayList<Integer>(i);
	    for (int j=0;j<=i;j++){
		int w = P.get(j);
		if (A[v][w] == 1) newP.add(w);
	    }
	    if (newP.isEmpty() && C.size() > maxSize) saveSolution(C);
	    if (!newP.isEmpty()) expand(C,newP);
	    C.remove(C.size()-1);
	    P.remove(i);
	}
    }
}
\end{lstlisting}

\subsubsection{A trace of MC}
\begin{figure}
\centering
\includegraphics[height=7.2cm,width=9.2cm]{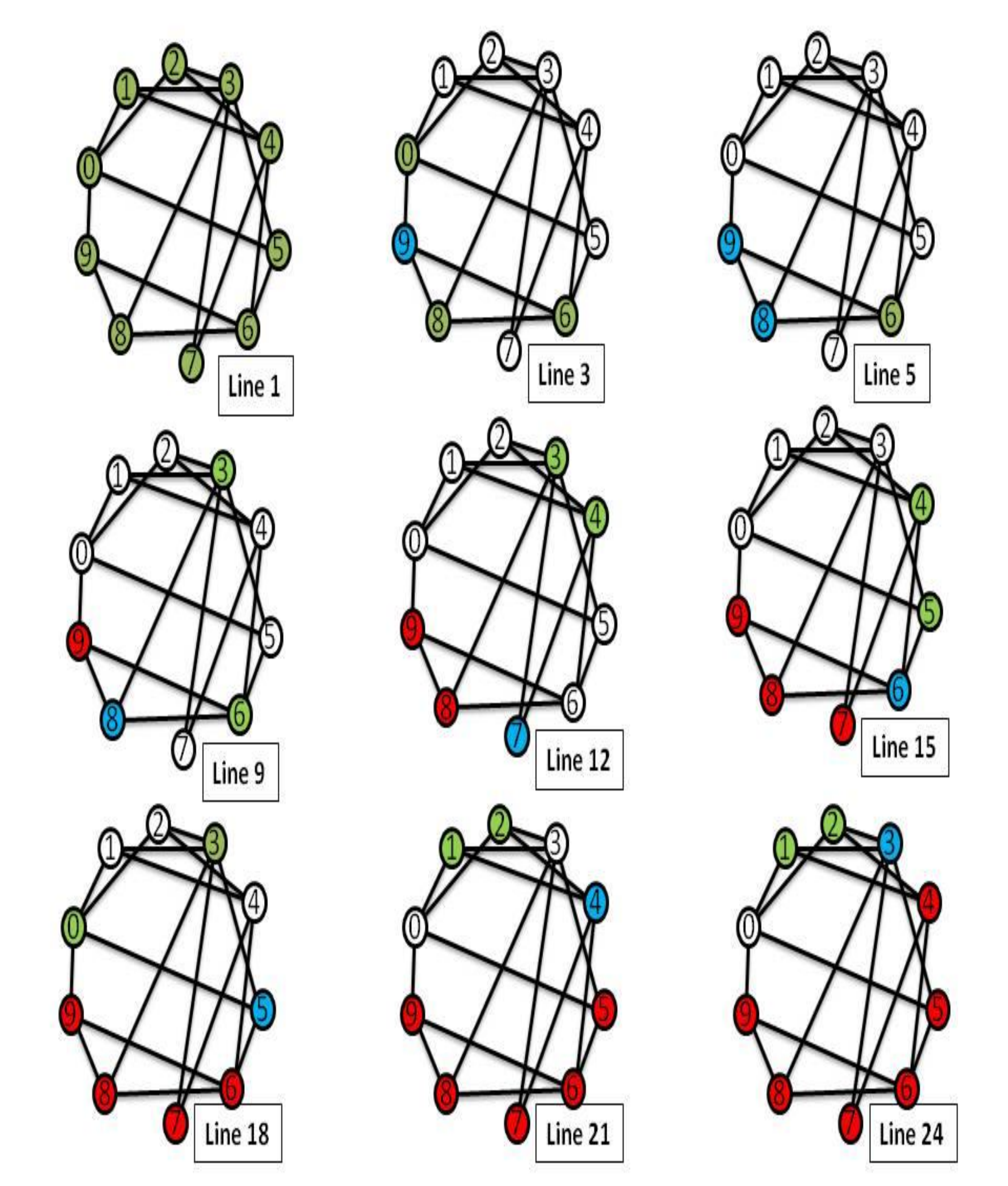}
\begin{scriptsize}
\begin{verbatim}
              1  > expand(C:[],P:[0, 1, 2, 3, 4, 5, 6, 7, 8, 9]
              2  > select 9 C:[] P:[0, 1, 2, 3, 4, 5, 6, 7, 8, 9] -> C:[9] & newP:[0, 6, 8]
              3  > > expand(C:[9],P:[0, 6, 8]
              4  > > select 8 C:[9] P:[0, 6, 8] -> C:[9, 8] & newP:[6]
              5  > > > expand(C:[9, 8],P:[6]
              6  > > > select 6 C:[9, 8] P:[6] -> SAVE: [9, 8, 6]
              7  > > FAIL: |C| + |P| <= 3  C:[9] P:[0, 6]
              8  > select 8 C:[] P:[0, 1, 2, 3, 4, 5, 6, 7, 8] -> C:[8] & newP:[3, 6]
              9  > > expand(C:[8],P:[3, 6]
             10  > > FAIL: |C| + |P| <= 3  C:[8] P:[3, 6]
             11  > select 7 C:[] P:[0, 1, 2, 3, 4, 5, 6, 7] -> C:[7] & newP:[3, 4]
             12  > > expand(C:[7],P:[3, 4]
             13  > > FAIL: |C| + |P| <= 3  C:[7] P:[3, 4]
             14  > select 6 C:[] P:[0, 1, 2, 3, 4, 5, 6] -> C:[6] & newP:[4, 5]
             15  > > expand(C:[6],P:[4, 5]
             16  > > FAIL: |C| + |P| <= 3  C:[6] P:[4, 5]
             17  > select 5 C:[] P:[0, 1, 2, 3, 4, 5] -> C:[5] & newP:[0, 3]
             18  > > expand(C:[5],P:[0, 3]
             19  > > FAIL: |C| + |P| <= 3  C:[5] P:[0, 3]
             20  > select 4 C:[] P:[0, 1, 2, 3, 4] -> C:[4] & newP:[1, 2]
             21  > > expand(C:[4],P:[1, 2]
             22  > > FAIL: |C| + |P| <= 3  C:[4] P:[1, 2]
             23  > select 3 C:[] P:[0, 1, 2, 3] -> C:[3] & newP:[1, 2]
             24  > > expand(C:[3],P:[1, 2]
             25  > > FAIL: |C| + |P| <= 3  C:[3] P:[1, 2]
             26  > FAIL: |C| + |P| <= 3  C:[] P:[0, 1, 2]
\end{verbatim}
\end{scriptsize}
\vspace{-3mm}
%
\centering
\includegraphics[height=7.2cm,width=9.2cm]{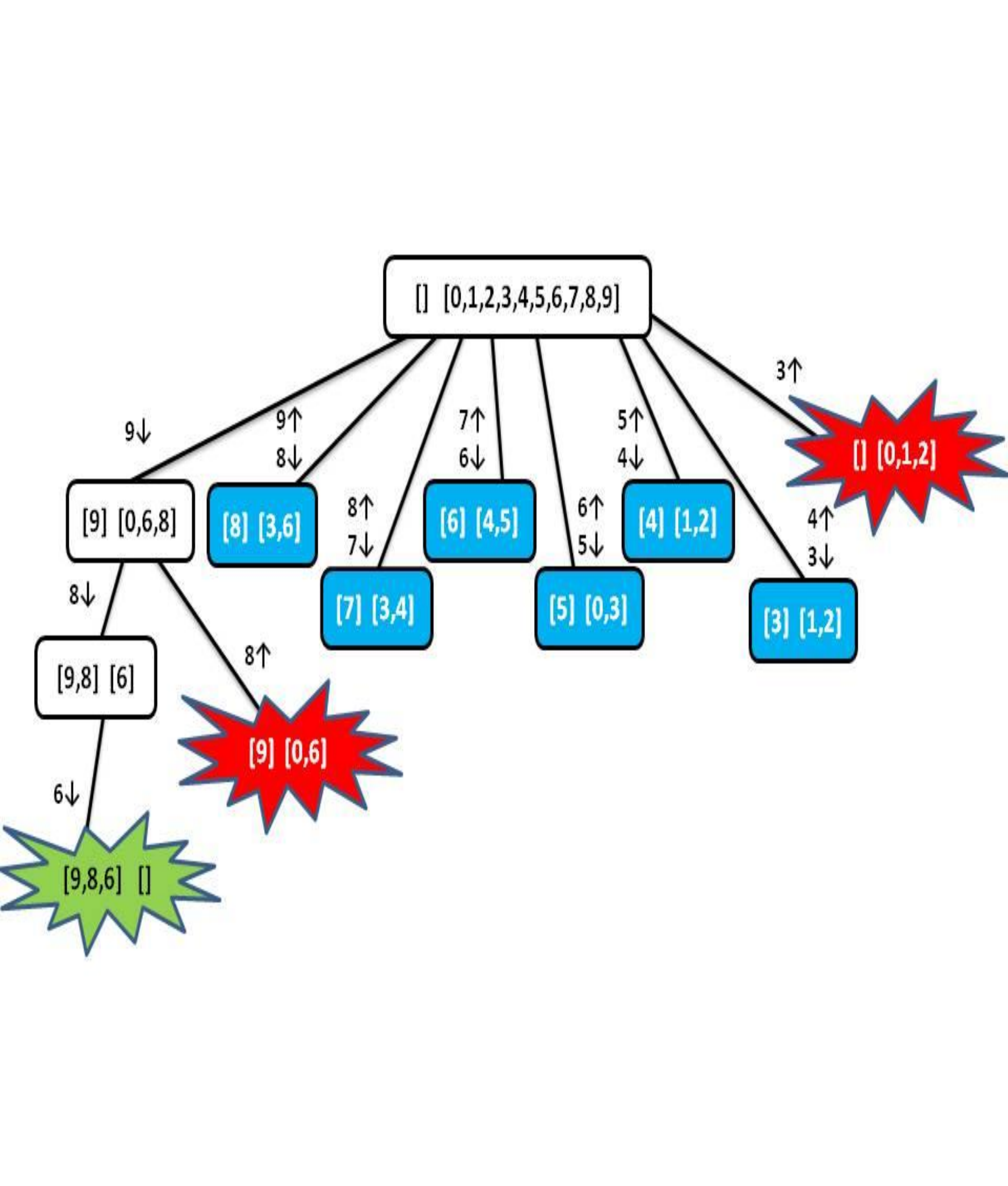}
\vspace{-15mm}
\caption{Cartoon, trace and backtrack-tree for MC on graph g10-50.}
\label{traceMC}
\end{figure}

We now present three views of the MC search process over a simple problem. The problem
is referred to as g10-50, and is a randomly generated graph with 10 vertices with edge probability 0.5. This is shown in
Figure \ref{traceMC} and has at top a cartoon of the search process and immediately below a trace of our program.
The program prints out the arguments $C$ and $P$ on each call to $expand$ (between lines 33 and 34), in line 37 if a FAIL occurs,
and between lines 38 and 39 when a vertex is selected. The indentation corresponds to the depth of recursion.
The \framebox{Line dd} boxes in the cartoon of Figure \ref{traceMC} corresponds
to the line numbers in the trace of Figure \ref{traceMC}, each of those a call to $expand$. Green coloured vertices are 
in $P$, blue vertices are those in $C$ and red vertices are those removed from $P$ and $C$ in lines 44 and 45 of 
Listing \ref{codeMC}. Also shown is the backtrack tree. The boxes correspond to calls to $expand$ and
contain $C$ and $P$. On arcs we have numbers with a down arrow $\downarrow$ if that vertex is added to $C$ and an up arrow
$\uparrow$ if that vertex is removed from $C$ and $P$. A clear white box is a call to $expand$ that is an \emph{interior} 
node of the backtrack tree
leading to further recursive calls or the creation of a new champion. The green shriek! is a champion clique and a red shriek! a
fail because $|C| + |P|$ was too small to unseat the champion. The blue boxes correspond to calls to $expand$ that fail first time on
entering the loop at line 36 of Listing \ref{codeMC}. By looking at the backtrack tree we get a feel for the nature of binomial search.

\subsection{MCQ}
\label{sec:mcqCode}
We now present Tomita's algorithm MCQ \cite{tomita2003} as Listings \ref{codeMCQa}, \ref{codeVertex}, \ref{codeMCRComparator} and \ref{codeMCQb}.
But first, a sketch of the algorithm.
MCQ is at heart an extension of MC, performing a binomial search, with two significant advances. First, the graph induced by the candidate set
is coloured using a greedy sequential colouring algorithm due to Welsh and Powell \cite{welshPowell}. This gives an upper bound
on the size of the clique in $P$ that can be added to $C$. Vertices in $P$ are then selected in decreasing colour order, that is $P$ is ordered
in non-decreasing colour order (highest colour last). And this is the second advance. Assume we select the $i^{th}$ entry in $P$ and call it $v$. 
We then know that we can colour
all the vertices in $P$ corresponding to the $0^{th}$ entry up to and including the $i^{th}$ entry using no more than the colour number
of $v$. Consequently that sub-graph can contain a clique no bigger than the colour number of $v$, and if this is too small to unseat the largest clique
search can be abandoned.

\subsubsection{MCQ in Java}
MCQ extends MC, Listing \ref{codeMCQa} line 3, and has an additional instance variable $colourClass$ (line 5) such that
$colourClass[i]$ is an ArrayList of integers (line 15) and will contain all the vertices 
of colour $i+1$ and is used when sorting vertices by their colour (lines 45 to 64).
At the top of search (method $search$, lines 12 to 21, Listing \ref{codeMCQa}) vertices are sorted (call to $orderVertices(P)$
at line 19) into some order, and this is described later.

Method $expand$ (line 23 to 43) corresponds to Figure 2 in \cite{tomita2003}. The array
$colour$ is local to the method and holds the colour of the $i^{th}$ vertex in $P$.
The candidate set $P$ is then sorted in
non-decreasing colour order by the call to $numberSort$ in line 28, and $colour[i]$ is then the colour
of integer vertex $P.get(i)$. The search then begins in the loop at line 29. We first test to see if the 
combined size of the candidate set plus the colour of
vertex $v$ is sufficient to unseat the champion (the largest clique found so far)\footnote{This is the terminology used by 
Pablo San Segundo in a private email communication.} (line 30). If it is insufficient the search terminates. Note that the loop starts
at $m-1$, the position of the last element in $P$, and counts down to zero. The $i^{th}$ element of $P$ is selected and assigned to $v$.
As in MC we create a new candidate set $newP$, the set of vertices (integers) in $P$ that are adjacent to $v$ (lines 33 to 37).
We then test to see if $C$ is maximal (line 38) and if it unseats the champion. If the new candidate set is not empty we recurse 
(line 39). Regardless, $v$ is removed from $P$ and from $C$ (lines 40 and 41) as in (lines 21 and 22 of MC0).

Method $numberSort$ can be compared to Figure 3 in \cite{tomita2003}. $numberSort$ takes as arguments an ordered $ArrayList$ of integers 
$ColOrd$ corresponding to vertices to be coloured in that order, an $ArrayList$ of integers $P$ that will correspond to the coloured vertices
in non-decreasing colour order, and an array of integers $colour$ such that if $v = P.get(i)$ ($v$ is the $i^{th}$ vertex in $P$)
then the colour of $v$ is $colour[i]$. Lines 45 to 64 differs from Tomita's NUMBER-SORT method because we use
the additional arguments $ColOrd$ and the growing clique $C$ as this allows us to easily implement our next algorithm MCS. 

Rather than assign colours to vertices explicitly $numberSort$ places vertices into colour classes, i.e.
if a vertex is not adjacent to any of the vertices in $colourClass[i]$ then that vertex can be placed into
that class and given colour number $i+1$ ($i+1$ so that colours range from 1 upwards). The vertices can then be sorted into
colour order via a pigeonhole sort, where colour classes are the pigeonholes.

$numberSort$ starts by clearing out the colour classes that might be used (line 48). In lines 49 to 55 vertices are selected
from $ColOrd$ and placed into the first colour class in which there are no conflicts, i.e. a class in which the vertex is not adjacent to any 
other vertex in that class (lines 51 to 53, and method $conflicts$). Variable $colours$ records the number of colours used. Lines
56 to 63 is in fact a pigeonhole sort, starting by clearing $P$ and then iterating over the 
colour classes (loop start at line 58) and in each colour class adding those vertices into $P$ (lines 59 to 63). The boolean method
$conflicts$, lines 66 to 72, takes a vertex $v$ and an ArrayList of vertices $colourClass$ where vertices in $colourClass$
are not pair-wise adjacent and have the same colour i.e. the vertices are an independent set. If vertex $v$ is adjacent to 
any vertex in $colourClass$ the method returns true (lines 67 to 70) otherwise false. Note that if vertex $v$ 
needs to be added into a new colour class in $numberSort$ the size of that $colourClass$ will be zero, the for loop
of lines 67 to 70 will not be performed and $conflicts$ returns true. The complexity of $numberSort$ is quadratic in the size of $P$.

\begin{figure}
\lstset{caption={MCQ (part 1), Tomita 2003},label=codeMCQa}
\begin{lstlisting}
import java.util.*;

class MCQ extends MC {

    ArrayList[] colourClass;

    MCQ (int n,int[][]A,int[] degree,int style) {
	super(n,A,degree);
	this.style = style;
    }
    
    void search(){
	cpuTime                              = System.currentTimeMillis();
	nodes                                = 0;
	colourClass                          = new ArrayList[n];
	ArrayList<Integer> C                 = new ArrayList<Integer>(n);
	ArrayList<Integer> P                 = new ArrayList<Integer>(n);
	for (int i=0;i<n;i++) colourClass[i] = new ArrayList<Integer>(n);
	orderVertices(P);
	expand(C,P);
    }

    void expand(ArrayList<Integer> C,ArrayList<Integer> P){
	if (timeLimit > 0 && System.currentTimeMillis() - cpuTime >= timeLimit) return;
	nodes++;
	int m = P.size();
	int[] colour = new int[m];
	numberSort(P,P,colour);
	for (int i=m-1;i>=0;i--){
	    if (C.size() + colour[i] <= maxSize) return;
	    int v = P.get(i);
	    C.add(v);
	    ArrayList<Integer> newP = new ArrayList<Integer>(i);
	    for (int j=0;j<=i;j++){
		int u = P.get(j);
		if (A[u][v] == 1) newP.add(u);
	    }
	    if (newP.isEmpty() && C.size() > maxSize) saveSolution(C);
	    if (!newP.isEmpty()) expand(C,newP);
	    C.remove(C.size()-1);
	    P.remove(i);
	}
    }

    void numberSort(ArrayList<Integer> C,ArrayList<Integer> ColOrd,ArrayList<Integer> P,int[] colour){
	int colours = 0;
	int m = ColOrd.size();
	for (int i=0;i<m;i++) colourClass[i].clear();
	for (int i=0;i<m;i++){
	    int v = ColOrd.get(i);
	    int k = 0;
	    while (conflicts(v,colourClass[k])) k++;
	    colourClass[k].add(v);
	    colours = Math.max(colours,k+1);
	}
	P.clear();
	int i = 0;
	for (int k=0;k<colours;k++)
	    for (int j=0;j<colourClass[k].size();j++){
		int v = (Integer)(colourClass[k].get(j));
		P.add(v); 
		colour[i++] = k+1;
	    }
    }

    boolean conflicts(int v,ArrayList<Integer> colourClass){
	for (int i=0;i<colourClass.size();i++){
	    int w = colourClass.get(i);
	    if (A[v][w] == 1) return true;
	}
	return false;
    }
\end{lstlisting}
\end{figure}

Vertices need to be sorted at the top of search, line 19. To do this we use the class Vertex in Listing \ref{codeVertex}
and the comparator MCRComparator in Listing \ref{codeMCRComparator}.
If $i$ is a vertex in $P$ then the corresponding Vertex $v$ has an integer $index$ equal to $i$. The Vertex also has attributes
$degree$ and $nebDeg$. $degree$ is the degree of the vertex $index$ and $nebDegree$ is the sum of the degrees of the vertices
in the neighbourhood of vertex $index$. Given an array
$V$ of class Vertex this can be sorted using Java's $Arrays.sort(V)$ method in $O(n.\log(n))$ time, and is ordered 
by default using the
$compareTo$ method in class Vertex. Our method forces a strict ordering of $V$ by non-increasing degree, tie-breaking on $index$. And
this is one step we take to ensure reproducibility of results. If we allowed the $compareToMethod$ to deliver $0$ when
two vertices have the same degree then $Arrays.sort$ would then break ties. If the sort method was unstable, i.e. did
not maintain the relative order of objects with equal keys \cite{qsort}, results may be unpredictable. 

\begin{figure}
\lstset{caption={Vertex},label=codeVertex}
\begin{lstlisting}
import java.util.*;

public class Vertex implements Comparable<Vertex> {

    int index, degree, nebDeg;

    public Vertex (int index,int degree) {
	this.index  = index;
	this.degree = degree;
	nebDeg      = 0;
    }

    public int compareTo(Vertex v){
	if (degree < v.degree || degree == v.degree && index > v.index) return 1;
	return -1;
    }
}
\end{lstlisting}
\end{figure}

The class MCRComparator (Listing \ref{codeMCRComparator}) allows us to sort vertices by non-increasing degree, tie breaking on
the sum of the neighbourhood degree $nebDeg$ and then on $index$, giving again a strict order. This is the MCR order given
in \cite{tomita2007}, where MCQ uses the simple degree ordering and MCR is MCQ with tie-breaking on neighbourhood degree. 

\begin{figure}
\lstset{caption={MCRComparator},label=codeMCRComparator}
\begin{lstlisting}
import java.util.*;

public class MCRComparator implements Comparator {

    public int compare(Object o1, Object o2){
	Vertex u = (Vertex) o1;
	Vertex v = (Vertex) o2;
	if (u.degree < v.degree || 
	    u.degree == v.degree && u.nebDeg < v.nebDeg ||
	    u.degree == v.degree && u.nebDeg == v.nebDeg && u.index > v.index) return 1;
	return -1;
    }
}
\end{lstlisting}
\end{figure}

Vertices can also be sorted into a minimum-width order via method $minWidthOrder$ of lines 86 to 98 of Listing \ref{codeMCQb}.
The minimum width order (mwo) was proposed by Freuder \cite{freuder} and also by Matula and Beck \cite{matulaBeck} where it was called 
``smallest last'', and more recently in \cite{eppstein2011} as a \emph{degeneracy ordering}.
The method $minWidthOrder$, lines 86 to 98 of Listing \ref{codeMCQb} sorts the array $V$ of Vertex into a mwo.
The vertices of $V$ are copied into an ArrayList $L$ (lines 87 and 89). The while loop starting at line 90 selects the vertex
in $L$ with smallest degree (lines 91 and 92) and calls it $v$. Vertex $v$ is pushed onto the stack $S$ and removed from $L$ (line
93) and all vertices in $L$ that are adjacent to $v$ have their degree reduced (line 94). On termination
of the while loop vertices are popped off the stack and placed back into $V$ giving a minimum width (smallest last) ordering.

\begin{figure}
\lstset{caption={MCQ (part 2), Tomita 2003},label=codeMCQb,firstnumber=73}
\begin{lstlisting}
    
    void orderVertices(ArrayList<Integer> ColOrd){
	Vertex[] V = new Vertex[n];
	for (int i=0;i<n;i++) V[i] = new Vertex(i,degree[i]);
	for (int i=0;i<n;i++)
	    for (int j=0;j<n;j++) 
		if (A[i][j] == 1) V[i].nebDeg = V[i].nebDeg + degree[j];
	if (style == 1) Arrays.sort(V);
	if (style == 2) minWidthOrder(V);
	if (style == 3) Arrays.sort(V,new MCRComparator());
	for (Vertex v : V) ColOrd.add(v.index);
    }

     void minWidthOrder(Vertex[] V){
	ArrayList<Vertex> L = new ArrayList<Vertex>(n);
	Stack<Vertex> S = new Stack<Vertex>();
	for (Vertex v : V) L.add(v);
	while (!L.isEmpty()){
	    Vertex v = L.get(0);
	    for (Vertex u : L) if (u.degree < v.degree) v = u;
	    S.push(v); L.remove(v);
	    for (Vertex u : L) if (A[u.index][v.index] == 1) u.degree--;
	}
	int k = 0;
	while (!S.isEmpty()) V[k++] = S.pop();
    }
}
\end{lstlisting}
\end{figure}

Method $orderVertices$ (Listing \ref{codeMCQb} lines 74 to 84) is then called once, at the top of search. The array
of Vertex $V$ is created for sorting in lines 75 and 76, and the sum of the neighbourhood degrees is 
computed in lines 77 to 79. $ColOrd$ is then sorted in one of three orders: $style == 1$ in non-increasing degree order,
$style == 2$ in minimum width order, $style == 3$ in non-increasing degree tie-breaking on sum of neighbourhood degree.
MCQ then uses the ordered candidate set $P$ for colouring, initially in one of the initial orders, thereafter in the 
order resulting from $numberSort$ and that is non-decreasing colour order. In \cite{tomita2003} it is claimed that this is
an improving order (however, no evidence was presented for this claim). In \cite{tomita2007} Tomita proposes a new algorithm,
MCR, where MCR is MCQ with a different initial ordering of vertices. Here MCR is MCQ with $style = 3$.

\subsubsection{A trace of MCQ}
\begin{figure}
\centering
\includegraphics[height=7.2cm,width=9.2cm]{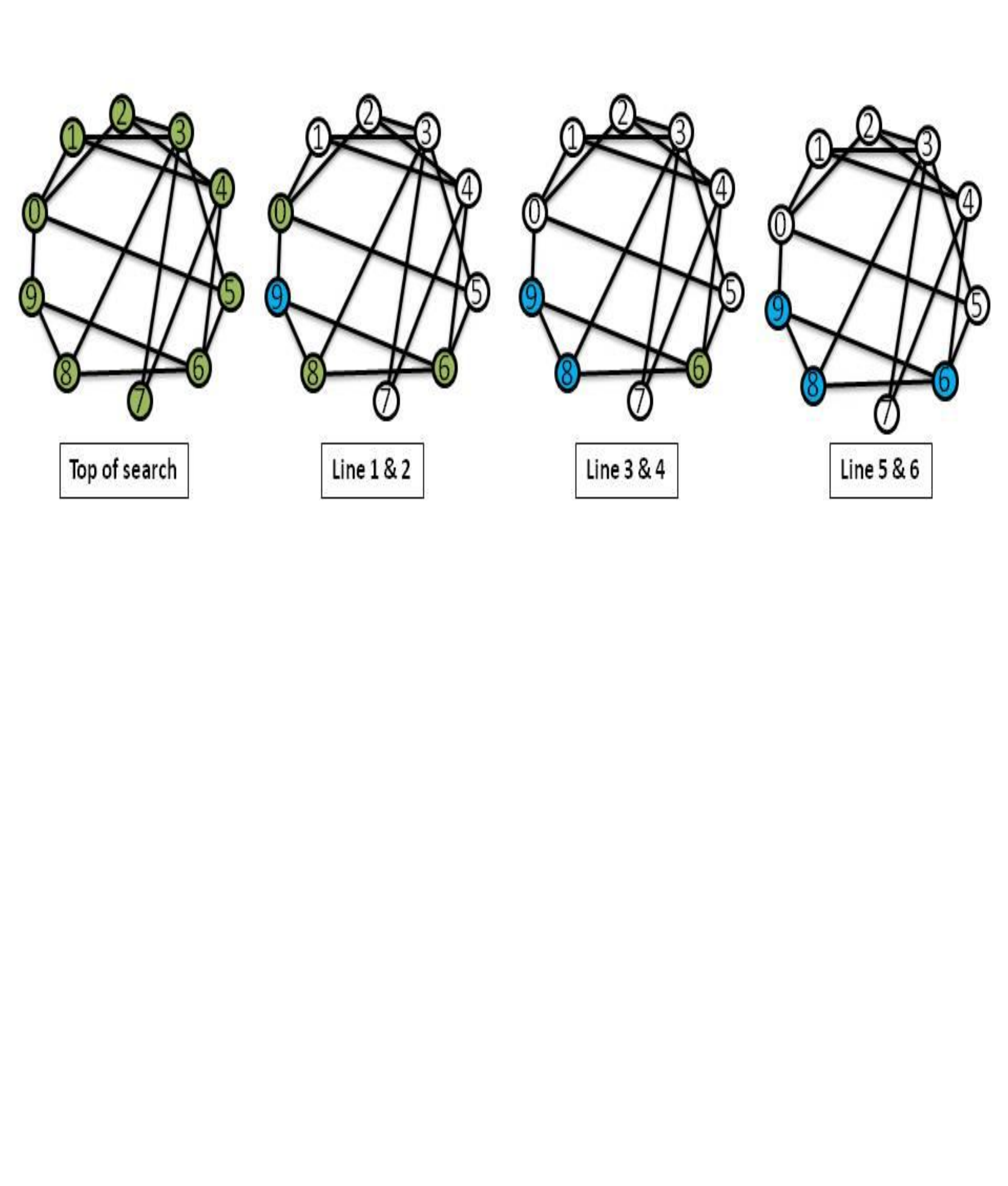}
\vspace{-40mm}

\begin{scriptsize}
\begin{verbatim}

                0 <3,5> <0,4> <4,4> <6,4> <1,3> <2,3> <5,3> <8,3> <9,3> <7,2> 
                1 > expand(C:[],P:[3, 0, 4, 6, 1, 2, 5, 8, 9, 7]
                2 > select 9 C:[] P:[3, 0, 4, 6, 1, 2, 7, 5, 8, 9] -> C:[9] & newP:[0, 6, 8]
                3 > > expand(C:[9],P:[0, 6, 8]
                4 > > select 8 C:[9] P:[0, 6, 8] -> C:[9, 8] & newP:[6]
                5 > > > expand(C:[9, 8],P:[6]
                6 > > > select 6 C:[9, 8] P:[6] -> SAVE: [9, 8, 6]
                7 > > > reject 6 C:[9, 8, 6] P:[6]
                8 > > reject 8 C:[9, 8] P:[0, 6, 8]
                9 > > select 6 C:[9] P:[0, 6] -> FAIL: vertex 6 colour too small (colour = 1)
               10 > reject 9 C:[9] P:[3, 0, 4, 6, 1, 2, 7, 5, 8, 9]
               11 > select 8 C:[] P:[3, 0, 4, 6, 1, 2, 7, 5, 8] -> FAIL: vertex 8 colour too small (colour = 3)

\end{verbatim}
\end{scriptsize}
\vspace{-5mm}
\caption{Trace of MCQ1 on graph g10-50.}
\label{traceMCQ}
\end{figure}

Figure \ref{traceMCQ} shows a cartoon and trace of MCQ over graph g10-50. Print statements were placed immediately after the call to 
$expand$ (Listing \ref{codeMCQa} line 24), after the selection of a vertex $v$ (line 31) and just before $v$ is rejected from
$P$ and $C$ (line 40). Each picture in the cartoon gives the corresponding line numbers in the trace immediately below.
Line 0 of the trace is a print out of the ordered array $V$ just after line 83 in method $orderVertices$ in
Listing \ref{codeMCQb}. This shows for each vertex the pair $<index,degree>$: the first call to 
$expand$ has $P = \{3,0,4,6,1,2,5,8,9,7\}$, i.e. non-decreasing degree order. MCQ makes 3 calls to $expand$
whereas MC makes 9 calls, and the MCQ colour bound
cut off in line 30 of Listing \ref{codeMCQa} is satisfied twice (listing lines 9 and 11).

\subsubsection{Observations on MCQ}
We noted above that MC can make recursive calls that immediately fail. Can this happen in MCQ? Looking at 
lines 39 of Listing \ref{codeMCQa}, $|C| + |newP|$ must be greater than $maxSize$. Since the colour of the vertex selected
$colour[i]$ was sufficient to satisfy the condition of line 30 it must be that integer vertex $v$ (line 31) is
adjacent to at least $colour[i]$ vertices in $P$ and thus in $newP$, therefore the next recursive call will not immediately fail.
Consequently each call to $expand$ corresponds to an internal node in the backtrack tree.

We also see again exactly what is measured as cpu time: it includes the creation of our data structures, the reordering
of vertices at the top of search and all recursive calls to $expand$ (lines 12 to 20).

Why is $colourClass$ an ArrayList$[]$ rather than an ArrayList$<$ArrayList$<$Integer$>>$ That would have done
away with the explicit cast in line 60. When using an ArrayList$<$ArrayList$<$Integer$>>$ we can do away with the cast
but Java generates an implicit cast, so nothing is gained.

At the top of MCQ's search Tomita sorts vertices in non-increasing degree order
and the first $\Delta$ vertices are given colours 1 to $\Delta$ respectively, where $\Delta$ is the maximum degree in the graph, thereafter
vertices are given colour $\Delta + 1$. This is done to \emph{prime} the candidate set for the initial call to EXPAND.
Thereafter Tomita calls NUMBER-SORT immediately before the recursive call to EXPAND. A simpler option is taken here:
colouring and sorting of the candidate set is done at the start of $expand$ (call to $numberSort$ at line 28). This strategy 
is adopted in all the algorithms here.

\subsection{MCS}
\label{sec:mcsCode}
Tomita's MCR \cite{tomita2007} is MCQ with a richer initial ordering, essentially 
non-decreasing degree with tie-breaking on the sum of the degrees of adjacent vertices. This ordering
is then modified during search via the colouring routine $numberSort$ as in MCQ. MCR is
compared to MCQ in \cite{tomita2007} over 8 of the 66 instances of the DIMACS benchmarks \cite{DIMACS}
showing an improvement in MCR over MCQ. As previously stated, MCR is our MCQ with $style = 3$.

MCS \cite{tomita2010} is MCR with two further modifications. 
The first modification is that MCS uses 
``\emph{... an adjunct ordered set of vertices for approximate coloring}''. This is an ordered list of vertices to
be used in the sequential colouring, and was called $V_{a}$. This order is static, set at the top of search. 
Therefore, rather than use the order in the candidate set $P$ for colouring the vertices in $P$ the
vertices in $P$ are coloured in the order of vertices in $V_{a}$.

The second modification is to use a repair mechanism when
colouring vertices (this is called a Re-NUMBER in Figure 1 of \cite{tomita2010}). When colouring vertices
an attempt is made to reduce the colours used by performing exchanges between vertices in
different colour classes. This is similar to the Sequential-X algorithm of Maffray and Preissmann
\cite{maffray1999}, i.e. a \emph{bi-chromatic exchange} is performed between a subset of vertices
in a pair of colour classes indifferent to a given vertex. In \cite{tomita2010} a recolouring of a vertex
$v$ occurs when a new colour class is about to be opened for $v$ and that colour class exceeds the search bound, i.e.
if the number of colours can be reduced this could result in search being cut off. 
In the context of colouring I will say that vertex $u$ and $v$ \emph{conflict} if they are adjacent, and that $v$ \emph{conflicts} with a colour 
class $C$ if there exists a vertex $u \in C$ that is in \emph{conflict} with $v$.
Assume vertex $v$ is in colour class $C_k$. If there exists a lower colour class $C_i$ ($i < k-1$) and $v$ conflicts only with a 
single vertex $w \in C_{i}$ and
there also exists a colour class $C_j$, where $i < j < k$, and $w$ does not conflict with any vertex in $C_j$ then we can
place $v$ in $C_i$ and $w$ in $C_j$, freeing up colour class $C_k$. This is given in Figure 1 of \cite{tomita2010} and 
the procedure is named Re-NUMBER. 

\begin{figure}
\centering
\includegraphics[height=7.2cm,width=9.2cm]{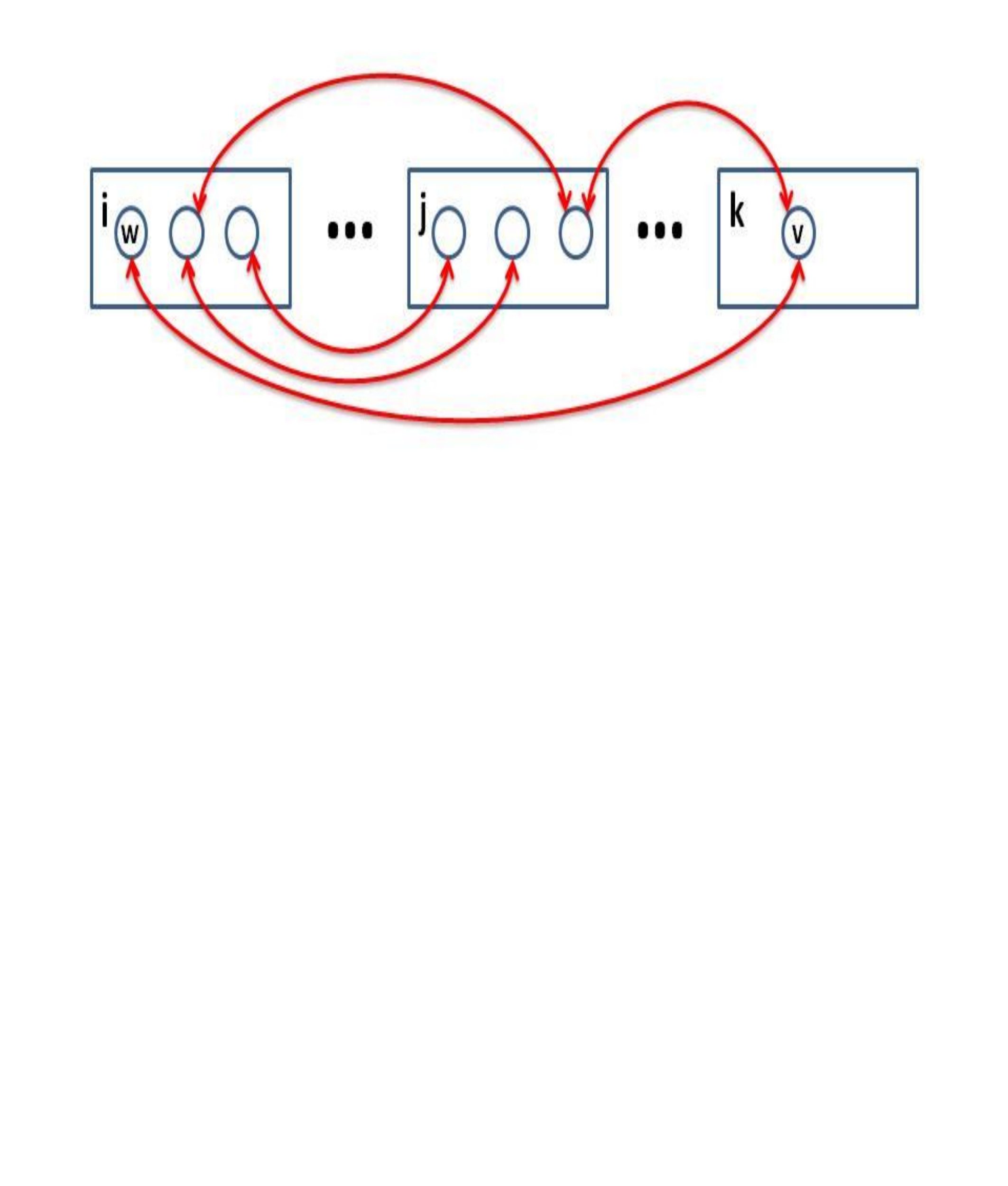}
\vspace{-45mm}
\caption{A repair scenario with colour classes i, j and k..}
\label{repair}
\end{figure}

Figure \ref{repair} illustrates this procedure. The boxes correspond to colour classes $i$, $j$ and $k$ where $i < j < k$. The circles
correspond to vertices in that colour class and the red arrowed lines as conflicts between pairs of vertices. Vertex $v$ has just been
added to colour class $k$, $v$ conflicts only with $w$ in colour class $i$ and $w$ has no conflicts in colour class $j$. We can then move 
$w$ to colour class $j$ and $v$ to colour class $i$.

\vspace{-5mm}

Experiments were then presented in \cite{tomita2010}
comparing MCR against MCS in which MCS is always the champion. But it is not clear where the advantage of
MCS comes from: does it come from the static colour order (the ``adjunct ordered set'') or does it 
come from the colour repair mechanism? 

I now present two versions of MCS. The first I call MCSa and this uses the static colouring order. The second, MCSb,
uses the static colouring ordering \emph{and} the colour repair mechanism (so MCSb \emph{is} Tomita's MCS). 
Consequently we will be able to determine where the
improvement in MCS comes from: static colour ordering or colour repair.

\subsubsection{MCSa in Java}

\begin{figure}
\lstset{caption={MCSa, Tomita 2010},label=codeMCSa,firstnumber=1}
\begin{lstlisting}
import java.util.*;

class MCSa extends MCQ {

    MCSa (int n,int[][]A,int[] degree,int style) {
	super(n,A,degree,style);
    }

    void search(){
	cpuTime                              = System.currentTimeMillis();
	nodes                                = 0;
	colourClass                          = new ArrayList[n];
	ArrayList<Integer> C                 = new ArrayList<Integer>(n);
	ArrayList<Integer> P                 = new ArrayList<Integer>(n);
	ArrayList<Integer> ColOrd            = new ArrayList<Integer>(n);
	for (int i=0;i<n;i++) colourClass[i] = new ArrayList<Integer>(n);
	orderVertices(ColOrd);
	expand(C,P,ColOrd);
    }

    void expand(ArrayList<Integer> C,ArrayList<Integer> P,ArrayList<Integer> ColOrd){
	if (timeLimit > 0 && System.currentTimeMillis() - cpuTime >= timeLimit) return;
	nodes++;
	int m = ColOrd.size();
	int[] colour = new int[m];
	numberSort(C,ColOrd,P,colour);
	for (int i=m-1;i>=0;i--){
	    int v = P.get(i);
	    if (C.size() + colour[i] <= maxSize) return;
	    C.add(v);
	    ArrayList<Integer> newP      = new ArrayList<Integer>(i);
	    ArrayList<Integer> newColOrd = new ArrayList<Integer>(i);
	    for (int j=0;j<=i;j++){
		int u = P.get(j);
		if (A[u][v] == 1) newP.add(u);
		int w = ColOrd.get(j);
		if (A[v][w] == 1) newColOrd.add(w);
	    }
	    if (newP.isEmpty() && C.size() > maxSize) saveSolution(C);
	    if (!newP.isEmpty()) expand(C,newP,newColOrd);
	    C.remove(C.size()-1);
	    P.remove(i);
	    ColOrd.remove((Integer)v);
	}
    }
}
\end{lstlisting}
\end{figure}

In Listing \ref{codeMCSa} we present MCSa as an extension to MCQ. Method $search$ creates an explicit colour ordering $ColOrd$ and the $expand$
method is called with this in line 18 (compare this to line 20 of $MCQ$). Method $expand$ now takes
three arguments: the growing clique $C$, the candidate set $P$ and the colouring order $ColOrd$. In line 26 $numberSort$
is called using $ColOrd$ (compare to line 28 in MCQ) and lines 27 to 45 are essentially the same as lines 29 to 42
in MCQ with the exception that $ColOrd$ must also be copied and updated (line 32, 36 and 37) prior to the recursive call to $expand$ (line 40)
and then down-dated after the recursive call (line 43). Therefore MCSa is a simple extension of MCQ and like MCQ has
three styles of ordering.

\subsubsection{MCSb in Java}

\begin{figure}
\lstset{caption={MCSb, Tomita 2010},label=codeMCSb,firstnumber=1}
\begin{lstlisting}
import java.util.*;

class MCSb extends MCSa {

    MCSb (int n,int[][]A,int[] degree,int style) {
	super(n,A,degree,style);
    }

    void numberSort(ArrayList<Integer> C,ArrayList<Integer> ColOrd,ArrayList<Integer> P,int[] colour){
	int delta   = maxSize - C.size();
	int colours = 0;
	int m       = ColOrd.size();
	for (int i=0;i<m;i++) colourClass[i].clear();
	for (int i=0;i<m;i++){
	    int v = ColOrd.get(i);
	    int k = 0;
	    while (conflicts(v,colourClass[k])) k++;
	    colourClass[k].add(v);
	    colours = Math.max(colours,k+1);
	    if (k+1 > delta && colourClass[k].size() == 1 && repair(v,k)) colours--;
	}
	P.clear();
	int i = 0;
	for (int k=0;k<colours;k++)
	    for (int j=0;j<colourClass[k].size();j++){
		int v = (Integer)(colourClass[k].get(j));
		P.add(v); 
		colour[i++] = k+1;
	    }
    }

    int getSingleConflictVariable(int v,ArrayList<Integer> colourClass){
	int conflictVar = -1;
	int count       = 0;
	for (int i=0;i<colourClass.size() && count<2;i++){
	    int w = colourClass.get(i);
	    if (A[v][w] == 1){conflictVar = w; count++;}
	}
	if (count > 1) return -count;
	return conflictVar;
    }

    boolean repair(int v, int k){
	for (int i=0;i<k-1;i++){
	    int w = getSingleConflictVariable(v,colourClass[i]);
	    if (w >= 0)
		for (int j=i+1;j<k;j++)
		    if (!conflicts(w,colourClass[j])){
			colourClass[k].remove((Integer)v);
			colourClass[i].remove((Integer)w);
			colourClass[i].add(v);
			colourClass[j].add(w);
			return true;
		    }
	}
	return false;
    }
}
\end{lstlisting}
\end{figure}

In Listing \ref{codeMCSb} we present MCSb as an extension to MCSa: the difference between MCSb and MCSa is in $numberSort$,
with the addition of lines 10 and  20. At line 10 we compute $delta$ as the minimum number of colour classes required to match 
the search bound. At line 20, if we have exceeded the number of colour classes required to exceed the search bound and
a new colour class $k$ has been opened for vertex $v$ and we can repair the colouring such that one less colour class is used
we can decrement the number of colours used. This repair is done in the boolean method $repair$ of lines 43 to 57.  The $repair$ method returns
true if vertex $v$ in colour class $k$ can be recoloured into a lower colour class, false otherwise, and can be compared to Tomita's Re-NUMBER
procedure. We search for a colour class $i$, where $i < k-1$, in which there exists only one vertex in conflict with $v$ and we call this
$w$ (line 45). The method $getSingleConflictVariable$, lines 32 to 41, takes as arguments a vertex $v$ and a colour class 
and counts and records the conflict vertex. If this exceeds 1 we return a negative integer otherwise we deliver the single 
conflicting vertex. The $repair$ method then proceeds at line 46 if a single conflicting vertex $w$ was found, searching
for a colour class $j$ \emph{above} $i$ (for loop of line 47) in which there are no conflicts with  $w$. If that was found (line 48)
vertex $v$ is removed from colour class $k$, $w$ is removed from colour class $i$, $v$ is added to colour class $i$ and $w$ to colour class $j$ 
(lines 49 to 52) and $repair$ delivers $true$ (line 53). Otherwise, no repair occurred (line 56).

\subsubsection{Observations on MCS}
Tomita did not investigate where MCS's improvement comes from and neither did \cite{carmoZuge},
coding up MCS in Python in one piece. We can also \emph{tune} MCS. In MCSb we repair colourings when we open a new colour class
that exceeds the search bound. We could instead repair unconditionally every time we open a new colour class, attempting to maintain a compact
colouring. We do not investigate this here.

\subsection{BBMC}
\label{sec:bbmc}
San Segundo's BB-MaxClique algorithm \cite{segundo2011} (BBMC) 
is similar to the earlier algorithms in that vertices are selected from the candidate set to add to the current clique
in non-increasing colour order, with a colour cut-off within a binomial search. BBMC is at heart a bit-set encoding of MCSa with
the following features.
\begin{enumerate}
\item The ``BB'' in ``BB-MaxClique'' is for ``Bit Board''. Sets are represented using bit strings.
\item BBMC colours the candidate set using a static sequential ordering, the ordering set at the top of search, the same as MCSa.
\item BBMC represents the neighbourhood of a vertex and its inverse neighbourhood as bit strings, rather than use
an adjacency matrix and its compliment
\item When colouring takes place a colour class perspective is taken, determining what vertices can be placed in a colour class together, 
before moving on to the next colour class. Other algorithms (e.g \cite{tomita2003,tomita2010}) takes a vertex perspective, deciding on the colour 
of a vertex.
\end{enumerate}

\noindent
The candidate set $P$ is encoded as a bit string, as is
the currently growing clique $C$. For a given vertex $v$ we also have its neighbourhood $N[v]$, again encoded as a bit-string, and
its inverse neighbourhood $invN[v]$. That is, $invN[v]$ defines the set of vertices that are \emph{not} adjacent to $v$ ($invN[v]$ is the
compliment of $N[v]$) and this is
used in colouring. When a vertex $v$ is selected from the candidate set $P$ to be added to $C$ a new candidate set 
is produced $newP$, where $newP$ is the set of vertices in the
candidate set $P$ that are adjacent to $v$ i.e. $newP = P \land N[v]$. 
For the set elements that reside in word boundaries this operation is fast.

BBMC takes a ``colour-class'' perspective. $BBColour$ starts off building the first colour class, i.e. the first 
set of vertices in $P$ that form an independent set. It then finds the next colour class, and so on until $P$ is exhausted.
Given a set of vertices $Q$, when a vertex $v$ is selected and removed from $Q$ and added to a colour class, 
$Q$ becomes the set of vertices that are 
in $Q$ but not adjacent to $v$. That is $Q = Q \land invN[v]$.
Colour classes are then combined using a pigeonhole sort delivering a list of vertices in non-decreasing colour order and
this is then used in the $BBMaxClique$ method (the BBMC equivalent of $expand$) to cut off search as in MCQ and MCSa. 

MCSa colours vertices in a static order. This is achieved in BBMC by a renaming of the vertices, and this is done by
re-ordering the adjacency matrix at the top of search.

\subsubsection{BBMC in Java}
We implement sets using Java's BitSet class, therefore from now on we refer to $P$ as \emph{the candidate BitSet} and
an ordered array of integers $U$ as the \emph{ordered candidate set}.
In Listing \ref{codeBBMCa}, lines 5 to 7, we have the an array of BitSet $N$ for representing neighbourhoods,
$invN$ as the inverse neighbourhoods (the compliment of $N$) and $V$ an array of Vertex.
$N[i]$ is then a BitSet representing the neigbourhood of the $i^{th}$ vertex in the array $V$, and $invN$ as its compliment.
The array $V$ is used at the top of search for renaming vertices (and we discuss this later).

The $search$ method (lines 16 to 30) creates the candidate BitSet $P$, current clique (as a BitSet) $C$, and Vertex array $V$. 
The $orderVertices$ method
renames the vertices and will be discussed later. The method $BBMaxClique$ corresponds to the procedure in Figure 3 of \cite{segundo2011}
and can be compared to the $expand$ method in Listing \ref{codeMCSa}. In a BitSet we use $cardinality$ rather than $size$ (line 35, 40 and 44).
The integer array $U$ (same name as in \cite{segundo2011} is essentially the colour ordered candidate set such that
if $v = U[i]$ then $colour[i]$ corresponds to the colour given to $v$ and $colour[i] \leq colour[i+1]$. The method call
of line 38 colours the vertices and delivers those colours in the array $colour$ and the sorted candidate set in $U$. The
for loop, lines 39 to 47 (again, counting down from $m-1$ to zero), first tests to see if the colour cut off occurs (line 40) and if it 
does the method returns. Otherwise a new candidate BitSet is created, $newP$ on line 41, as a clone of $P$. The current vertex $v$ is 
then selected (line 42) and in line 43 $v$ is added to the growing clique $C$ and $newP$ becomes the BitSet corresponding to
the vertices in the candidate BitSet that are in the neighbourhood of $v$. The operation $newP.and(N[v])$ (line 43) is equivalent to the for loop
in lines 34 to 37 of Listing \ref{codeMCQa} of MCQ. If the current clique is both maximal and a maximum it is saved via BBMC's specialised
save method (described later) otherwise if $C$ is not maximal (i.e. $newP$ is not empty) a recursive call is made
to $BBMaxClique$. Regardless, $v$ is removed from the current candidate BitSet and the current clique (line 46) and the for loop continues.

\begin{figure}
\lstset{caption={San Segundo's BB-MaxClique in Java (part 1)},label=codeBBMCa}
\begin{lstlisting}
import java.util.*;

public class BBMC extends MCQ {

    BitSet[] N;    // neighbourhood
    BitSet[] invN; // inverse neighbourhood
    Vertex[] V;    // mapping bits to vertices

    BBMC (int n,int[][]A,int[] degree,int style) {
	super(n,A,degree,style);
	N    = new BitSet[n];
	invN = new BitSet[n];
	V    = new Vertex[n];
    }
    
    void search(){
	cpuTime  = System.currentTimeMillis();
	nodes    = 0;
	BitSet C = new BitSet(n);
	BitSet P = new BitSet(n);
	for (int i=0;i<n;i++){	  
	    N[i]    = new BitSet(n);
	    invN[i] = new BitSet(n);
	    V[i]    = new Vertex(i,degree[i]);
	}
	orderVertices();
	C.set(0,n,false);
	P.set(0,n,true);
	BBMaxClique(C,P);
    }
    
    void BBMaxClique(BitSet C,BitSet P){
	if (timeLimit > 0 && System.currentTimeMillis() - cpuTime >= timeLimit) return;
	nodes++;
	int m = P.cardinality();
	int[] U = new int[m];
	int[] colour = new int[m];
	BBColour(P,U,colour);
	for (int i=m-1;i>=0;i--){
	    if (colour[i] + C.cardinality() <= maxSize) return;
	    BitSet newP = (BitSet)P.clone();	    
	    int v = U[i];
	    C.set(v,true); newP.and(N[v]);
	    if (newP.isEmpty() && C.cardinality() > maxSize) saveSolution(C);
	    if (!newP.isEmpty()) BBMaxClique(C,newP);
	    P.set(v,false); C.set(v,false);
	}
    }
    
    void BBColour(BitSet P,int[] U,int[] colour){
	BitSet copyP    = (BitSet)P.clone();
	int colourClass = 0;
	int i           = 0;
       	while (copyP.cardinality() != 0){
	    colourClass++;
	    BitSet Q = (BitSet)copyP.clone();
	    while (Q.cardinality() != 0){
		int v = Q.nextSetBit(0);
		copyP.set(v,false);
		Q.set(v,false);
		Q.and(invN[v]);
		U[i] = v;     
		colour[i++] = colourClass;
	    }
	}
    }
\end{lstlisting}
\end{figure}

Method $BBColour$ corresponds to the procedure of the same name in Figure 2 of \cite{segundo2011} but differs
in that it does not explicitly represent colour classes and therefore does not require a pigeonhole sort as in San Segundo's description. Our
method takes the candidate BitSet $P$ (see line 38), ordered candidate set $U$ and array of $colour$ as parameters. 
Due to the nature of Java's BitSet the $and$ operation is not functional but actually modifies bits, consequently cloning is 
required (line 51) and we take a copy of $P$. In line 52 $colourClass$
records the current colour class, initially zero, and $i$ is used as a counter for adding coloured vertices into the array $U$.
The while loop, lines 54 to 64, builds up colour classes whilst consuming vertices in $copyP$. The BitSet $Q$ (line 56) is 
the candidate BitSet as we are about to start a new colour class. The while loop of
lines 57 to 64 builds a colour class: the first vertex in $Q$ is selected (line 58) and is removed
from the candidate BitSet $copyP$ (line 59) and BitSet $Q$ (line 60), $Q$ then becomes the set of vertices that 
are in the current candidate BitSet ($Q$) and in the inverse neighborhood of $v$ (line 61), i.e. $Q$ becomes the BitSet 
of vertices that can join the same colour class with $v$. We then add $v$ to the ordered candidate set $U$ (line 62),
record its colour and increment our counter (line 63). When $Q$ is exhausted (line 57) the outer while loop (line 54) starts a 
new colour class (lines 55 to 64).

Listing \ref{codeBBMCb} shows how the candidate BitSet is renamed/reordered. In fact it is not
the candidate BitSet that is reordered, rather it is the description of the neighbourhood $N$ and its inverse $invN$ that
is reordered. Again, as in MCQ and MCSa a Veretx array is
created (lines 69 to 73) and is sorted into one of three possible orders (lines 74 to 76). Once sorted,
a bit in position $i$ of the candidate BitSet $P$ corresponds to the integer vertex $v = V[i].index$.
The neighbourhood and its inverse are then reordered in the loop of lines 77 to 83. For all pairs $(i,j)$
we select the corresponding vertices $u$ and $v$ from $V$ (lines 79 and 80) and if they are adjacent then the $j^{th}$ bit of $N[i]$ is set
true, otherwise false (line 81). Similarly, the inverse neighbourhood is updated in line 82. The loop could be made
twice as fast by exploiting symmetries in the adjacency matrix $A$.
In any event, this method is called once, at the top of search and is generally an insignificant contribution to run time.

BBMC requires its own $saveSolution$ method (lines 86 to 90 of Listing \ref{codeBBMCb}) due to $C$ being a BitSet. Again the
solution is saved into the integer array $solution$ and again we need to use the Vertex array $V$ to map bits to vertices.
This is done in line 88: if the $i^{th}$ bit of $C$ is true then integer vertex $V[i].index$ is in the solution. This explains
why $V$ is global to the $BBMC$ class.

\begin{figure}
\lstset{caption={San Segundo's BB-MaxClique in Java (part 2)},label=codeBBMCb,firstnumber=67}
\begin{lstlisting}

    void orderVertices(){
	for (int i=0;i<n;i++){
	    V[i] = new Vertex(i,degree[i]);
	    for (int j=0;j<n;j++) 
		if (A[i][j] == 1) V[i].nebDeg = V[i].nebDeg + degree[j];
	}
        if (style == 1) Arrays.sort(V); 
	if (style == 2) minWidthOrder(V);    
	if (style == 3) Arrays.sort(V,new MCRComparator());
	for (int i=0;i<n;i++)
	    for (int j=0;j<n;j++){
		int u = V[i].index;
		int v = V[j].index;
		N[i].set(j,A[u][v] == 1);
		invN[i].set(j,A[u][v] == 0);
	    }
    }

    void saveSolution(BitSet C){
	Arrays.fill(solution,0);
	for (int i=0;i<C.size();i++) if (C.get(i)) solution[V[i].index] = 1;
	maxSize = C.cardinality();
    }
}
\end{lstlisting}
\end{figure}

\subsubsection{Observations on BBMC}
In our Java implementation we might expect a speed up if
we did away with the in-built BitSet and did our own bit-sting manipulations explicitly. 
It is also worth noting that in \cite{segundo2011b} comparisons are made with Tomita's results in \cite{tomita2010}
by re-scaling tabulated results, i.e. Tomita's code was not actually run. This is not unusual.

In \cite{segundo2011b} there is the bit-board version, $BB\_ReCol$ in Fig 1, of Tomita's Re-NUMBER.
I believe that version reported is flawed and can result in colour classes not being pair-wise disjoint, and that Fig. 1 should have a return
statement between lines 7 and 8, similar to the return statement in Re-NUMBER. As it stands
$BB\_ReCol$ can result in the candidate set becoming a multi-set resulting in redundant re-exploration of the search space with subsequent
poor performance.

\subsection{Summary of MCQ, MCR, MCS and BBMC}
\label{sec:algSummary}
\begin{figure}
\centering
\includegraphics[height=9.2cm,width=13.2cm]{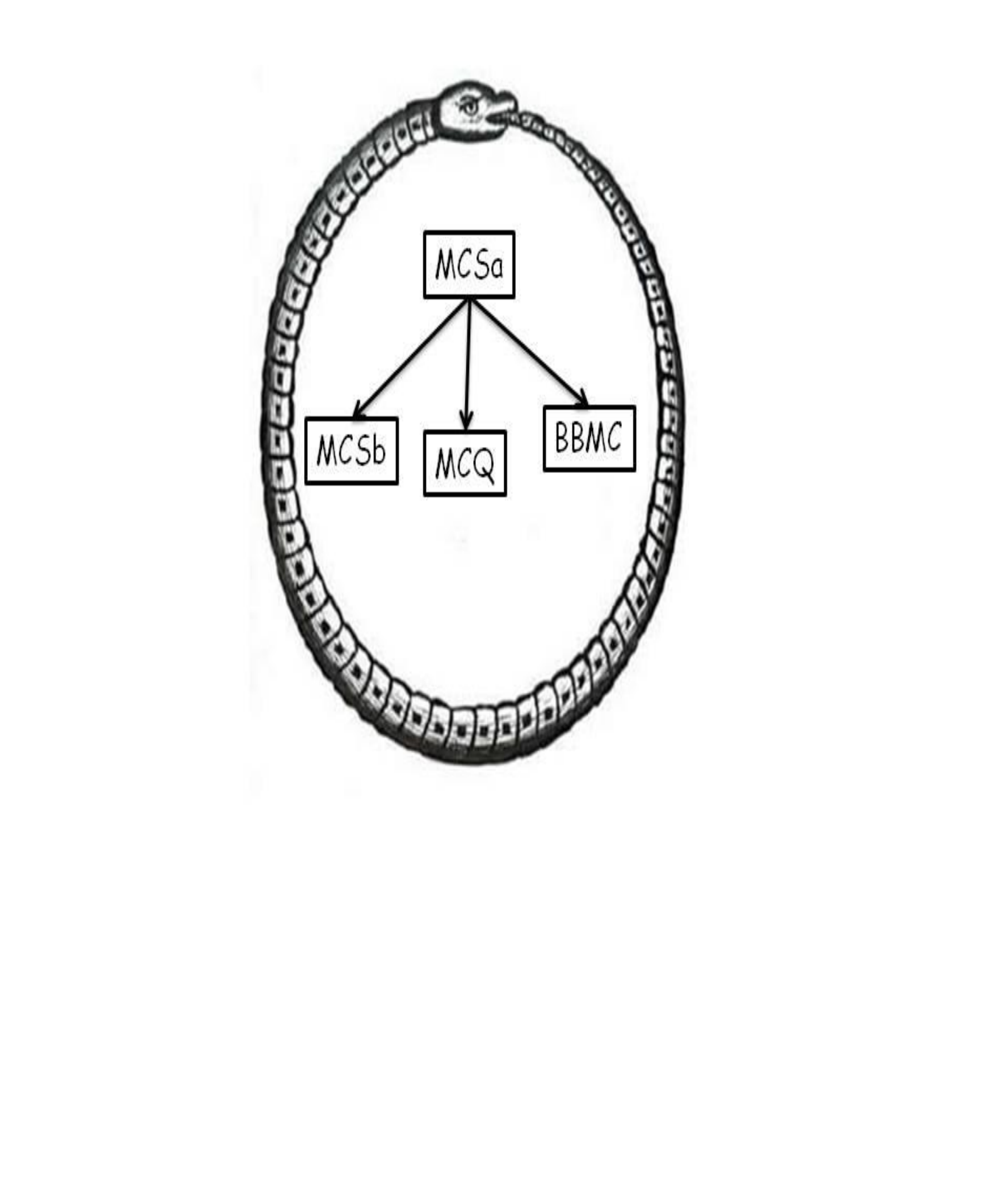}
\vspace{-30mm}
\caption{An alternative hierarchy of the algorithms.}
\label{uroborus}
\end{figure}

Putting aside the chronology \cite{tomita2003,tomita2007,tomita2010,segundo2011}, MCSa is the most general algorithm presented here. 
BBMC is in essence MCSa with a BitSet encoding of sets.
MCQ is MCSa except that we do away with the static colour ordering and allow MCQ to colour and sort the candidate set
using the candidate set, somewhat in the manner of Uroborus the serpent that eats itself. And MCSb is 
MCSa with an additional colour repair step. Therefore we might have an alternative hierarchy of the algorithms, such as the 
hierarchy of Figure \ref{uroborus} within the image of Uroborus.

\vspace{-7mm}

\section{Exact Algorithms for Maximum Clique: a brief history}
We now present a brief history of complete algorithms for the maximum clique problems, starting from 1990. The algorithms are presented
in chronological order. \\

\noindent
{\bf 1990:} In 1990 \cite{carraghanPardalos90} Carraghan and Pardalos present a branch and bound algorithm. Vertices are ordered
in non-decreasing degree order at each depth in the binomial search with a cut-off based on the size of the largest clique
found so far. Their algorithm is presented in Fortran 77 along with code to generate random graphs, consequently
their empirical results are entirely reproducible. Their algorithm is similar to MC (Listing \ref{codeMC}) but 
sorts the candidate set $P$ using current degree in each call to $expand$. \\

\noindent
{\bf 1992:}
In \cite{pardalosRodgers92}
Pardalos and Rodgers present a zero-one encoding of the problem where a vertex $v$ is represented by a variable
$x_{v}$ that takes the value 1 if search decides that $v$ is in the clique and 0 if it is rejected. Pruning
takes place via the constraint $\neg adjacent(u,v) \rightarrow x_{u} + x_{v} \leq 1$ (Rule 4).  
In addition, a candidate vertex adjacent to all vertices in 
the current clique is forced into the clique (Rule 5) and a vertices of degree too low to
contribute to the growing clique is rejected (Rule 7). The branch and
bound search selects variables dynamically based on current degree in the candidate set: a \emph{non-greedy}
selection chooses a vertex of lowest degree and \emph{greedy} selects highest degree. The computational results showed that
greedy was good for (easy) sparse graphs and non-greedy good for (hard) dense graphs.\\

\noindent
{\bf 1994:}
In \cite{maxClique} Pardalos and Xue reviewed algorithms for the enumeration problem (counting maximal cliques) 
and exact algorithms for the maximum clique problem. Although dated, it continues to be an excellent review. \\

\noindent
{\bf 1997:}
In \cite{wood97} graph colouring and fractional colouring is used to bound search. Comparing again to MC (Listing \ref{codeMC})
the candidate set is coloured greedily and if the size of the current clique plus the number of colours used is
less than or equal to the size of the largest clique found so far that branch of search is cut off. In \cite{wood97}
vertices are selected in non-increasing degree order, the opposite of that proposed by \cite{pardalosRodgers92}.
We can get a similar effect to \cite{wood97}
in MC if we allow free selection of vertices, colour $newP$ between lines 42 and 43 and make the recursive call to $expand$ in line 43 
conditional on the colour bound.\\

\noindent
{\bf 2002:}
Patric R. J. \"{O}sterg\aa{}rd proposed an algorithm that has a dynamic programming flavour \cite{prjo2002}. The search
process starts by finding the largest clique containing vertices drawn from the set $S_{n} = \{v_{n}\}$ and records it size in $c[n]$. Search
then proceeds to find the largest clique in the set $S_{i} = \{v_{i},v_{i+1},...,v_{n}\}$ using the value in $c[i+1]$ as a bound.
The vertices are ordered at the top of search in colour order, i.e. the vertices are coloured greedily and then ordered
in non-decreasing colour order, similar to that in $numberSort$ Listing \ref{codeMCQa}.
\"{O}sterg\aa{}rd's algorithm is available as Cliquer\footnote{Available from http://users.tkk.fi/pat/cliquer.html/}.
In the same year Torsten Fahle \cite{fahle} presented a simple algorithm (Algorithm 1) that is essentially MC but with a free selection of 
vertices rather than the fixed iteration in line 36 of Listing \ref{codeMC} and dynamic maintenance
of vertex degree in the candidate set. This is then enhanced (Algorithm 2) with forced accept and 
forced reject steps similar to Rules 4, 5 and 7 of \cite{pardalosRodgers92} and the algorithm is named \emph{DF} 
(\emph{D}omain \emph{F}iltering). DF is then enhanced to incorporate a colouring bound, similar to that in Wood \cite{wood97}.\\

\noindent
{\bf 2003:}
Jean-Charles R\'{e}gin proposed a constraint programming model for the maximum clique problem \cite{regin2003}. 
His model uses a matching in a duplicated graph to deliver a bound within search, a \emph{Not Set} as used in the Bron Kerbosch
enumeration Algorithm 457 \cite{bk73} and vertex selection using the pivoting strategy similar to that in 
\cite{bk73,akk73,tomita2006,eppstein2011}. That same year Tomita reported MCQ \cite{tomita2003}.\\

\noindent
{\bf 2007:}
Tomita proposed MCR \cite{tomita2007} and in the same year
Janez Konc and Du\u{s}anka Jane\u{z}i\u{c} proposed the \emph{MaxCliqueDyn} algorithm 
\cite{Konc_Janezic_2007}\footnote{Available from http://www.sicmm.org/$\sim$konc/}. The algorithm is essentially
MCQ \cite{tomita2003} with dynamic reordering of vertices in the candidate set, using current degree, prior to colouring. This reordering is
expensive and takes place high up in the backtrack tree and is controlled by a parameter $T_{limit}$. Varying this parameter
influences the cost of the search process and $T_{limit}$ must be tuned on an instance-by-instance basis.\\

\noindent
{\bf 2010:}
Pablo San Segundo and Crist{\'o}bal Tapia presented an early version of BBMC (BB-MCP) \cite{segundoT10} and
Tomita presented MCS \cite{tomita2010}. In the same year Li and Quan proposed new max-SAT encodings for maximum clique 
\cite{aaai2010,tai2010}. \\

\noindent
{\bf 2011:}
Pablo San Segundo proposed BBMC \cite{segundo2011} and in \cite{segundo2011b} a version of BBMC with the colour repair
steps from Tomita's MCS. \\

\noindent
{\bf 2012:}
Renato Carmo and Alexandre P. Z{\"u}ge \cite{carmoZuge} reported an empirical
study of 8 algorithms including those from \cite{carraghanPardalos90} and \cite{fahle}
along with MCQ, MCR (equivalent to MCQ3), MCS (equivalent to MCSb1) and MaxCliqueDyn. The claim is made that the Bron Kerbosch 
algorithm provides a unified framework for all the algorithms studied, although
a Not Set is not used, neither do they use pivoting as described in \cite{bk73,akk73,tomita2006,eppstein2011}. 
Their algorithms can be viewed as an iterative (non-recursive) version of MC. 
Referring to algorithm MaxClique(G) in \cite{carmoZuge}, the stack $S$ holds pairs (Q,K) where Q is the currently growing clique 
and K is the candidate set. In line 4 a pair (Q,K) is popped from $S$. The while loop lines 5 to 8 selects and removes a vertex v from K (line 6), 
pushes the pair (Q,K) onto $S$ (line 7), and then updates Q and K (line 8) such that Q has vertex v added to it and K becomes all vertices in 
K adjacent to v, i.e. K becomes the updated candidate set for updated Q. Therefore line 6 and 7 give a deferred iteration over the candidate 
set with v removed from K and v not added to Q. Therefore a pop in line 4 is equivalent to going once round the for loop in method $expand$ in MC.
All algorithms are coded in Python, therefore the study is \emph{objective} (the authors include none of their own algorithms) 
and \emph{fair} (all algorithms are coded by the authors and run in the same environment) and in that regard is both exceptional and laudable.

\section{The Computational Study}
\vspace{-1.5mm}
The computational study attempts to answer the following questions.
\begin{enumerate}
\item Where does the improvement in MCS come from? By comparing MCQ with MCSa we can
measure the contribution due to static colouring and by comparing MCSa with MCSb 
we can measure the contribution due to colour repair.
\item How much benefit can be had from the BitSet encoding? We compare MCSa with BBMC over a variety of problems.
\item We have three possible initial orderings (styles). Is any one of them better than the others and is this algorithm independent? 
\item The candidate set is ordered in non-decreasing colour order. Could a tie-breaking rule
influence performance?
\item Most papers use only random problems and the DIMACS benchmarks. What other problems might we use in our investigation?
\item Is it safe to recalibrate published results?
\end{enumerate}
Throughout our study we use a reference machine (named Cyprus), a machine with two Intel E5620 2.4GHz quad-core processors with 
48 GB of memory, running linux centos 5.3 and Java version 1.6.0\_07. 

\subsection{MCQ versus MCS: static ordering and colour repair}
Is MCS faster than MCQ, and if so why? Just to recap, MCSa is MCQ with a static colour ordering set at the top
of search and MCSb is MCSa with the colour repair mechanism. By comparing these algorithms we can determine
if indeed MCSb is faster than MCQ and where that gain comes from: the static colouring order or the colour repair.
We start our investigation with Erd\'{o}s-R\"{e}nyi random graphs $G(n,p)$ where $n$ is the number of vertices and
each edge is included in the graph with probability $p$ independent from every other edge. The code for generating
these graphs is given in Appendix 2. 

\begin{figure}
\vspace{-6cm}
\begin{center}
\hspace{-1.5cm}
\begin{minipage}[t]{0.49\textwidth}
\includegraphics[height=11.5cm]{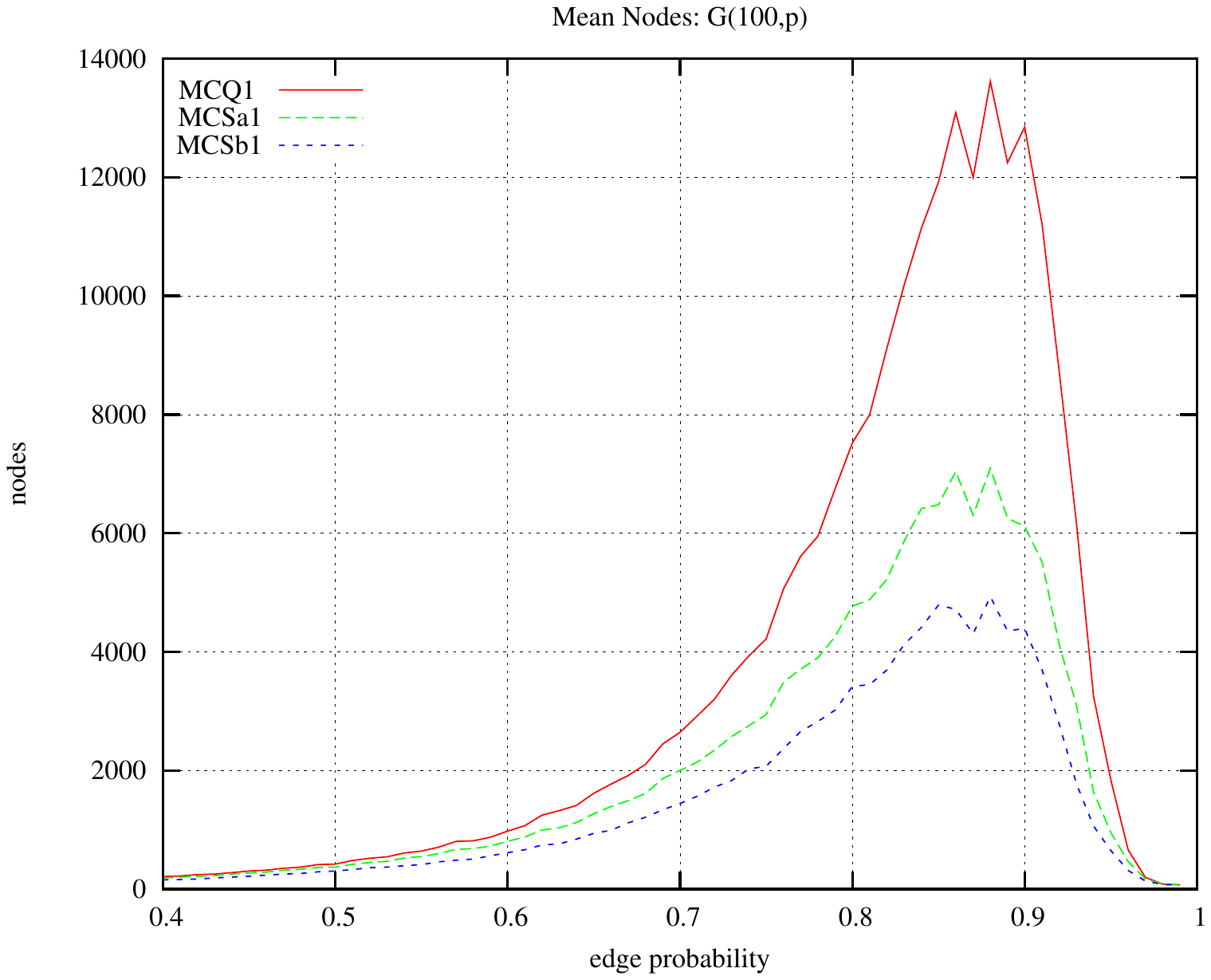}
\end{minipage}
\hfill
\begin{minipage}[t]{0.49\textwidth}
\includegraphics[height=11.5cm]{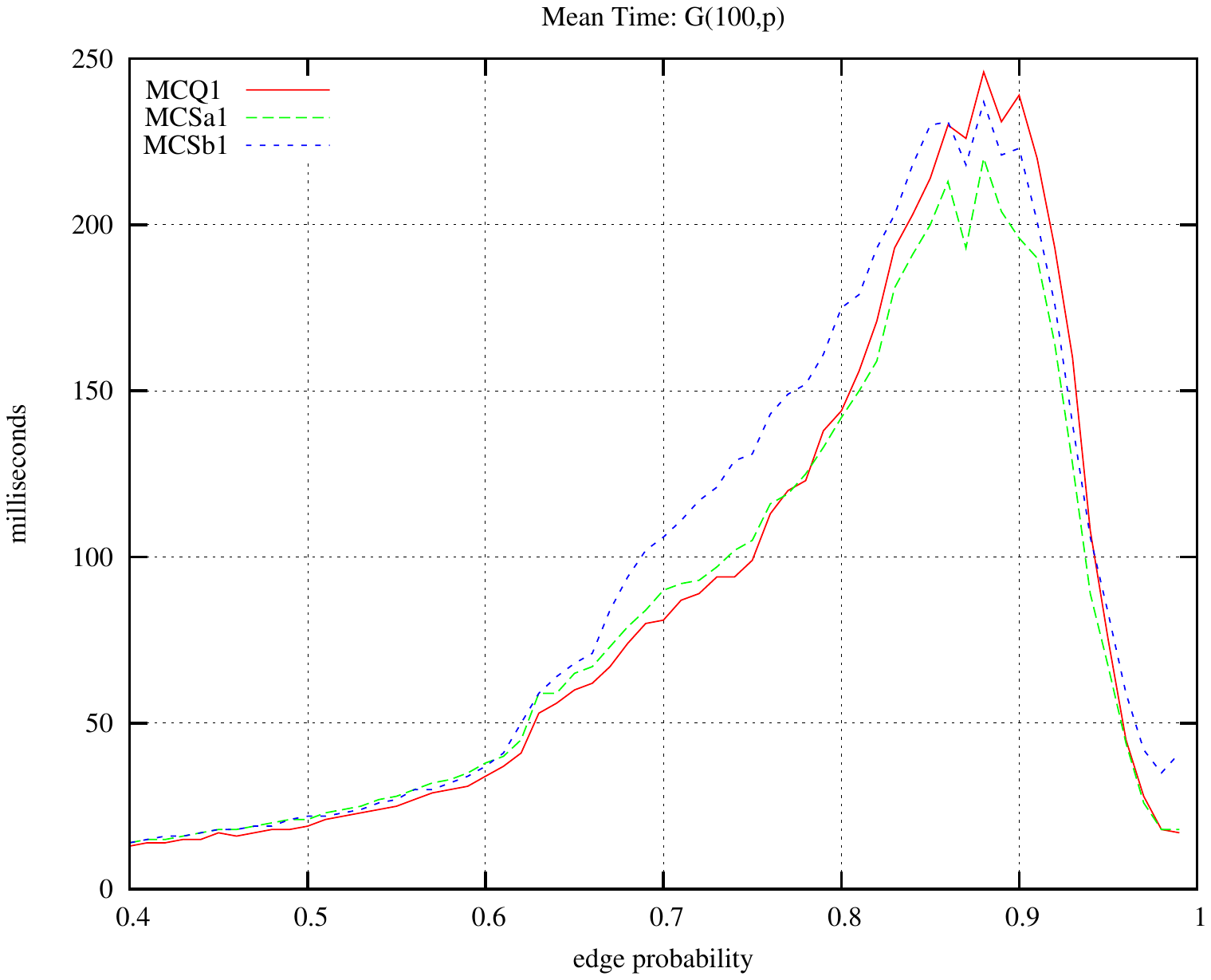}
\end{minipage}
\end{center}
\vspace{-6cm}
\begin{center}
\hspace{-1.5cm}
\begin{minipage}[t]{0.49\textwidth}
\includegraphics[height=11.5cm]{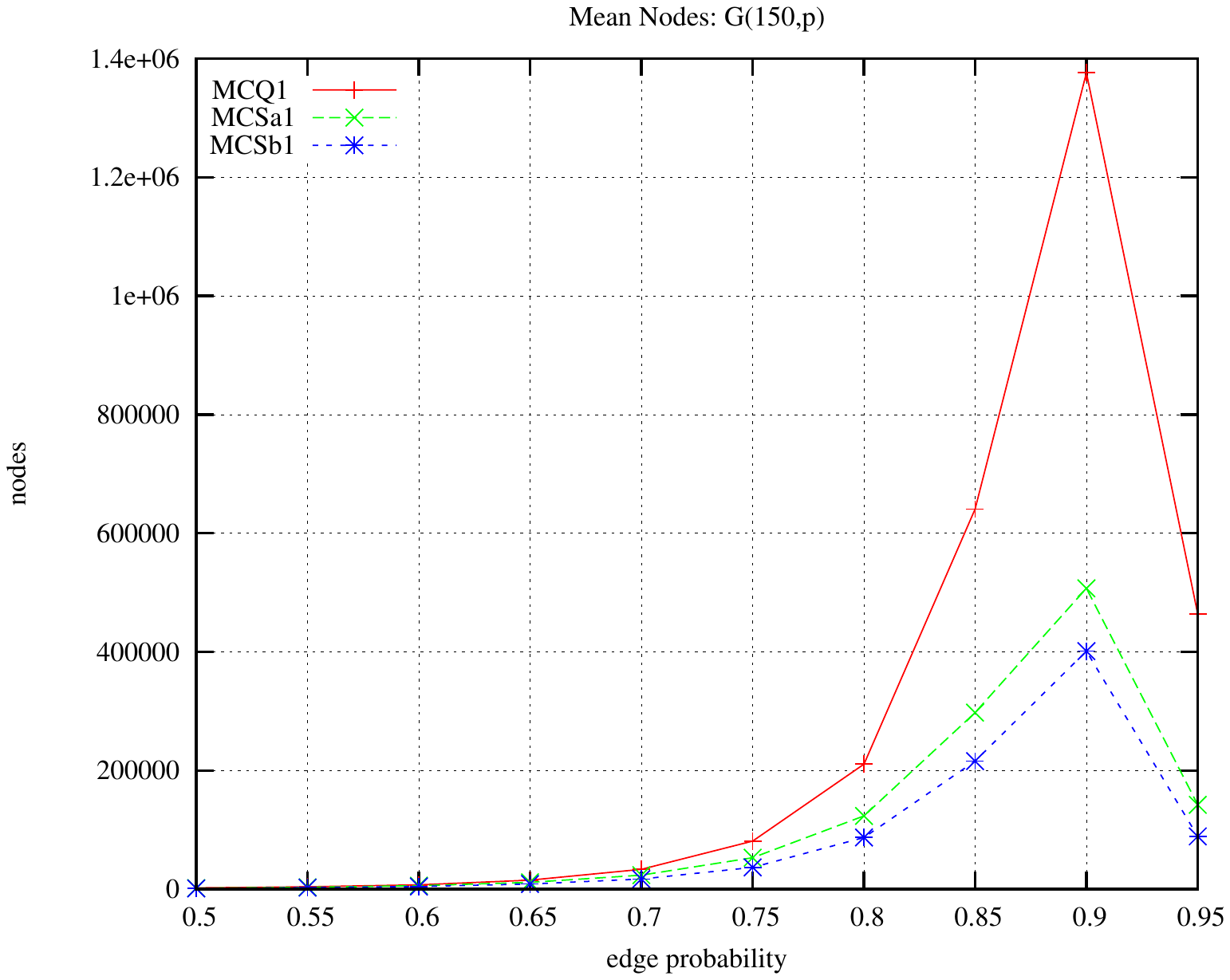}
\end{minipage}
\hfill
\begin{minipage}[t]{0.49\textwidth}
\includegraphics[height=11.5cm]{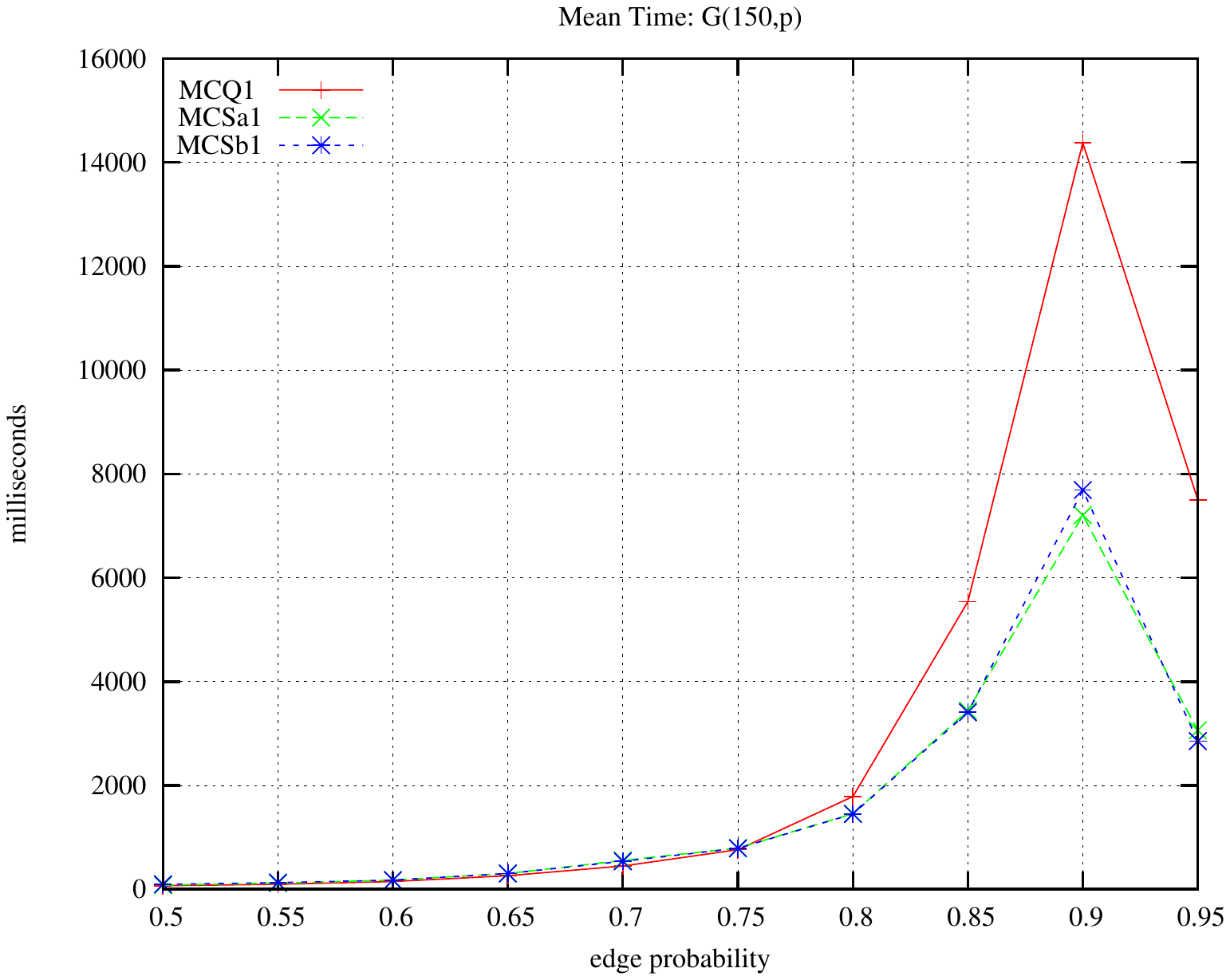}
\end{minipage}
\end{center}
\vspace{-6cm}
\begin{center}
\hspace{-1.5cm}
\begin{minipage}[t]{0.49\textwidth}
\includegraphics[height=11.5cm]{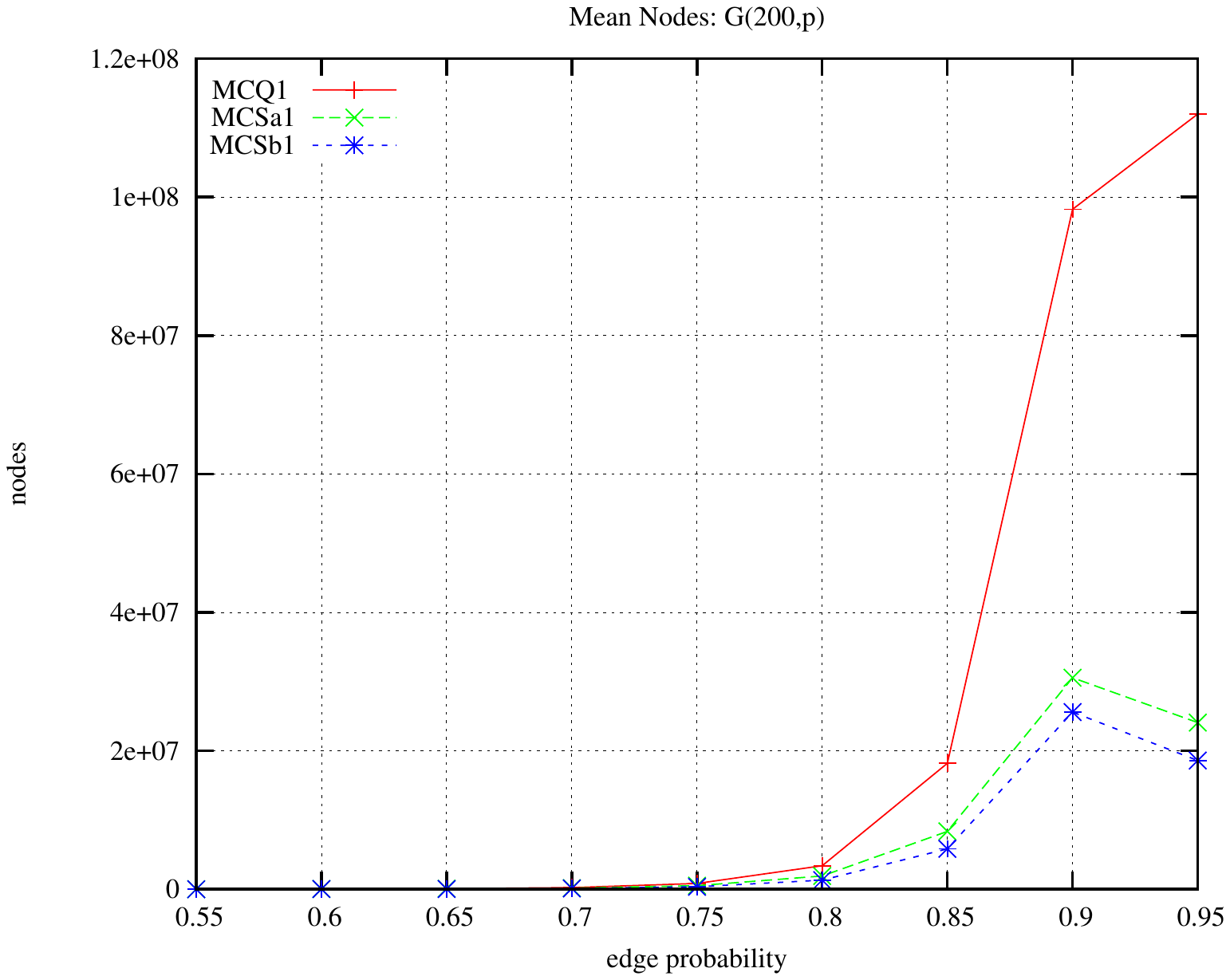}
\end{minipage}
\hfill
\begin{minipage}[t]{0.49\textwidth}
\includegraphics[height=11.5cm]{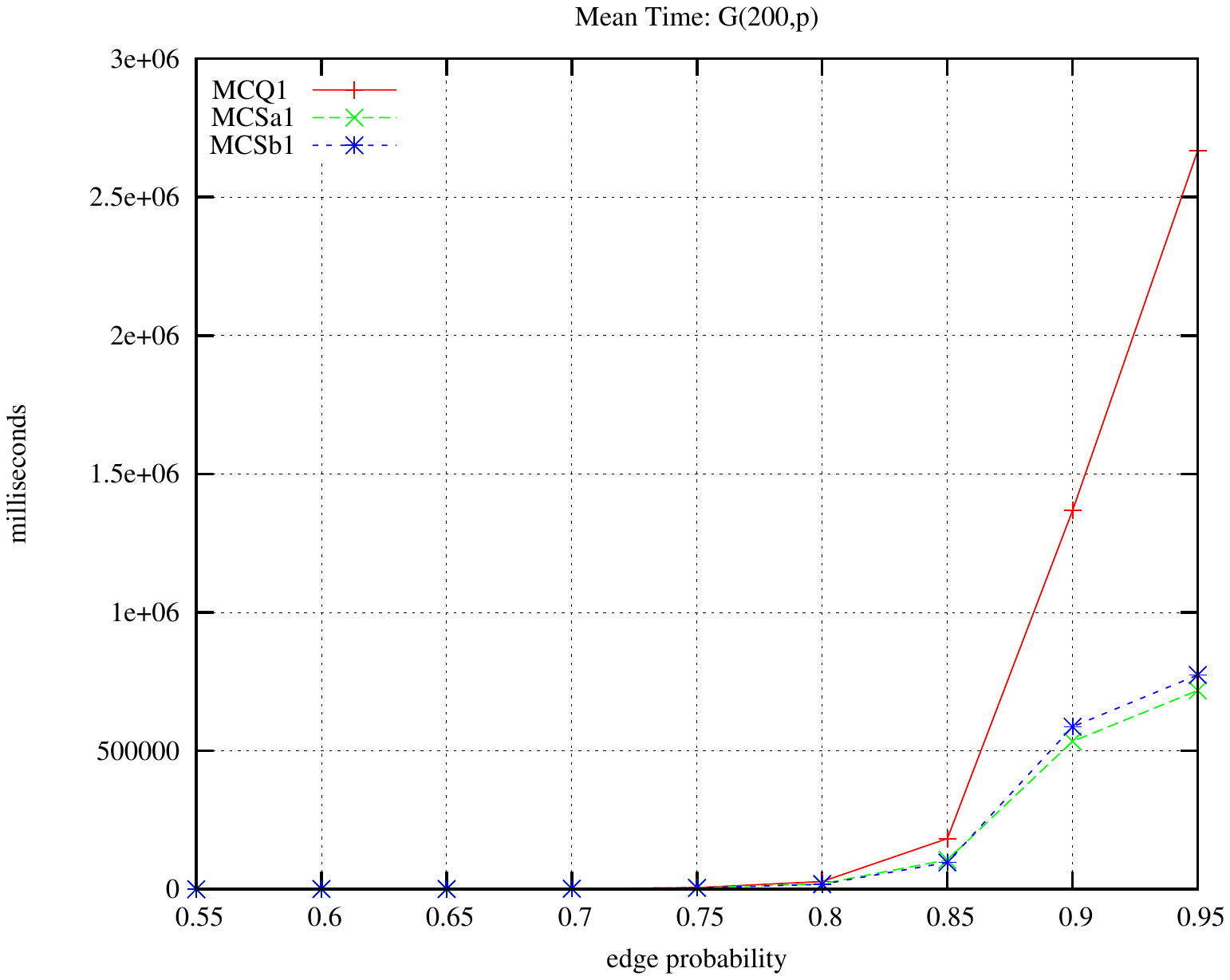}
\end{minipage}
\end{center}
\caption{$G(n,p)$, sample size 100.
MCQ versus MCS, where's the win? On the left search effort in nodes visited (i.e. decisions made by the search 
process) and on the right run time in milliseconds.}
\label{mcqVmcs}
\end{figure}

The first experiments are on random $G(n,p)$, 
first with $n = 100$, $0.40 \leq p \leq 0.99$, $p$ varying in steps of 0.01, sample size of 100,
then with $n = 150$, $0.50 \leq p \leq 0.95$, $p$ varying in steps of 0.10, sample size of 100, and
$n = 200$, $0.55 \leq p \leq 0.95$, $p$ varying in steps of 0.10, sample size of 100.
Unless otherwise stated, all experiments are carried out on our reference machine. The algorithms MCQ, MCSa and MCSb all use
$style = 1$ (i.e. MCQ1, MCSa1, MCSb1). Figure \ref{mcqVmcs} shows on the left average number of nodes against edge probability and on the right
average run time in milliseconds against edge probability for MCQ1, MCSa1 and MCSb1. 
The top row has $n = 100$, middle row $n = 150$ and bottom row $n = 200$\footnote{For MCQ1 the sample size at $G(200,0.95)$
was reduced to 28, i.e. the MCQ1-200 job was terminated after 60 hours.}.
As we apply the modifications to MCQ we
see a reduction in nodes with MCSb1 exploring less states than MCSa1 and MCSa1 less than MCQ1. 
But on the right we see that reduction in search space
does not always result in a reduction in run time. MCSb1 is always slower than MCSa1, i.e. the colour repair is too expensive
and when $n = 100$ MCSb1 is often more expensive to run than MCQ! Therefore it appears that MCS gets its advantage just
from the static colour ordering and that the colour repair slows it down.

We also see a region where problems are hard for all our algorithms, at $n = 100$ and $n = 150$, both in terms of nodes and run time. 
However at $n = 200$ there is a different picture. We see a hard region in terms of nodes but an ever increasing run time. That
is, even though nodes are falling cpu time is climbing. This agrees with the tabulated results in \cite{segundo2011} (Tables 4
and 5 on page 580) for BB-MaxClique. It is a conjecture that run time increases because the cost of each node (call to
expand) incurs more cost in the colouring of the relatively larger candidate set. 
In going from $G(200,0.90)$ to $G(200,0.95)$ maximum clique size increased on average from 41 to 62, a 50\% increase,
and for MCSa1 the average number of nodes fell by 20\% (30\% for MCSb1): search space has fallen, clique size has increased, this 
increases the cost of colouring and this results in an overall increase in run time.

\begin{figure}
\vspace{-8cm}
\begin{center}
\hspace{-1.5cm}
\begin{minipage}[t]{0.49\textwidth}
\includegraphics[height=12.5cm]{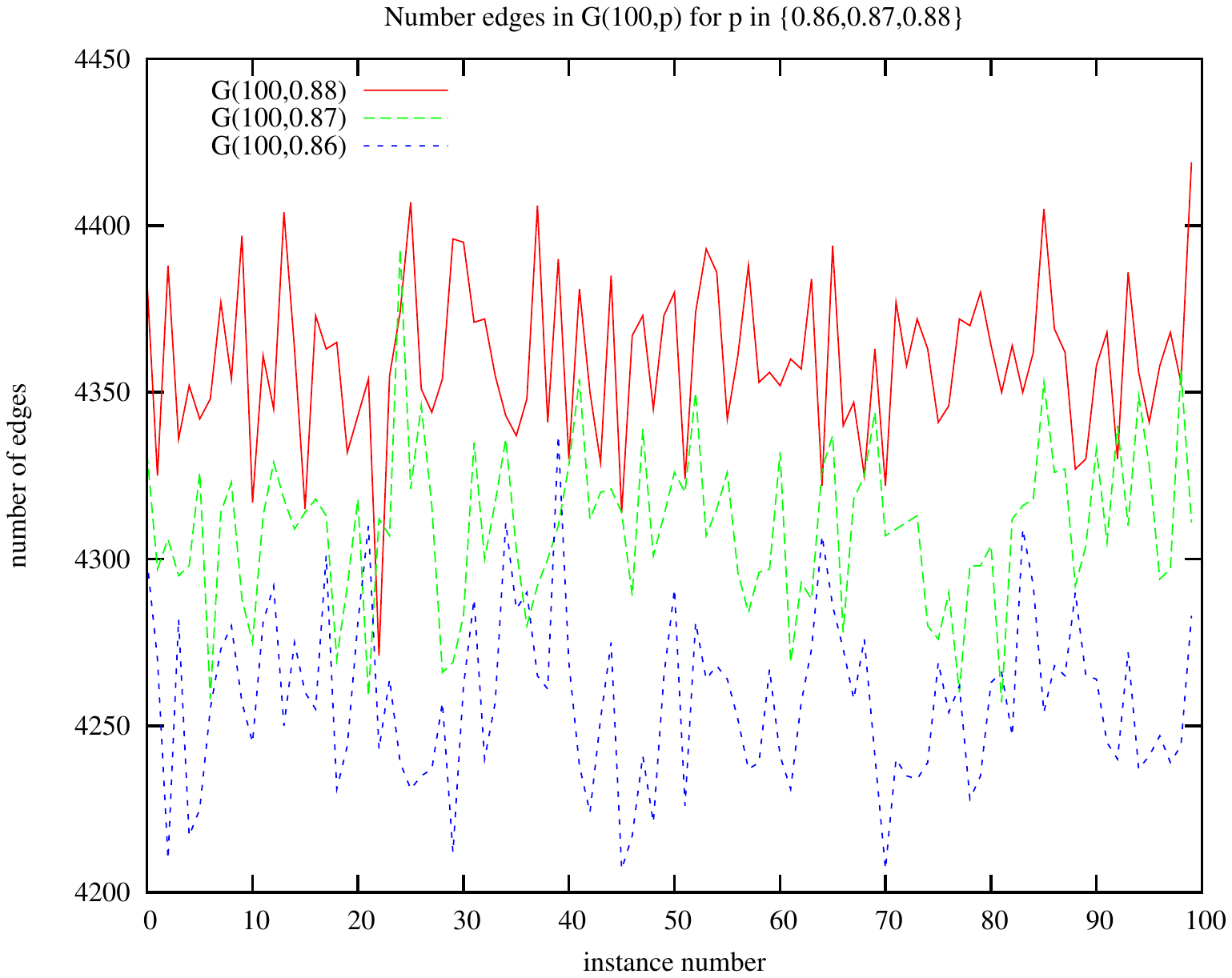}
\end{minipage}
\hfill
\begin{minipage}[t]{0.49\textwidth}
\includegraphics[height=12.5cm]{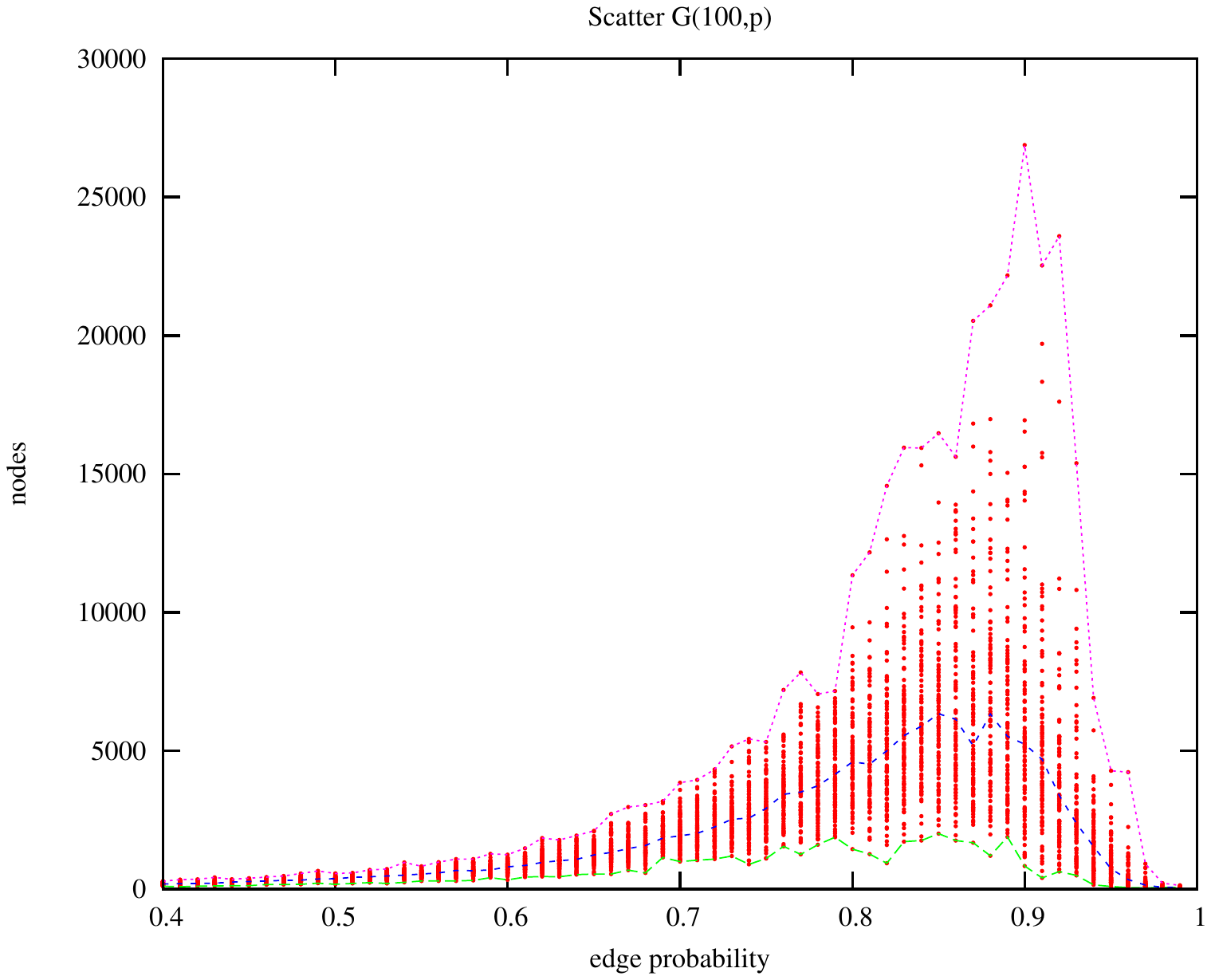}
\end{minipage}
\end{center}
\caption{How many edges are there in $G(100,p)$ with $p \in \{0.86,0.87,0.88\}$ and how does search cost vary?}
\label{edges}
\end{figure}

Figure \ref{mcqVmcs} shows erratic behaviour on $G(100,p)$ with $0.85 \leq p \leq 0.90$. Why is this? Figure \ref{edges}
shows on the left a plot of the number of edges in each of the 100 instances generated for $G(100,p)$ with $p \in \{0.86,0.87,0.88\}$,
one contour for each. $G(100,0.87)$ instances are very much a mix between $G(100,0.86)$ and $G(100,0.88)$ instances. On the right 
is a scatter plot of search effort against edge probability for MCSa1 on $G(100,p)$, sample size 100. We see a large variation
and it is this that gives us the erratic behaviour with $\leq 0.85 p \leq 0.90$. Most likely, a larger
sample size would smooth out the average.

We now report on the 66 DIMACS instances \cite{DIMACS} in Table \ref{tableMCQvMCS}. 
For each algorithm we have 3 entries: the number of nodes, cpu time in seconds, and in brackets the size of the largest clique found.
Each algorithm was allowed 14,400 cpu seconds, i.e. 4 hours, and if that was exceeded we have a table entry of ``---''. The best cpu
time in a row is in {\bf bold} font, and when cpu time limit is exceeded the largest maximum clique size is {\bf emboldened}.  A time entry
of 0 corresponds to a run time of less than a second and we then consider the problem as being too easy to be of
interest. Overall, we see that MCQ1 is rarely the best choice with MCSa1 or MCSb1 performing better. There are 11 problems where
MCSb1 beats MCSa1 and 8 problems where MCSa1 beats MCSb1. Therefore the DIMACS benchmarks don't significantly separate 
the behaviour of these two algorithms.

\begin{table}
\begin{center}
\begin{scriptsize}
\begin{tabular}{|l|r r r|r r r|r r r|} \hline 
\multicolumn{1}{|c|}{instance} & \multicolumn{3}{|c|}{MCQ1} & \multicolumn{3}{|c|}{MCSa1} & \multicolumn{3}{|c|}{MCSb1} \\ \hline
brock200-1 & 868,213 & 7 & (21) & 524,723 & 4 & (21) & 245,146 & \bf{3} & (21) \\ 
brock200-2 & 4,330 & 0 & (12) & 3,826 & 0 & (12) & 3,229 & 0 & (12) \\ 
brock200-3 & 17,817 & 0 & (15) & 14,565 & 0 & (15) & 11,234 & 0 & (15) \\ 
brock200-4 & 80,828 & 0 & (17) & 58,730 & 0 & (17) & 41,355 & 0 & (17) \\ 
brock400-1 & 342,473,950 & 4,471 & (27) & 198,359,829 & 2,888 & (27) & 142,253,319 & \bf{2,551} & (27) \\ 
brock400-2 & 224,839,070 & 2,923 & (29) & 145,597,994 & 2,089 & (29) & 61,327,056 & \bf{1,199} & (29) \\ 
brock400-3 & 194,403,055 & 2,322 & (31) & 120,230,513 & 1,616 & (31) & 70,263,846 & \bf{1,234} & (31) \\ 
brock400-4 & 82,056,086 & 1,117 & (33) & 54,440,888 & \bf{802} & (33) & 68,252,352 & 1,209 & (33) \\ 
brock800-1 & 1,247,519,247 & --- & \bf{(23)} & 1,055,945,239 & --- & \bf{(23)} & 911,465,283 & --- & (21) \\ 
brock800-2 & 1,387,973,191 & --- & (21) & 1,171,057,646 & --- & \bf{(24)} & 914,638,570 & --- & (21) \\ 
brock800-3 & 1,332,309,827 & --- & (21) & 1,159,165,900 & --- & (21) & 914,235,793 & --- & (21) \\ 
brock800-4 & 804,901,115 & --- & (26) & 640,444,536 & \bf{12,568} & (26) & 659,145,642 & 13,924 & (26) \\ 
c-fat200-1 & 24 & 0 & (12) & 24 & 0 & (12) & 23 & 0 & (12) \\ 
c-fat200-2 & 24 & 0 & (24) & 24 & 0 & (24) & 24 & 0 & (24) \\ 
c-fat200-5 & 139 & 0 & (58) & 139 & 0 & (58) & 139 & 0 & (58) \\ 
c-fat500-10 & 126 & 0 & (126) & 126 & 0 & (126) & 126 & 0 & (126) \\ 
c-fat500-1 & 14 & 0 & (14) & 14 & 0 & (14) & 14 & 0 & (14) \\ 
c-fat500-2 & 26 & 0 & (26) & 26 & 0 & (26) & 26 & 0 & (26) \\ 
c-fat500-5 & 64 & 0 & (64) & 64 & 0 & (64) & 64 & 0 & (64) \\ 
hamming10-2 & 512 & \bf{0} & (512) & 512 & \bf{0} & (512) & 512 & 8 & (512) \\ 
hamming10-4 & 636,203,658 & --- & \bf{(40)} & 950,939,457 & --- & (37) & 858,347,653 & --- & (37) \\ 
hamming6-2 & 32 & 0 & (32) & 32 & 0 & (32) & 32 & 0 & (32) \\ 
hamming6-4 & 82 & 0 & (4) & 82 & 0 & (4) & 82 & 0 & (4) \\ 
hamming8-2 & 128 & 0 & (128) & 128 & 0 & (128) & 128 & 0 & (128) \\ 
hamming8-4 & 41,492 & 0 & (16) & 36,452 & 0 & (16) & 33,629 & 0 & (16) \\ 
johnson16-2-4 & 323,036 & 0 & (8) & 256,100 & 0 & (8) & 256,100 & 0 & (8) \\ 
johnson32-2-4 & 10,447,210,976 & --- & (16) & 8,269,639,389 & --- & (16) & 7,345,343,221 & --- & (16) \\ 
johnson8-2-4 & 36 & 0 & (4) & 24 & 0 & (4) & 24 & 0 & (4) \\ 
johnson8-4-4 & 144 & 0 & (14) & 126 & 0 & (14) & 126 & 0 & (14) \\ 
keller4 & 13,113 & 0 & (11) & 13,725 & 0 & (11) & 10,470 & 0 & (11) \\ 
keller5 & 603,233,453 & --- & (27) & 596,150,386 & --- & (27) & 523,346,613 & --- & (27) \\ 
keller6 & 285,704,599 & --- & (48) & 226,330,037 & --- & (52) & 240,958,450 & --- & \bf{(54)} \\ 
MANN-a27 & 38,019 & 9 & (126) & 38,019 & \bf{6} & (126) & 38,597 & 8 & (126) \\ 
MANN-a45 & 2,851,572 & 4,989 & (345) & 2,851,572 & \bf{3,766} & (345) & 2,545,131 & 4,118 & (345) \\ 
MANN-a81 & 550,869 & --- & (1100) & 631,141 & --- & (1100) & 551,612 & --- & (1100) \\ 
MANN-a9 & 71 & 0 & (16) & 71 & 0 & (16) & 39 & 0 & (16) \\ 
p-hat1000-1 & 237,437 & 2 & (10) & 176,576 & 2 & (10) & 151,033 & 2 & (10) \\ 
p-hat1000-2 & 466,616,845 & --- & (45) & 34,473,978 & \bf{1,401} & (46) & 166,655,543 & 7,565 & (46) \\ 
p-hat1000-3 & 440,569,803 & --- & (52) & 345,925,712 & --- & (55) & 298,537,771 & --- & \bf{(56)} \\ 
p-hat1500-1 & 1,642,981 & 16 & (12) & 1,184,526 & 14 & (12) & 990,246 & 14 & (12) \\ 
p-hat1500-2 & 414,514,960 & --- & (52) & 231,498,292 & --- & \bf{(60)} & 259,771,137 & --- & (57) \\ 
p-hat1500-3 & 570,637,417 & --- & (56) & 220,823,126 & --- & (69) & 176,987,047 & --- & (69) \\ 
p-hat300-1 & 1,727 & 0 & (8) & 1,480 & 0 & (8) & 1,305 & 0 & (8) \\ 
p-hat300-2 & 13,814 & 0 & (25) & 4,256 & 0 & (25) & 4,877 & 0 & (25) \\ 
p-hat300-3 & 3,829,005 & 74 & (36) & 624,947 & \bf{13} & (36) & 713,107 & 21 & (36) \\ 
p-hat500-1 & 12,907 & 0 & (9) & 9,777 & 0 & (9) & 8,608 & 0 & (9) \\ 
p-hat500-2 & 1,022,190 & 23 & (36) & 114,009 & \bf{3} & (36) & 137,568 & 5 & (36) \\ 
p-hat500-3 & 515,071,375 & --- & (47) & 39,260,458 & \bf{1,381} & (50) & 104,684,054 & 4,945 & (50) \\ 
p-hat700-1 & 36,925 & 0 & (11) & 26,649 & 0 & (11) & 22,811 & 0 & (11) \\ 
p-hat700-2 & 18,968,155 & 508 & (44) & 750,903 & \bf{27} & (44) & 149,0522 & 74 & (44) \\ 
p-hat700-3 & 570,423,439 & --- & (48) & 255,745,746 & --- & (62) & 243,836,191 & --- & (62) \\ 
san1000 & 302,895 & 20 & (15) & 150,725 & 10 & (15) & 53,215 & \bf{3} & (15) \\ 
san200-0.7-1 & 12,355 & 0 & (30) & 13,399 & 0 & (30) & 232 & 0 & (30) \\ 
san200-0.7-2 & 767 & 0 & (18) & 464 & 0 & (18) & 343 & 0 & (18) \\ 
san200-0.9-1 & 981 & 0 & (70) & 87329 & 1 & (70) & 74 & 0 & (70) \\ 
san200-0.9-2 & 1,149,564 & 20 & (60) & 229,567 & 5 & (60) & 62,776 & \bf{1} & (60) \\ 
san200-0.9-3 & 8,260,345 & 154 & (44) & 6,815,145 & 111 & (44) & 1,218,317 & \bf{32} & (44) \\ 
san400-0.5-1 & 3,960 & 0 & (13) & 2,453 & 0 & (13) & 1,386 & 0 & (13) \\ 
san400-0.7-1 & 55,010 & \bf{1} & (40) & 119,356 & 2 & (40) & 134,772 & 3 & (40) \\ 
san400-0.7-2 & 606,159 & \bf{14} & (30) & 889,125 & 19 & (30) & 754,146 & 16 & (30) \\ 
san400-0.7-3 & 582,646 & 11 & (22) & 521,410 & 10 & (22) & 215,785 & \bf{5} & (22) \\ 
san400-0.9-1 & 523,531,417 & --- & (56) & 4,536,723 & 422 & (100) & 582,445 & \bf{54} & (100) \\ 
sanr200-0.7 & 206,262 & 1 & (18) & 152,882 & 1 & (18) & 100,977 & 1 & (18) \\ 
sanr200-0.9 & 44,472,276 & 892 & (42) & 14,921,850 & 283 & (42) & 9,730,778 & \bf{245} & (42) \\ 
sanr400-0.5 & 380,151 & 2 & (13) & 320,110 & 2 & (13) & 190,706 & 2 & (13) \\ 
sanr400-0.7 & 101,213,527 & 979 & (21) & 64,412,015 & 711 & (21) & 46,125,168 & \bf{650} & (21) \\ \hline
\end{tabular}
\end{scriptsize}
\end{center}
\caption{DIMACS instances: MCQ versus MCS, nodes, run time and (clique size)}
\label{tableMCQvMCS}
\end{table}

\subsection{BBMC versus MCSa: a change of representation}
What advantage is to be had from the change of representation between MCSa and BBMC, i.e. representing
sets as $ArrayList$ in MCSa and as a BitSet in BBMC? MCSa and BBMC are at heart the same algorithm. They both
produce the same colourings, order the candidate set in the same way and explore the same backtrack tree. The only difference
is in the implementation of sets and how these are exploited.

\begin{figure}
\vspace{-8cm}
\begin{center}
\hspace{-1.5cm}
\begin{minipage}[t]{0.49\textwidth}
\includegraphics[height=12.5cm]{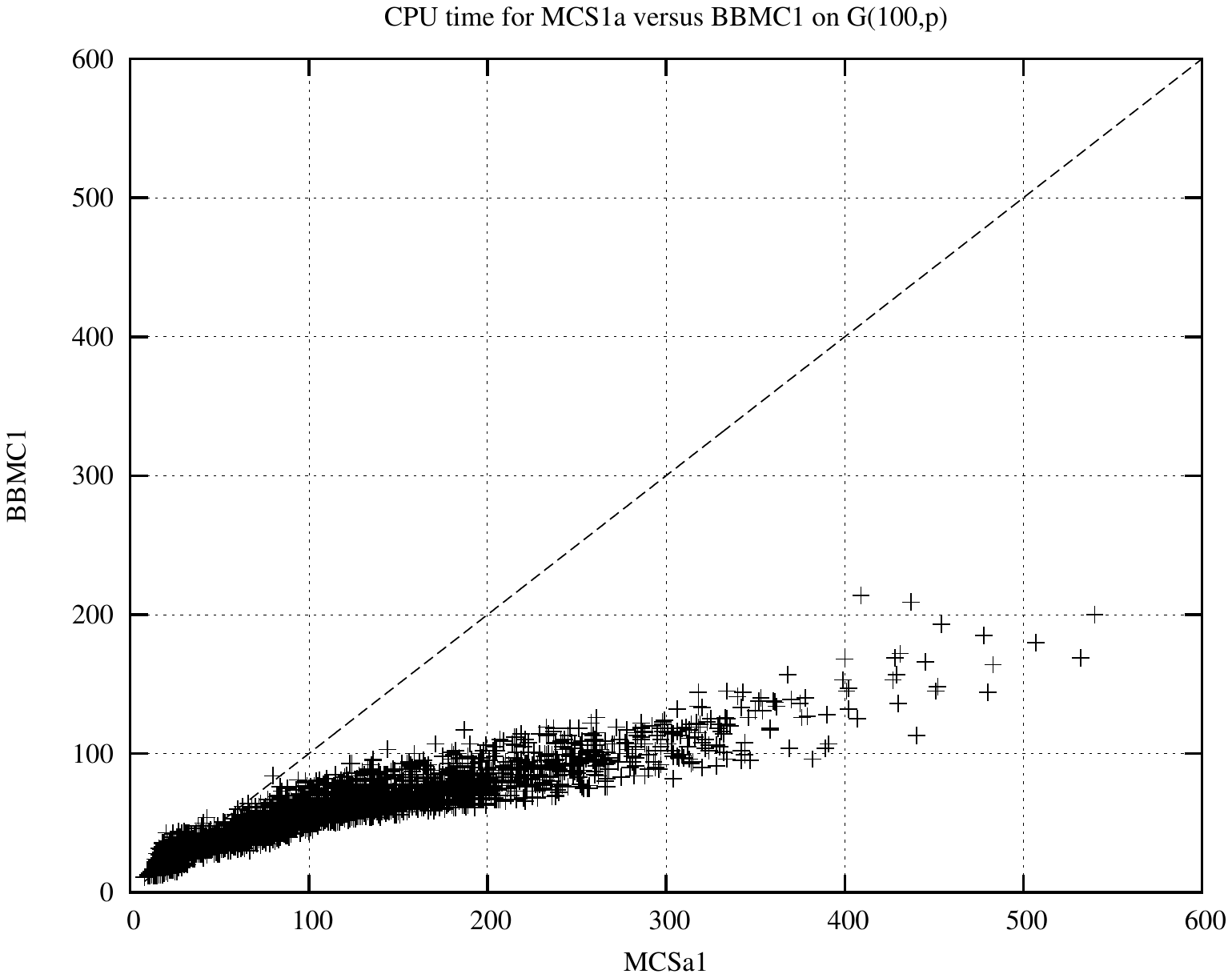}
\end{minipage}
\hfill
\begin{minipage}[t]{0.49\textwidth}
\includegraphics[height=12.5cm]{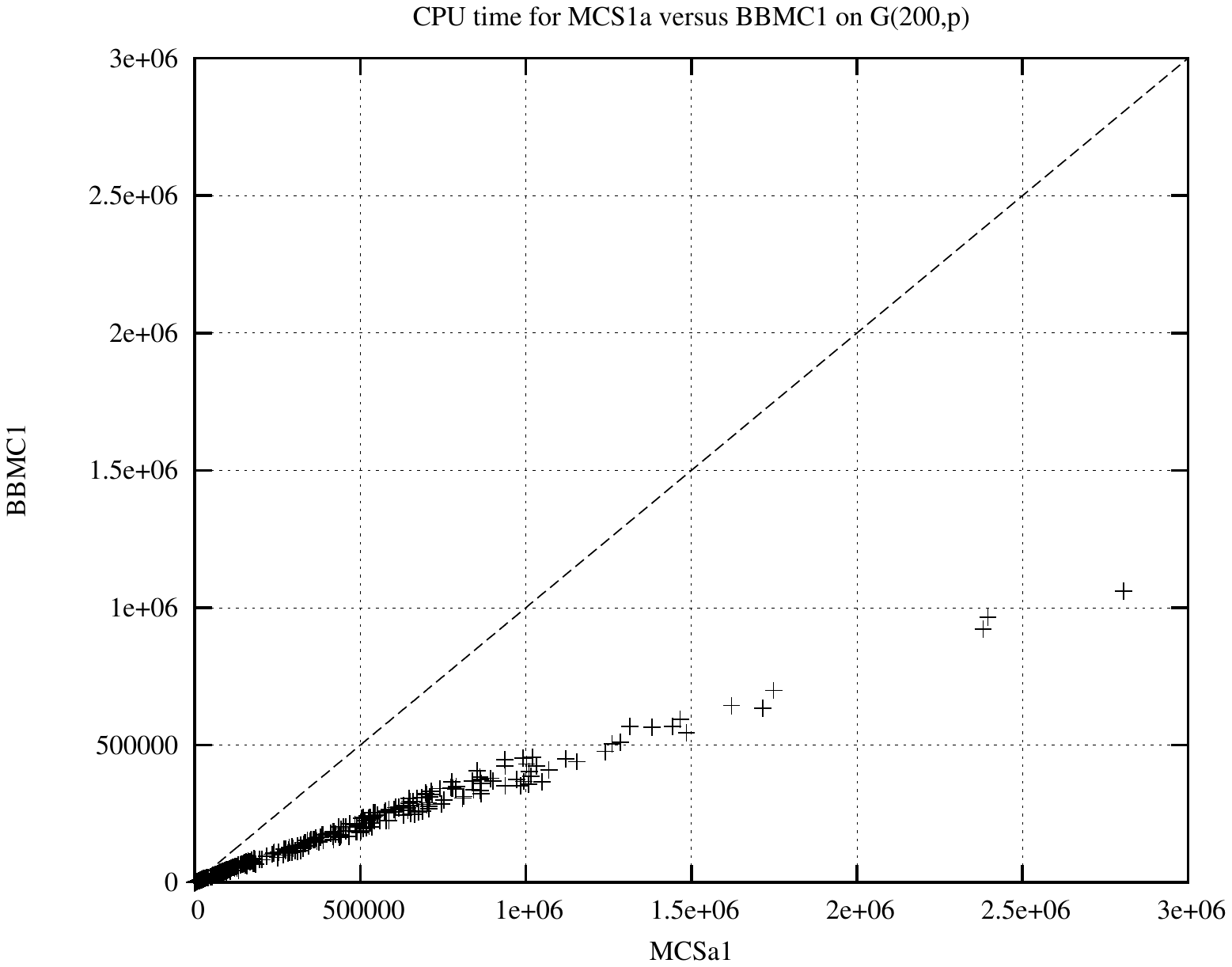}
\end{minipage}
\end{center}
\caption{Run time of MCSa1 against BBMC1, on the left $(G100,p)$ and on the right $G(200,p)$.}
\label{mcsVbbmc}
\end{figure}

Figure \ref{mcsVbbmc} shows on the left run time of MCSa1 (x-axis) against run time of BBMC1 (y-axis) in milliseconds on each of the
$G(100,p)$ random instances. On the right we have the same but for $G(200,p)$. The dotted line is the reference $x = y$. If points
are below the line then BBMC1 is faster than MCSa1. BBMC1 is typically twice as fast as MCSa1.

In Table \ref{tableMCSvBBMC} we tabulate \emph{Goldilocks} instances from the DIMACS benchmark suite: we remove the instances that are too
easy (take less than a second) and those that are too hard (take more than 4 hours) leaving those that are ``just right''
for both algorithms.
Under each algorithm we have: nodes visited (and this is the same for both algorithms), run time in seconds and in brackets
size of the maximum clique. The column on the far right is the ratio of MCSa1's run time over BBMC1's run time, and a value greater
than 1 shows that BBMC1 was faster by that amount. Again we see similar behaviour to that observed over the random problems: 
BBMC1 is typically twice as fast as MCSa1.

\begin{table}
\begin{center}
\begin{scriptsize}
\begin{tabular}{|l|r r r|r r r|c|} \hline 
\multicolumn{1}{|c|}{instance} & \multicolumn{3}{|c|}{MCSa1} & \multicolumn{3}{|c|}{BBMC1} & \multicolumn{1}{|c|}{MCSa1/BBMC1} \\ \hline
brock200-1 & 524,723 & 4 & (21) & 524,723 & 2 & (21) & 2.03   \\ 
brock400-1 & 198,359,829 & 2,888 & (27) & 198,359,829 & 1,421 & (27) & 2.03   \\ 
brock400-2 & 145,597,994 & 2,089 & (29) & 145,597,994 & 1,031 & (29) & 2.03   \\ 
brock400-3 & 120,230,513 & 1,616 & (31) & 120,230,513 & 808 & (31) & 2.00   \\ 
brock400-4 & 54,440,888 & 802 & (33) & 54,440,888 & 394 & (33) & 2.03   \\ 
brock800-4 & 640,444,536 & 12,568 & (26) & 640,444,536 & 6,908 & (26) & 1.82   \\ 
MANN-a27 & 38,019 & 6 & (126) & 38,019 & 1 & (126) & 4.12   \\ 
MANN-a45 & 2,851,572 & 3,766 & (345) & 2,851,572 & 542 & (345) & 6.94   \\ 
p-hat1000-1 & 176,576 & 2 & (10) & 176,576 & 1 & (10) & 1.80   \\ 
p-hat1000-2 & 34,473,978 & 1,401 & (46) & 34,473,978 & 720 & (46) & 1.95   \\ 
p-hat1500-1 & 1,184,526 & 14 & (12) & 1,184,526 & 9 & (12) & 1.52   \\ 
p-hat300-3 & 624,947 & 13 & (36) & 624,947 & 5 & (36) & 2.36   \\ 
p-hat500-2 & 114,009 & 3 & (36) & 114,009 & 1 & (36) & 2.56   \\ 
p-hat500-3 & 39,260,458 & 1,381 & (50) & 39,260,458 & 606 & (50) & 2.28   \\ 
p-hat700-2 & 750,903 & 27 & (44) & 750,903 & 12 & (44) & 2.20   \\ 
san1000 & 150,725 & 10 & (15) & 150,725 & 5 & (15) & 1.76   \\ 
san200-0.9-2 & 229,567 & 5 & (60) & 229,567 & 2 & (60) & 2.36   \\ 
san200-0.9-3 & 6,815,145 & 111 & (44) & 6,815,145 & 50 & (44) & 2.20   \\ 
san400-0.7-1 & 119,356 & 2 & (40) & 119,356 & 1 & (40) & 2.04   \\ 
san400-0.7-2 & 889,125 & 19 & (30) & 889,125 & 9 & (30) & 2.12   \\ 
san400-0.7-3 & 521,410 & 10 & (22) & 521,410 & 5 & (22) & 2.10   \\ 
san400-0.9-1 & 4,536,723 & 422 & (100) & 4,536,723 & 125 & (100) & 3.37   \\ 
sanr200-0.9 & 14,921,850 & 283 & (42) & 14,921,850 & 123 & (42) & 2.30   \\ 
sanr400-0.5 & 320,110 & 2 & (13) & 320,110 & 1 & (13) & 1.85   \\ 
sanr400-0.7 & 64,412,015 & 711 & (21) & 64,412,015 & 365 & (21) & 1.95   \\ \hline
\end{tabular}
\end{scriptsize}
\end{center}
\caption{DIMACS \emph{Goldilocks} instances: MCSa1 versus BBMC1}
\label{tableMCSvBBMC}
\end{table}

\subsection{MCQ and MCS: the effect of style}
\begin{figure}
\vspace{-6cm}
\begin{center}
\hspace{-1.5cm}
\begin{minipage}[t]{0.49\textwidth}
\includegraphics[height=11.5cm]{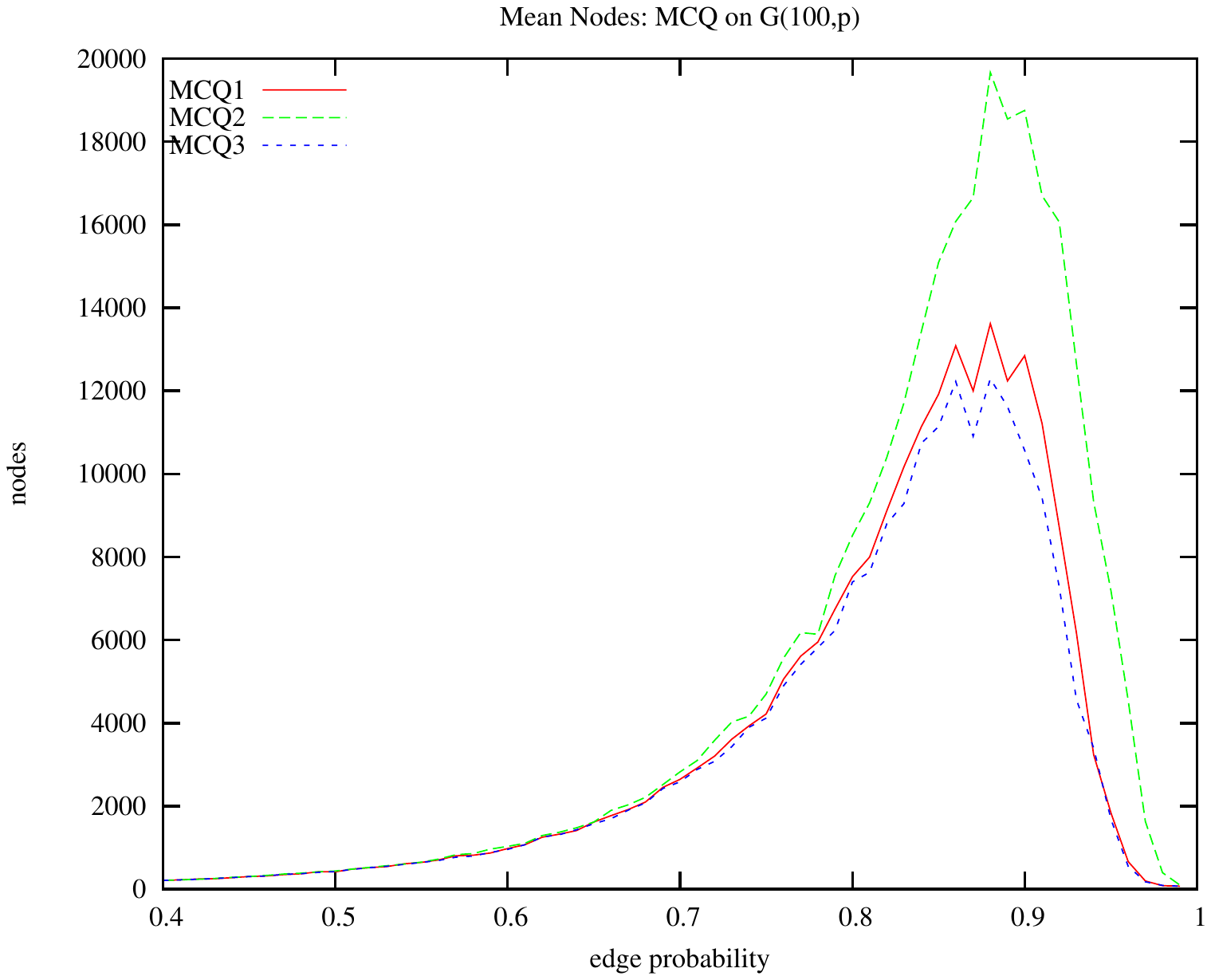}
\end{minipage}
\hfill
\begin{minipage}[t]{0.49\textwidth}
\includegraphics[height=11.5cm]{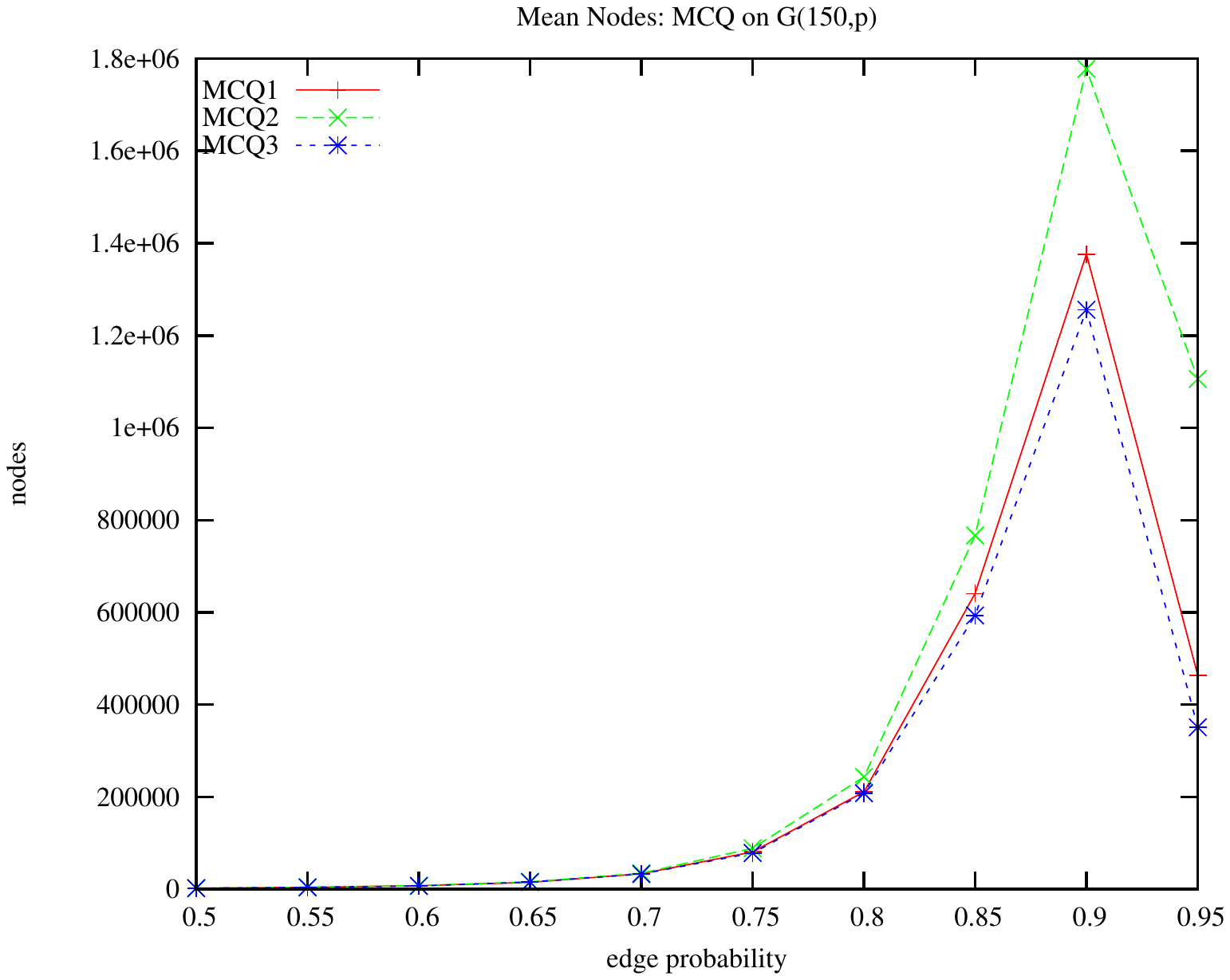}
\end{minipage}
\end{center}
\vspace{-6cm}
\begin{center}
\hspace{-1.5cm}
\begin{minipage}[t]{0.49\textwidth}
\includegraphics[height=11.5cm]{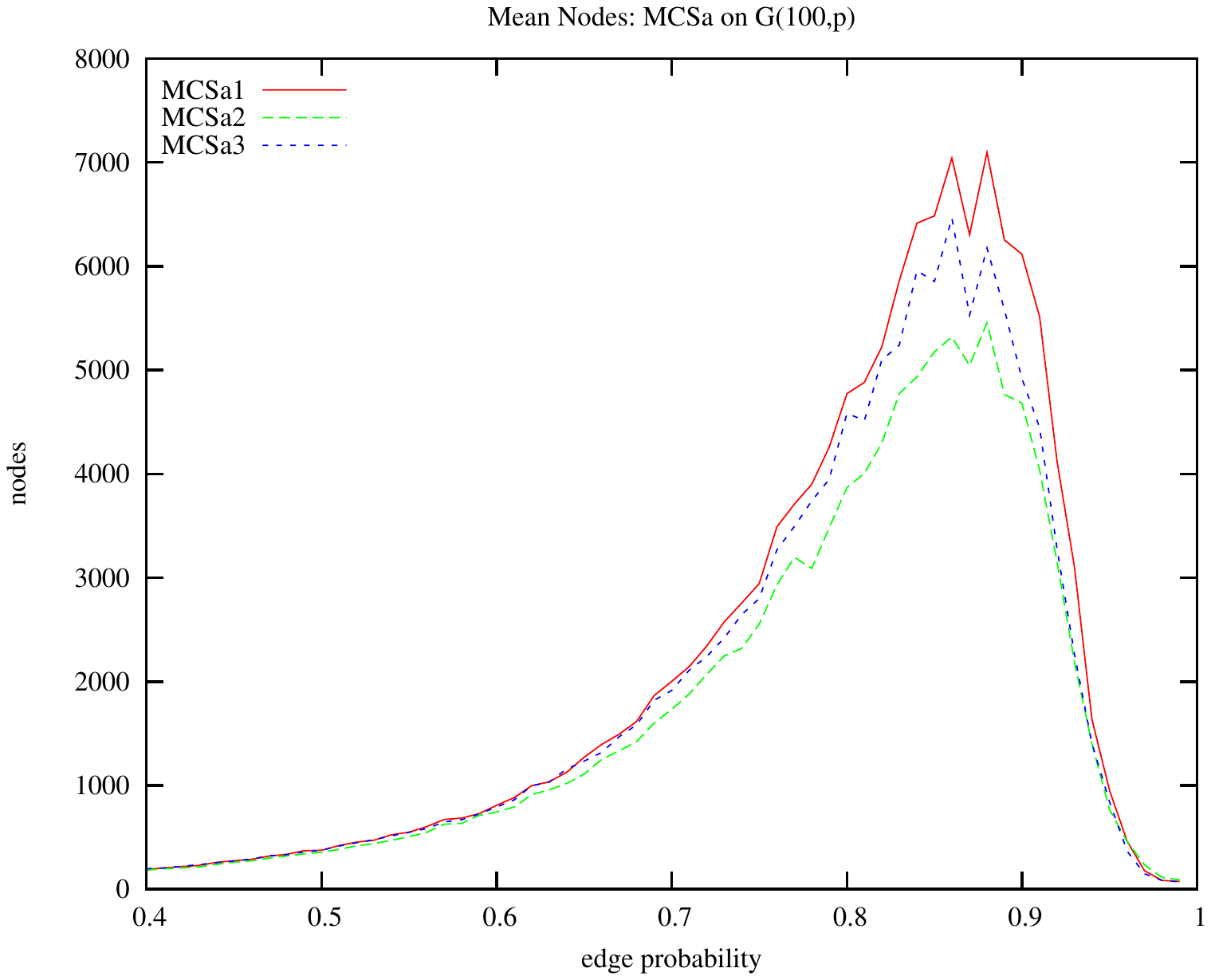}
\end{minipage}
\hfill
\begin{minipage}[t]{0.49\textwidth}
\includegraphics[height=11.5cm]{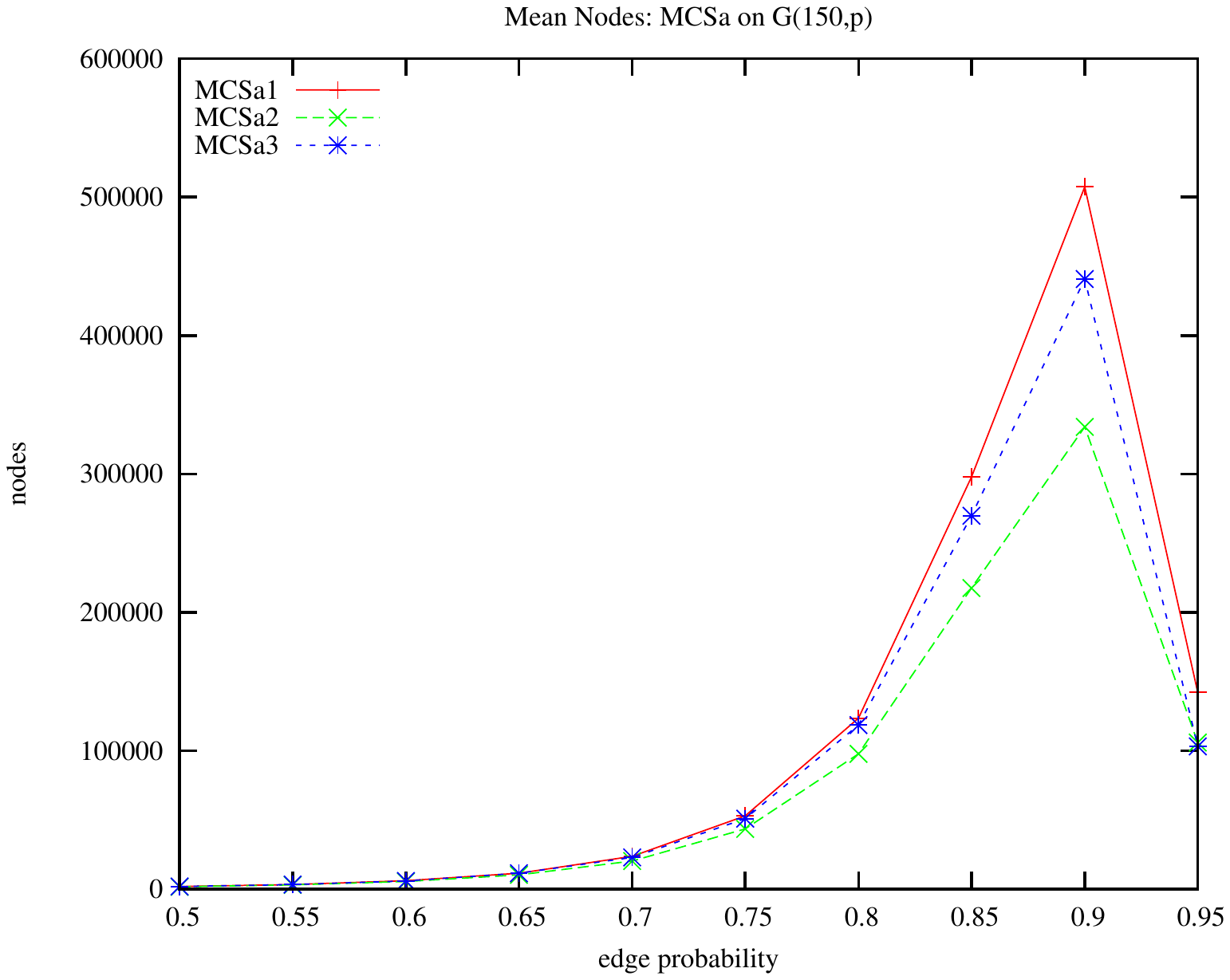}
\end{minipage}
\end{center}
\vspace{-6cm}
\begin{center}
\hspace{-1.5cm}
\begin{minipage}[t]{0.49\textwidth}
\includegraphics[height=11.5cm]{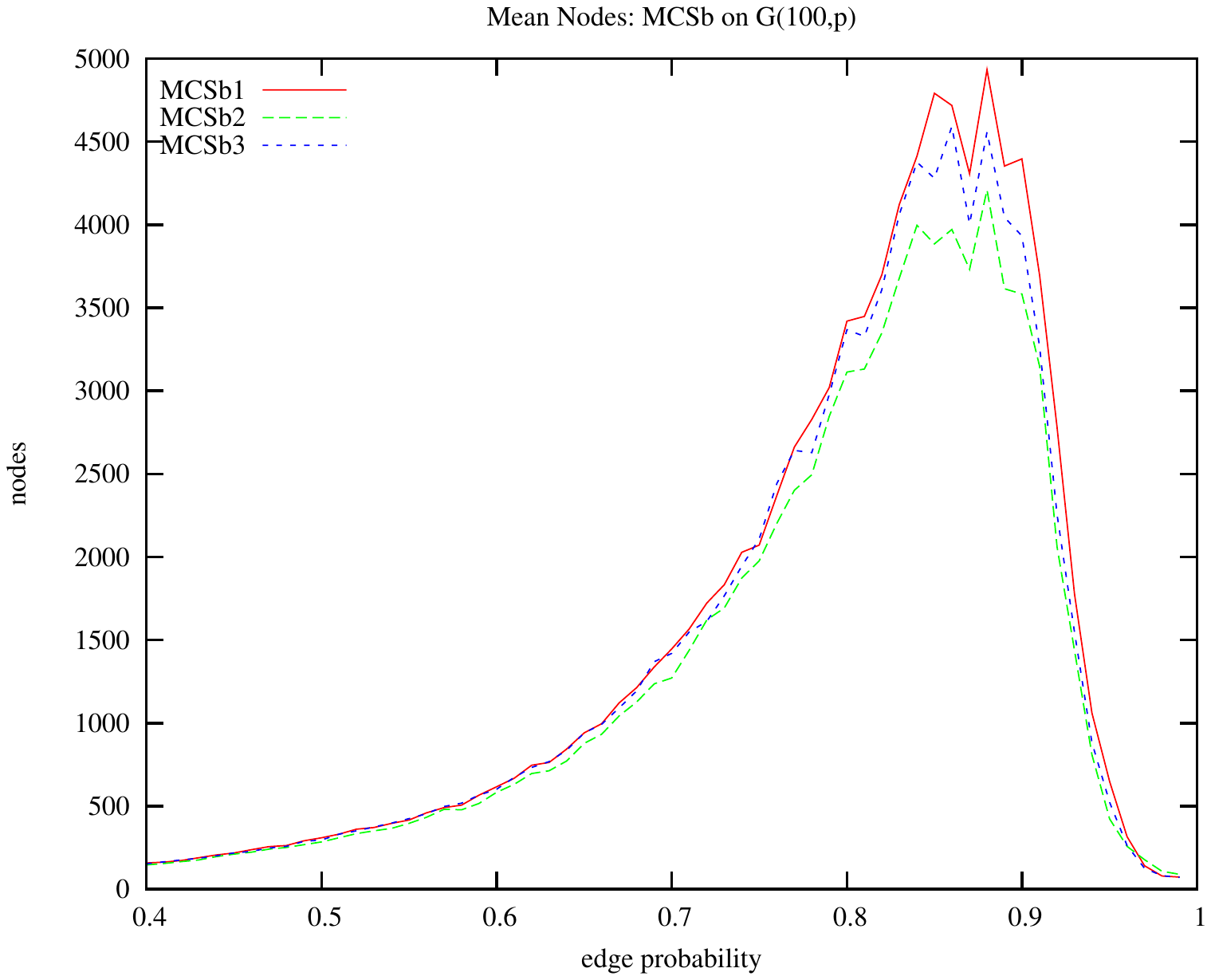}
\end{minipage}
\hfill
\begin{minipage}[t]{0.49\textwidth}
\includegraphics[height=11.5cm]{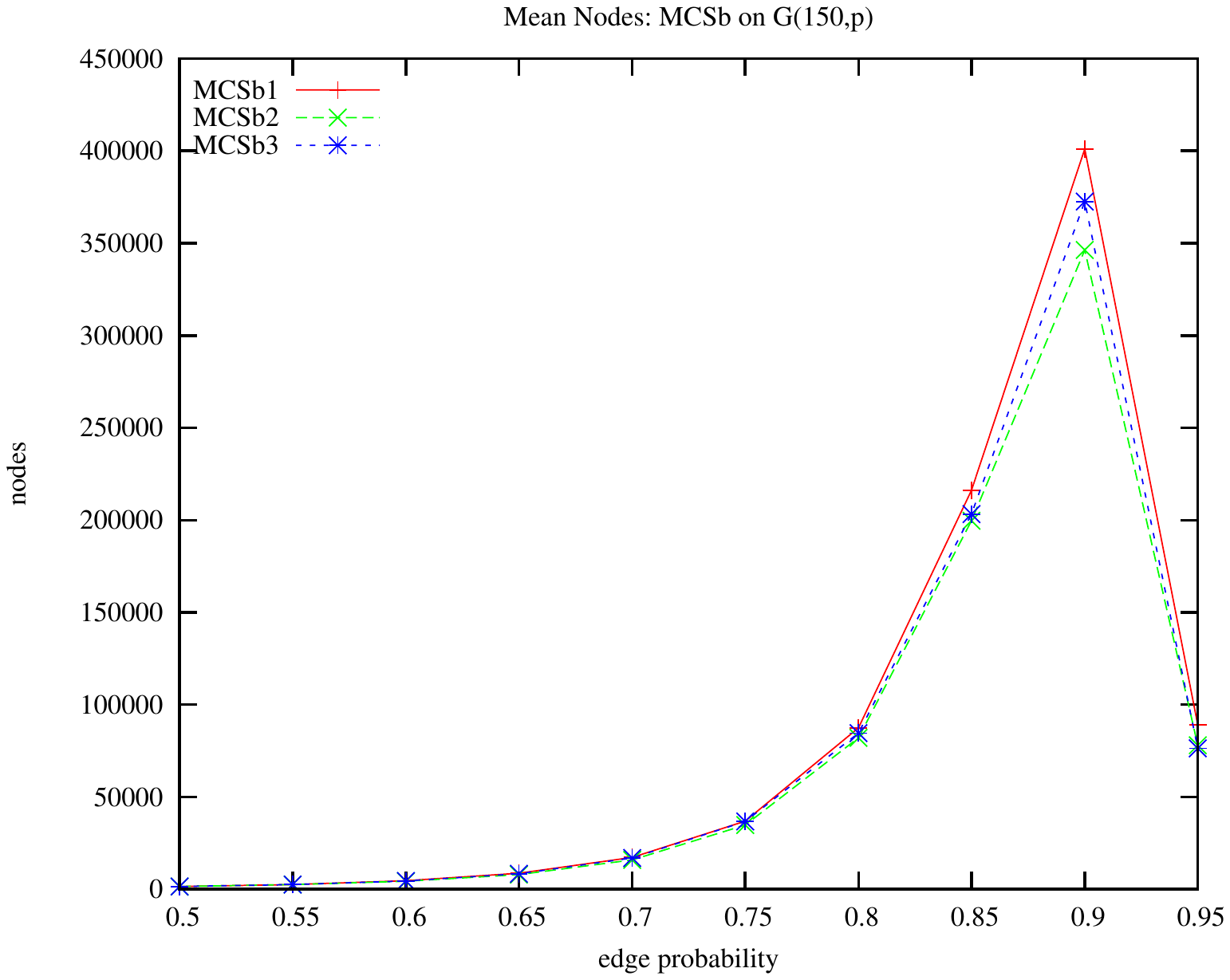}
\end{minipage}
\end{center}
\caption{The effect of style on MCQ, MCSa and MCSb. On the left $G(100,p)$ and on the right $G(150,p)$. Plotted is search effort in nodes against
edge probability. The top two plots are for MCQ, middle plots MCSa and bottom MCSb.}
\label{mc-style}
\end{figure}

What effect does the initial ordering of vertices have on performance? First, we investigate MCQ, MCSa and MCSb with our three orderings:
style 1 being non-decreasing degree, style 2 a minimum width ordering, style 3 non-decreasing degree tie-breaking
on the accumulated degree of neighbours. At this stage we do not consider BBMC as it is just a BitSet encoding of MCSa.
We use random problems $G(n,p)$ with $n$ equal to 100 and 150 with sample size of 100.
This is shown graphically in Figure \ref{mc-style}: on the left $G(100,p)$ and on the right $G(150,p)$ with average nodes visited plotted against
edge probability. Plots on the first row are for MCQ, middle row MCSa and bottom MCSb. For MCQ style 3 is the winner and 
style 2 is worst, whereas in MCSa and MCSb style 2 is always best. Why is this? In MCQ the candidate set is
ordered as the result of colouring and this order is then used in the next colouring. Therefore MCQ gradually disrupts the initial 
minimum width ordering but MCSa and MCSb do not (and neither does BBMC). 
The minimum width ordering (style 2) is best for MCSa, MCSb and BBMC. Note that MCQ3 \emph{is} Tomita's MCR \cite{tomita2007} and our
experiments on $G(n,p)$ show that MCR (MCQ3) beats MCQ (MCQ1).

\begin{table}
\begin{center}
\begin{scriptsize}
\begin{tabular}{|l|c c c|c c c|c c c|c c c|} \hline 
\multicolumn{1}{|c|} {} & \multicolumn{3}{|c|}{MCQ} & \multicolumn{3}{|c|}{MCSa} & \multicolumn{3}{|c|}{MCSb} & \multicolumn{3}{|c|}{BBMC}\\
\multicolumn{1}{|c|}{instance} & $s_1$ & $s_2$ & $s_3$ & $s_1$ & $s_2$ & $s_3$ & $s_1$ & $s_2$ & $s_3$ & $s_1$ & $s_2$ & $s_3$ \\ \hline
brock200-1 & 7 & 5 & 4 & 4 & 3 & 3 & 3 & 3 & 3 & 2 & 1 & 1 \\ 
brock200-2 & 0 & 0 & 0 & 0 & 0 & 0 & 0 & 0 & 0 & 0 & 0 & 0 \\ 
brock200-3 & 0 & 0 & 0 & 0 & 0 & 0 & 0 & 0 & 0 & 0 & 0 & 0 \\ 
brock200-4 & 0 & 0 & 0 & 0 & 0 & 0 & 0 & 0 & 0 & 0 & 0 & 0 \\ 
brock400-1 & 4,471 & \bf{3,640} & 5,610 & 2,888 & \bf{1,999} & 3,752 & \bf{2,551} & 3,748 & 2,152 & 1,421 & \bf{983} & 1,952 \\ 
brock400-2 & 2,923 & 4,573 & \bf{1,824} & 2,089 & 2,415 & \bf{1,204} & \bf{1,199} & 2,695 & 2,647 & 1,031 & 1,230 & \bf{616} \\ 
brock400-3 & 2,322 & 2,696 & \bf{1,491} & 1,616 & 1,404 & \bf{1,027} & \bf{1,234} & 2,817 & 2,117 & 808 & 711 & \bf{534} \\ 
brock400-4 & 1,117 & \bf{574} & 1,872 & 802 & \bf{338} & 1,283 & 1,209 & 1,154 & \bf{607} & 394 & \bf{158} & 651 \\ 
brock800-1 & --- & --- & --- & --- & --- & --- & --- & --- & --- & --- & --- & --- \\ 
brock800-2 & --- & --- & --- & --- & --- & --- & --- & --- & --- & --- & --- & --- \\ 
brock800-3 & --- & --- & --- & --- & --- & --- & --- & --- & --- & --- & \bf{9,479} & 12,815 \\ 
brock800-4 & --- & --- & --- & \bf{12,568} & 13,502 & --- & \bf{13,924} & --- & --- & \bf{6,908} & 7,750 & 12,992 \\ 
c-fat200-1 & 0 & 0 & 0 & 0 & 0 & 0 & 0 & 0 & 0 & 0 & 0 & 0 \\ 
c-fat200-2 & 0 & 0 & 0 & 0 & 0 & 0 & 0 & 0 & 0 & 0 & 0 & 0 \\ 
c-fat200-5 & 0 & 0 & 0 & 0 & 0 & 0 & 0 & 0 & 0 & 0 & 0 & 0 \\ 
c-fat500-10 & 0 & 0 & 0 & 0 & 0 & 0 & 0 & 0 & 0 & 0 & 0 & 0 \\ 
c-fat500-1 & 0 & 0 & 0 & 0 & 0 & 0 & 0 & 0 & 0 & 0 & 0 & 0 \\ 
c-fat500-2 & 0 & 0 & 0 & 0 & 0 & 0 & 0 & 0 & 0 & 0 & 0 & 0 \\ 
c-fat500-5 & 0 & 0 & 0 & 0 & 0 & 0 & 0 & 0 & 0 & 0 & 0 & 0 \\ 
hamming10-2 & 0 & 0 & 0 & 0 & 0 & 0 & 8 & 13 & 8 & 0 & 0 & 0 \\ 
hamming10-4 & --- & --- & --- & --- & --- & --- & --- & --- & --- & --- & --- & --- \\ 
hamming6-2 & 0 & 0 & 0 & 0 & 0 & 0 & 0 & 0 & 0 & 0 & 0 & 0 \\ 
hamming6-4 & 0 & 0 & 0 & 0 & 0 & 0 & 0 & 0 & 0 & 0 & 0 & 0 \\ 
hamming8-2 & 0 & 0 & 0 & 0 & 0 & 0 & 0 & 0 & 0 & 0 & 0 & 0 \\ 
hamming8-4 & 0 & 0 & 0 & 0 & 0 & 0 & 0 & 0 & 0 & 0 & 0 & 0 \\ 
johnson16-2-4 & 0 & 0 & 0 & 0 & 0 & 0 & 0 & 0 & 0 & 0 & 0 & 0 \\ 
johnson32-2-4 & --- & --- & --- & --- & --- & --- & --- & --- & --- & --- & --- & --- \\ 
johnson8-2-4 & 0 & 0 & 0 & 0 & 0 & 0 & 0 & 0 & 0 & 0 & 0 & 0 \\ 
johnson8-4-4 & 0 & 0 & 0 & 0 & 0 & 0 & 0 & 0 & 0 & 0 & 0 & 0 \\ 
keller4 & 0 & 0 & 0 & 0 & 0 & 0 & 0 & 0 & 0 & 0 & 0 & 0 \\ 
keller5 & --- & --- & --- & --- & --- & --- & --- & --- & --- & --- & --- & --- \\ 
keller6 & --- & --- & --- & --- & --- & --- & --- & --- & --- & --- & --- & --- \\ 
MANN-a27 & 9 & 9 & 9 & 6 & 7 & 6 & 8 & 7 & 8 & 1 & 1 & 1 \\ 
MANN-a45 & \bf{4,989} & 5,369 & 4,999 & 3,766 & \bf{3,539} & 3,733 & 4,118 & \bf{3,952} & 4,242 & \bf{542} & 580 & 554 \\ 
MANN-a81 & --- & --- & --- & --- & --- & --- & --- & --- & --- & --- & --- & --- \\ 
MANN-a9 & 0 & 0 & 0 & 0 & 0 & 0 & 0 & 0 & 0 & 0 & 0 & 0 \\ 
p-hat1000-1 & 2 & 2 & 1 & 2 & 2 & 2 & 2 & 2 & 2 & 1 & 1 & 1 \\ 
p-hat1000-2 & --- & --- & --- & 1,401 & \bf{861} & 1,481 & 7,565 & 8,459 & \bf{6,606} & 720 & \bf{431} & 763 \\ 
p-hat1000-3 & --- & --- & --- & --- & --- & --- & --- & --- & --- & --- & --- & --- \\ 
p-hat1500-1 & 16 & 16 & 15 & 14 & 15 & 15 & 14 & 14 & 16 & 9 & 9 & 10 \\ 
p-hat1500-2 & --- & --- & --- & --- & --- & --- & --- & --- & --- & --- & --- & --- \\ 
p-hat1500-3 & --- & --- & --- & --- & --- & --- & --- & --- & --- & --- & --- & --- \\ 
p-hat300-1 & 0 & 0 & 0 & 0 & 0 & 0 & 0 & 0 & 0 & 0 & 0 & 0 \\ 
p-hat300-2 & 0 & 0 & 0 & 0 & 0 & 0 & 0 & 0 & 0 & 0 & 0 & 0 \\ 
p-hat300-3 & 74 & 127 & \bf{69} & 13 & \bf{10} & 12 & 21 & 24 & \bf{18} & 5 & \bf{4} & 5 \\ 
p-hat500-1 & 0 & 0 & 0 & 0 & 0 & 0 & 0 & 0 & 0 & 0 & 0 & 0 \\ 
p-hat500-2 & 23 & 22 & 21 & 3 & 2 & 3 & 5 & 5 & 4 & 1 & 1 & 1 \\ 
p-hat500-3 & --- & --- & --- & 1,381 & \bf{660} & 1,122 & \bf{4,945} & 6,982 & 5,167 & 606 & \bf{282} & 500 \\ 
p-hat700-1 & 0 & 0 & 0 & 0 & 0 & 0 & 0 & 0 & 0 & 0 & 0 & 0 \\ 
p-hat700-2 & 508 & 551 & \bf{353} & 27 & 25 & \bf{24} & \bf{74} & 93 & 108 & 12 & 11 & 11 \\ 
p-hat700-3 & --- & --- & --- & --- & \bf{12,244} & --- & --- & --- & --- & 6,754 & \bf{5,693} & 7,000 \\ 
san1000 & 20 & 19 & 18 & 10 & 10 & 10 & 3 & 3 & 3 & 5 & 5 & 5 \\ 
san200-0.7-1 & 0 & 0 & 0 & 0 & 0 & 0 & 0 & 0 & 0 & 0 & 0 & 0 \\ 
san200-0.7-2 & 0 & 0 & 0 & 0 & 0 & 0 & 0 & 0 & 0 & 0 & 0 & 0 \\ 
san200-0.9-1 & 0 & 0 & 0 & 1 & 0 & 1 & 0 & 0 & 0 & 0 & 0 & 0 \\ 
san200-0.9-2 & \bf{20} & 73 & 35 & 5 & \bf{1} & 5 & 1 & 1 & 1 & 2 & \bf{0} & 2 \\ 
san200-0.9-3 & 154 & \bf{4} & 59 & 111 & \bf{0} & 65 & 32 & \bf{3} & 8 & 50 & \bf{0} & 27 \\ 
san400-0.5-1 & 0 & 0 & 0 & 0 & 0 & 0 & 0 & 0 & 0 & 0 & 0 & 0 \\ 
san400-0.7-1 & 1 & 5 & 2 & 2 & 17 & 4 & 3 & 0 & 1 & 1 & 8 & 1 \\ 
san400-0.7-2 & \bf{14} & 47 & 16 & \bf{19} & 26 & 23 & 16 & 9 & \bf{4} & \bf{9} & 11 & 10 \\ 
san400-0.7-3 & \bf{11} & 38 & 41 & \bf{10} & 22 & 39 & \bf{5} & 13 & 19 & \bf{5} & 9 & 18 \\ 
san400-0.9-1 & --- & --- & --- & \bf{422} & --- & 8,854 & 54 & \bf{0} & --- & \bf{125} & --- & 3,799 \\ 
sanr200-0.7 & 1 & 2 & 1 & 1 & 1 & 1 & 1 & 1 & 1 & 0 & 0 & 0 \\ 
sanr200-0.9 & \bf{892} & 1,782 & 1,083 & 283 & \bf{229} & 364 & 245 & \bf{227} & 444 & 123 & \bf{104} & 164 \\ 
sanr400-0.5 & 2 & 2 & 2 & 2 & 2 & 2 & 2 & 2 & 2 & 1 & 1 & 1 \\ 
sanr400-0.7 & 979 & 1,075 & \bf{975} & 711 & \bf{608} & 719 & \bf{650} & 660 & 674 & 365 & \bf{326} & 369 \\ \hline
\end{tabular}
\end{scriptsize}
\end{center}
\caption{DIMACS instances: the effect of style on run time in seconds}
\label{tabTime}
\end{table}

We now report on the 66 DIMACS instances \cite{DIMACS}, Tables \ref{tabTime} and \ref{tabNodes}.
Table \ref{tabTime} gives run times in seconds. An entry of ``---'' corresponds to the cpu time limit of 14,400 seconds being exceeded and
search terminating early and an entry of 0 (zero) when run time is less than a second. For each algorithm we have three columns, 
one for each \emph{style}: first column $s_1$ is
style 1 with vertices in non-increasing degree order, $s_2$ is style 2 with vertices in minimum width order, $s_3$ is style 3 with
vertices in non-increasing degree order tie-breaking on sum of neighbouring degrees. Table \ref{tabNodes} is the number of
nodes, in thousands, for the experiments in Table \ref{tabTime}. In Table \ref{tabTime} a {\bf bold} entry is the best 
run time for that algorithm against the problem instance, and this is done only when run times differ significantly.
For MCQ there is no particular style that is a consistent winner. This is a surprise as MCQ3 is Tomita's MCR 
and in \cite{tomita2007} it is claimed that MCR was faster than MCQ. The evidence that supports this claim is
Table 2 of \cite{tomita2007}, 8 of the 66 DIMACS instances. For MCSa and BBMC style 2 is best more often than not, 
and in MCSb style 1 is best more often than not. Overall we see that 
the BBMC2 is our best algorithm, i.e. BBMC with a minimum width ordering. 

\begin{table}
\begin{center}
\begin{scriptsize}
\begin{tabular}{|l|c c c|c c c|c c c|c c c|} \hline 
\multicolumn{1}{|c|} {} & \multicolumn{3}{|c|}{MCQ} & \multicolumn{3}{|c|}{MCSa} & \multicolumn{3}{|c|}{MCSb} & \multicolumn{3}{|c|}{BBMC}\\
\multicolumn{1}{|c|}{instance} & $s_1$ & $s_2$ & $s_3$ & $s_1$ & $s_2$ & $s_3$ & $s_1$ & $s_2$ & $s_3$ & $s_1$ & $s_2$ & $s_3$ \\ \hline
brock200-1 & 868 & 592 & 510 & 524 & 301 & 320 & 245 & 257 & 267 & 524 & 301 & 320 \\ 
brock200-2 & 4 & 4 & 4 & 3 & 3 & 3 & 3 & 2 & 3 & 3 & 3 & 3 \\ 
brock200-3 & 17 & 18 & 17 & 14 & 13 & 14 & 11 & 15 & 16 & 14 & 13 & 14 \\ 
brock200-4 & 80 & 85 & 63 & 58 & 53 & 49 & 41 & 35 & 39 & 58 & 53 & 49 \\ 
brock400-1 & 342,473 & 266,211 & 455,317 & 198,359 & 132,762 & 278,967 & 142,253 & 208,625 & 114,839 & 198,359 & 13,2762 & 278,967 \\ 
brock400-2 & 224,839 & 381,976 & 125,166 & 145,597 & 178,500 & 76,368 & 61,327 & 151,834 & 154,293 & 145,597 & 178,500 & 76,368 \\ 
brock400-3 & 194,403 & 213,988 & 114,716 & 120,230 & 101,550 & 72,814 & 70,263 & 163,468 & 125,495 & 120,230 & 101,550 & 72,814 \\ 
brock400-4 & 82,056 & 36,456 & 148,294 & 54,440 & 19,306 & 90,918 & 68,252 & 62,725 & 31,887 & 54,440 & 19,306 & 90,918 \\ 
brock800-1 & --- & --- & --- & --- & --- & --- & --- & --- & --- & --- & --- & --- \\ 
brock800-2 & --- & --- & --- & --- & --- & --- & --- & --- & --- & --- & --- & --- \\ 
brock800-3 & --- & --- & --- & --- & --- & --- & --- & --- & --- & --- & 949,447 & 1,369,115 \\ 
brock800-4 & --- & --- & --- & 640,444 & 773,255 & --- & 659,145 & --- & --- & 640,444 & 773,255 & 1,440,844 \\ 
c-fat200-1 & 0 & 0 & 0 & 0 & 0 & 0 & 0 & 0 & 0 & 0 & 0 & 0 \\ 
c-fat200-2 & 0 & 0 & 0 & 0 & 0 & 0 & 0 & 0 & 0 & 0 & 0 & 0 \\ 
c-fat200-5 & 0 & 0 & 0 & 0 & 0 & 0 & 0 & 0 & 0 & 0 & 0 & 0 \\ 
c-fat500-10 & 0 & 0 & 0 & 0 & 0 & 0 & 0 & 0 & 0 & 0 & 0 & 0 \\ 
c-fat500-1 & 0 & 0 & 0 & 0 & 0 & 0 & 0 & 0 & 0 & 0 & 0 & 0 \\ 
c-fat500-2 & 0 & 0 & 0 & 0 & 0 & 0 & 0 & 0 & 0 & 0 & 0 & 0 \\ 
c-fat500-5 & 0 & 0 & 0 & 0 & 0 & 0 & 0 & 0 & 0 & 0 & 0 & 0 \\ 
hamming10-2 & 0 & 0 & 0 & 0 & 0 & 0 & 0 & 0 & 0 & 0 & 0 & 0 \\ 
hamming10-4 & --- & --- & --- & --- & --- & --- & --- & --- & --- & --- & --- & --- \\ 
hamming6-2 & 0 & 0 & 0 & 0 & 0 & 0 & 0 & 0 & 0 & 0 & 0 & 0 \\ 
hamming6-4 & 0 & 0 & 0 & 0 & 0 & 0 & 0 & 0 & 0 & 0 & 0 & 0 \\ 
hamming8-2 & 0 & 0 & 0 & 0 & 0 & 0 & 0 & 0 & 0 & 0 & 0 & 0 \\ 
hamming8-4 & 41 & 27 & 41 & 36 & 18 & 36 & 33 & 6 & 33 & 36 & 18 & 36 \\ 
johnson16-2-4 & 323 & 218 & 323 & 256 & 365 & 256 & 256 & 288 & 256 & 256 & 365 & 256 \\ 
johnson32-2-4 & --- & --- & --- & --- & --- & --- & --- & --- & --- & --- & --- & --- \\ 
johnson8-2-4 & 0 & 0 & 0 & 0 & 0 & 0 & 0 & 0 & 0 & 0 & 0 & 0 \\ 
johnson8-4-4 & 0 & 0 & 0 & 0 & 0 & 0 & 0 & 0 & 0 & 0 & 0 & 0 \\ 
keller4 & 13 & 11 & 13 & 13 & 11 & 13 & 10 & 8 & 10 & 13 & 11 & 13 \\ 
keller5 & --- & --- & --- & --- & --- & --- & --- & --- & --- & --- & --- & --- \\ 
keller6 & --- & --- & --- & --- & --- & --- & --- & --- & --- & --- & --- & --- \\ 
MANN-a27 & 38 & 38 & 38 & 38 & 38 & 38 & 38 & 34 & 38 & 38 & 38 & 38 \\ 
MANN-a45 & 2,851 & 2,952 & 2,851 & 2,851 & 2,952 & 2,851 & 2,545 & 2,428 & 2,545 & 2,851 & 2,952 & 2,851 \\ 
MANN-a81 & --- & --- & --- & --- & --- & --- & --- & --- & --- & --- & --- & --- \\ 
MANN-a9 & 0 & 0 & 0 & 0 & 0 & 0 & 0 & 0 & 0 & 0 & 0 & 0 \\ 
p-hat1000-1 & 237 & 252 & 241 & 176 & 171 & 178 & 151 & 148 & 148 & 176 & 171 & 178 \\ 
p-hat1000-2 & --- & --- & --- & 34,473 & 19,211 & 36,870 & 166,655 & 177,879 & 142,038 & 34,473 & 19,211 & 36,870 \\ 
p-hat1000-3 & --- & --- & --- & --- & --- & --- & --- & --- & --- & --- & --- & --- \\ 
p-hat1500-1 & 1,642 & 1,792 & 1,888 & 1,184 & 1,201 & 1,350 & 990 & 957 & 1,153 & 1,184 & 1,201 & 1,350 \\ 
p-hat1500-2 & --- & --- & --- & --- & --- & --- & --- & --- & --- & --- & --- & --- \\ 
p-hat1500-3 & --- & --- & --- & --- & --- & --- & --- & --- & --- & --- & --- & --- \\ 
p-hat300-1 & 1 & 2 & 1 & 1 & 1 & 1 & 1 & 1 & 1 & 1 & 1 & 1 \\ 
p-hat300-2 & 13 & 28 & 19 & 4 & 3 & 5 & 4 & 5 & 4 & 4 & 3 & 5 \\ 
p-hat300-3 & 3,829 & 7,117 & 3,997 & 624 & 488 & 639 & 713 & 821 & 640 & 624 & 488 & 639 \\ 
p-hat500-1 & 12 & 13 & 13 & 9 & 9 & 9 & 8 & 8 & 9 & 9 & 9 & 9 \\ 
p-hat500-2 & 1,022 & 910 & 1,085 & 114 & 83 & 125 & 137 & 117 & 95 & 114 & 83 & 125 \\ 
p-hat500-3 & --- & --- & --- & 39,260 & 16,963 & 30,908 & 104,684 & 152,786 & 111,639 & 39,260 & 16,963 & 30,908 \\ 
p-hat700-1 & 36 & 41 & 34 & 26 & 28 & 24 & 22 & 21 & 19 & 26 & 28 & 24 \\ 
p-hat700-2 & 18,968 & 19,066 & 12,782 & 750 & 634 & 593 & 1490 & 1,964 & 2,244 & 750 & 634 & 593 \\ 
p-hat700-3 & --- & --- & --- & --- & 216,516 & ---  & --- & --- & --- & 282,412 & 216,516 & 297,114 \\ 
san1000 & 302 & 307 & 297 & 150 & 150 & 147 & 53 & 50 & 50 & 150 & 150 & 147 \\ 
san200-0.7-1 & 12 & 1 & 3 & 13 & 11 & 11 & 0 & 0 & 0 & 13 & 11 & 11 \\ 
san200-0.7-2 & 0 & 0 & 0 & 0 & 0 & 0 & 0 & 0 & 0 & 0 & 0 & 0 \\ 
san200-0.9-1 & 0 & 0 & 6 & 87 & 0 & 84 & 0 & 0 & 0 & 87 & 0 & 84 \\ 
san200-0.9-2 & 1,149 & 4,305 & 2,081 & 229 & 65 & 243 & 62 & 49 & 31 & 229 & 65 & 243 \\ 
san200-0.9-3 & 8,260 & 233 & 3,169 & 6,815 & 13 & 3,570 & 1,218 & 119 & 238 & 6,815 & 13 & 3,570 \\ 
san400-0.5-1 & 3 & 1 & 1 & 2 & 2 & 2 & 1 & 1 & 1 & 2 & 2 & 2 \\ 
san400-0.7-1 & 55 & 115 & 93 & 119 & 662 & 147 & 134 & 15 & 51 & 119 & 662 & 147 \\ 
san400-0.7-2 & 606 & 1,684 & 670 & 889 & 881 & 929 & 754 & 312 & 155 & 889 & 881 & 929 \\ 
san400-0.7-3 & 582 & 1,882 & 2,290 & 521 & 921 & 1,913 & 215 & 550 & 998 & 521 & 921 & 1,913 \\ 
san400-0.9-1 & --- & --- & --- & 4,536 & --- & 220,183 & 582 & 15 & --- & 4536 & --- & 220,183 \\ 
sanr200-0.7 & 206 & 290 & 224 & 152 & 175 & 164 & 100 & 123 & 111 & 152 & 175 & 164 \\ 
sanr200-0.9 & 44,472 & 101,008 & 62,243 & 14,921 & 12,513 & 20,607 & 9,730 & 8,100 & 19,042 & 14,921 & 12,513 & 20,607 \\ 
sanr400-0.5 & 380 & 418 & 351 & 320 & 320 & 299 & 190 & 183 & 203 & 320 & 320 & 299 \\ 
sanr400-0.7 & 101,213 & 106,688 & 101,531 & 64,412 & 54,359 & 64,131 & 46,125 & 44,880 & 48,664 & 64,412 & 54,359 & 64,131 \\ \hline
\end{tabular}
\end{scriptsize}
\end{center}
\caption{DIMACS instances: the effect of style on search nodes in 1,000's}
\label{tabNodes}
\end{table}

\subsection{MCSa: tie breaking in colour classes}
\begin{figure}
\vspace{-7cm}
\begin{center}
\includegraphics[height=15.0cm]{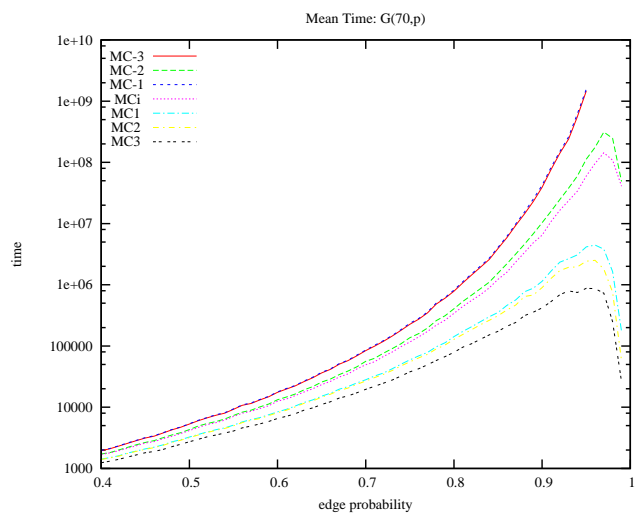}
\end{center}
\vspace{-1cm}
\caption{MC applied to G(70,p) with different initial orderings.}
\label{mc-70}
\end{figure}

It has often been reported that the order that vertices are chosen for expansion can have a profound effect on search effort.
This is demonstrated in Figure \ref{mc-70} where the algorithm MC0 was applied to $G(70,p)$ with a sample
size of 100 using all three styles (MC1, MC2, MC3), using index order (MCi) and the reverse orderings (MC-1, MC-2, MC-3).
Plotted is edge probability against logarithm of average run time. MC (and MC0) expand vertices in the candidate
set from last to first therefore with a style of 1 vertices are expanded in non-decreasing degree order (smallest first).
Figure \ref{mc-70} shows that the order can have an enormous effect. Styles -1 and -3 (largest degree first) are orders of magnitude
worse than styles 1 and 3 (smallest degree first). In fact the MC-1 and MC-3 experiments were abandoned after 72 hours with
$G(70,0.96)$ incomplete. This suggests that ordering vertices in colour classes may be worthwhile.

The $numberSort$ method used by MCS delivers a colour-ordered candidate set. Vertices are picked out of colour classes
and added to the candidate set in the order they were added to the colour class. If vertices are coloured in 
non-increasing degree order, vertices in a colour class will also be in non-increasing degree order. Consequently the candidate set
will be in non-decreasing colour order tie-breaking on non-increasing degree. The expand method iterates over the candidate
set from last to first and expands vertices of the same colour class in non-increasing degree order. What might happen if 
this was reversed so that we visit vertices from the same colour class in non-decreasing degree order? 

\begin{figure}
\vspace{-7cm}
\begin{center}
\includegraphics[height=15.0cm]{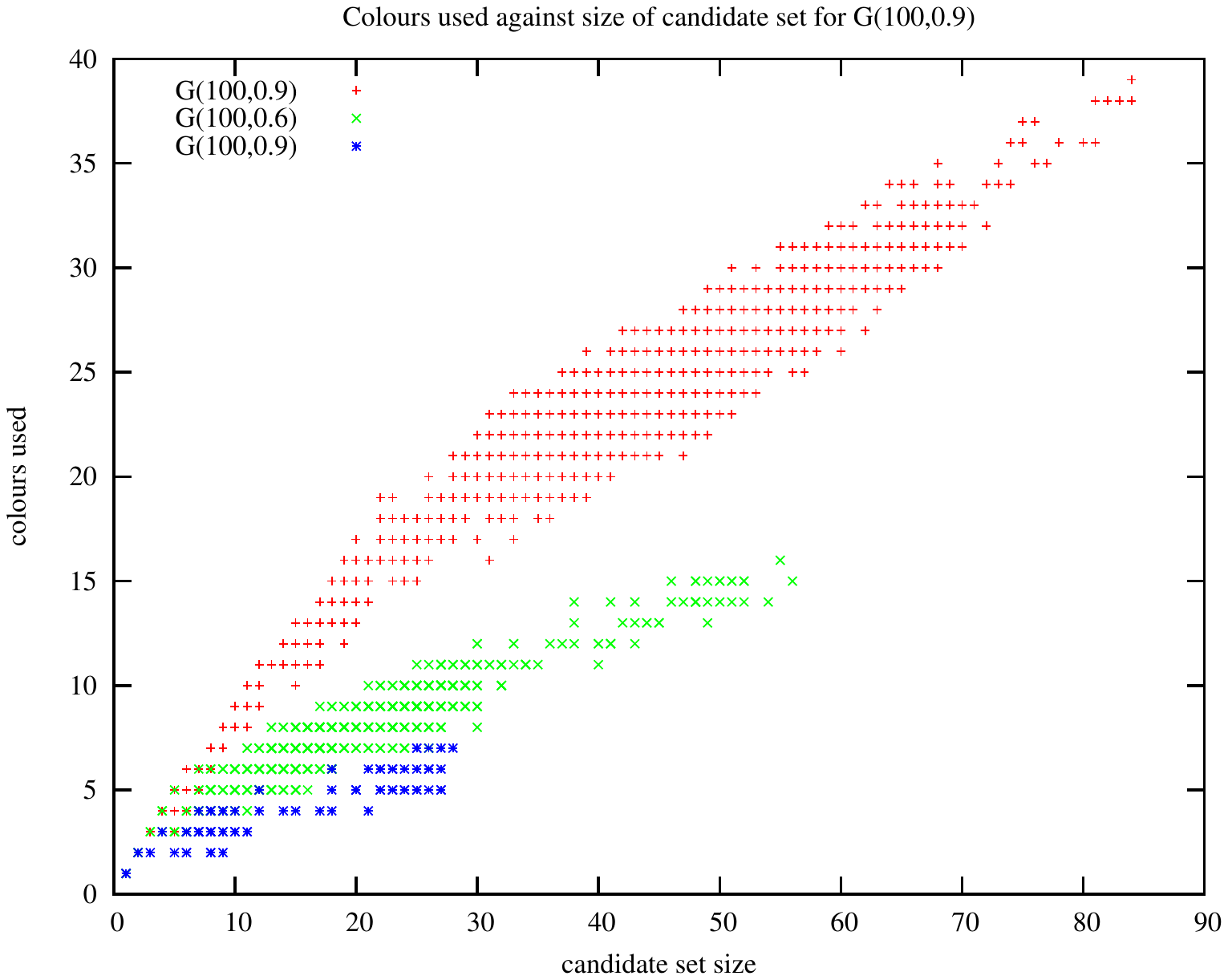}
\end{center}
\vspace{-1cm}
\caption{How many vertices are in a colour class? We plot colours used against size of candidate set for MCSa1 and an instance of $G(100,0.9)$,
$G(100,0.6)$ and $G(100,0.3)$}
\label{colClass}
\end{figure}

This was done by altering the for loop of line 59 in $numberSort$ Listing \ref{codeMCQa} so that colour classes are
processed in reverse order. Experiments were then run on $G(100,p)$ and
$G(150,p)$ using MCSa with each of our three orderings. The effect was insignificant. 
Why might that be? Three individual problems were analysed from $G(100,30)$, $G(100,0.6)$ and $G(100,0.9)$. 
Within each call to expand the size of the candidate set and the number of colours used was logged. This is 
presented in Figure \ref{colClass}. What we see is that
for $G(100,0.9)$ a candidate set of size of $m$ typically required $m/2$ colours, therefore a colour class typically
contained two vertices and there is little scope for tie-breaking to have an effect. When problems are sparse
colour classes get larger (typically 4 to 6 vertices per colour class in $G(100,0.6)$)
but they are easy and again tie-breaking makes little if any gain.

\subsection{More Benchmarks (not DIMACS)}
In \cite{eppstein2011} experiments are performed on counting maximal cliques in exceptionally
large sparse graphs, such as the Pajek data sets 
(graphs with hundreds of thousands of 
vertices\footnote{Available from http://vlado.fmf.uni-lj.si/pub/networks/data/}) and 
SNAP data sets  
(graphs with vertices in the millions\footnote{Available from http://snap.stanford.edu/data/index.html}). Those
graphs are out of the reach of the exact algorithms reported here. The initial reason for this is space consumption.
To tackle such large sparse problems we require a change of representation, away from the adjacency matrix
and towards the adjacency lists as used in \cite{eppstein2011}. Therefore we explore large random
instances as in \cite{segundo2011,tomita2010} to further investigate ordering and the effect of the BitSet representation,
the hard solvable instances in BHOSLIB to see how far we can go, and structured graphs produced via the SNAP 
(Stanford Network Analysis Project) graph generator. But first, we start with BHOSLIB.

In Table \ref{bhoslib} we have the only instances from the BHOSLIB suite 
(Benchmarks with Hidden Optimum Solutions\footnote{Available from http://www.nisde.buaa.edu.cn/$\sim$kexu/benchmarks/graph-benchmarks.htm})
that could be solved in 4 hours. Each instance has a maximum clique of size 30.
A {\bf bold} entry is the best run time. For this suite we see that with respect to style there is no clear winner.

\begin{table}
\begin{center}
\begin{tabular}{|l|c|c|c c|c c|c c|} \hline 
instance & n & edges & \multicolumn{2}{|c|}{BBMC1} & \multicolumn{2}{|c|}{BBMC2} & \multicolumn{2}{|c|}{BBMC3} \\ \hline
frb30-15-1 & 450 & 83,198 & 292,095 & \bf{3,099} & 626,833 & 6,503 & 361,949 & 3,951  \\ 
frb30-15-2 & 450 & 83,151 & 557,252 & 5,404 & 599,543 & 6,136 & 436,110 & \bf{4,490}  \\ 
frb30-15-3 & 450 & 83,126 & 167,116 & 1,707 & 265,157 & 2,700 & 118,495 & \bf{1,309}  \\ 
frb30-15-4 & 450 & 83,194 & 991,460 & 9,663 & 861,391 & \bf{8,513} & 1,028,129 & 9,781  \\ 
frb30-15-5 & 450 & 83,231& 282,763 & 2,845 & 674,987 & 7,033 & 281,152 & \bf{2,802}  \\ \hline
\end{tabular}
\end{center}
\caption{BHOSLIB using BBMC: 1,000's of nodes and run time in seconds. Problems have 450 vertices and graph density 0.82}
\label{bhoslib}
\end{table}

Table \ref{tableBig} shows results on large random problems. Similar experiments are reported in Table 4 and 5 of \cite{segundo2011} and
Table 2 in \cite{tomita2010}. The first three columns are the nodes visited, and this is the same for MCSa and BBMC. Run times are then given in
seconds for MCSa and BBMC using each of the tree styles. Highlighted in {\bf bold} is the search of fewest nodes and this 
is style 2 (minimum width ordering) in all but one case. Comparing the run times we see that as problems get larger, involving more
vertices, the relative speed difference between BBMC and MCSa diminishes and at $n=15,000$ MCSa and BBMC's performances are substantially the same.

\begin{table}
\begin{center}
\begin{scriptsize}
\begin{tabular}{|c c|c c c|c c c|c c c|} \hline 
\multicolumn{2}{|c|} {instance} & \multicolumn{3}{|c|}{nodes} & \multicolumn{3}{|c|}{MCSa} & \multicolumn{3}{|c|}{BBMC}\\
n & p & $s_1$ & $s_2$ & $s_3$ & $s_1$ & $s_2$ & $s_3$ & $s_1$ & $s_2$ & $s_3$ \\ \hline
1,000 & 0.1 & 4,536 & \bf{4,472} & 4,563 & 0 & 0 & 0 & 0 & 0 & 0  \\
      & 0.2 & 39,478 & \bf{38,250} & 38,838& 0 & 0 & 0 & 0 & 0 & 0 \\
      & 0.3 & 400,018 & \bf{371,360} & 404,948 & 4 & 4 & 4 & 2 & 2 & 2 \\
      & 0.4 & 3,936,761 & \bf{3,780,737} & 4,052,677 & 40 & 39 & 38 & 26 & 25 & 26 \\
      & 0.5 & 79,603,712 & \bf{75,555,478} & 80,018,645 & 860 & 910 & 859 & 570 & 574 & 604 \\
3,000 & 0.1 & 144,375 & \bf{142,719} & 145,487 & 3 & 3 & 3 & 2 & 2 & 2 \\
      & 0.2 & 2,802,011 & \bf{2,723,443} & 2,804,830 & 38 & 38 & 38 & 32 & 32 & 32 \\
      & 0.3 & 73,086,978 & \bf{71,653,889} & 73,354,584 & 964 & 960 & 978 & 926 & 930 & 931 \\
10,000 & 0.1 & 5,351,591 & \bf{5,303,615} & 5,432,812 & 236 & 252 & 245 & 212 & 216 & 214 \\
15,000 & 0.1 & 22,077,212 & 21,751,100 & \bf{21,694,036} & 1,179 & 1,117 & 1,081 & 1,249 & 1,235 & 1,208 \\ \hline
\end{tabular}
\end{scriptsize}
\end{center}
\caption{Large random graphs, sample size 10}
\label{tableBig}
\end{table}

The graphgen program was downloaded from the SNAP web site and modified to use a random seed so that generated graphs
with the same parameters were actually different. This allows us to generate a variety of graphs, such as 
complete graphs, star graphs, 2D grid graphs, Erd\'{o}s-R\"{e}nyi random graphs with an exact number of edges, 
k-regular graphs (each vertex with degree k), Albert-Barbasi graphs, power law graphs, Klienberg copying model
graphs and small-world graphs. Finding maximum cliques in a complete graph, star graph and 2D grid graph is trivial.
Similarly, and surprisingly, small scale experiments suggested that Albert-Barbasi and Klienberg's graphs are also easy with respect
to maximum clique. However k-regular and small world are a challenge.

\begin{figure}
\vspace{-7cm}
\begin{center}
\includegraphics[height=15.0cm]{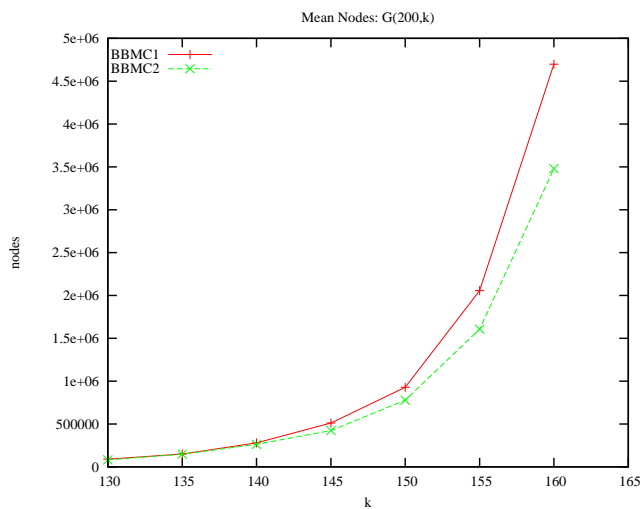}
\end{center}
\vspace{-1cm}
\caption{k-regular SNAP instances $KR(200,k)$, $130\leq k \leq 160$, sample size 20.}
\label{bbmc-regular-200}
\end{figure}

The SNAP graphgen program was used to generated k-regular graphs $KR(n,k)$, i.e. random graphs 
with $n$ vertices each with degree $k$. Graphs were generated with $n=200$ and $50 \leq k \leq 160$,
with $k$ varying in steps of 5, 20 instances at each point. BBMC1 and BBMC2 were then applied to each instance.
Obviously, with style equal to 1 or 3, there is no heuristic information to be exploited at the top of search.
But would a minimum width ordering, style 2, have an advantage? Figure \ref{bbmc-regular-200} shows average search effort
in nodes plotted against uniform degree $k$. We see that minimum width ordering does indeed have an advantage. What is also of
interest is that $KR(n,k)$ instances tend to be harder than their $G(n,p)$ equivalents. For example, we can compare
$KR(200,160)$ with $G(200,0.8)$ in Figure \ref{mcqVmcs}: MCSa1 took on average 1.9 million nodes for $G(200,0.8)$ and
BBMC1 took on average 4.7 million nodes on the twenty $KR(200,160)$ instances.

\begin{figure}
\vspace{-7cm}
\begin{center}
\hspace{-3.5cm}
\begin{minipage}[t]{0.49\textwidth}
\includegraphics[height=14.0cm]{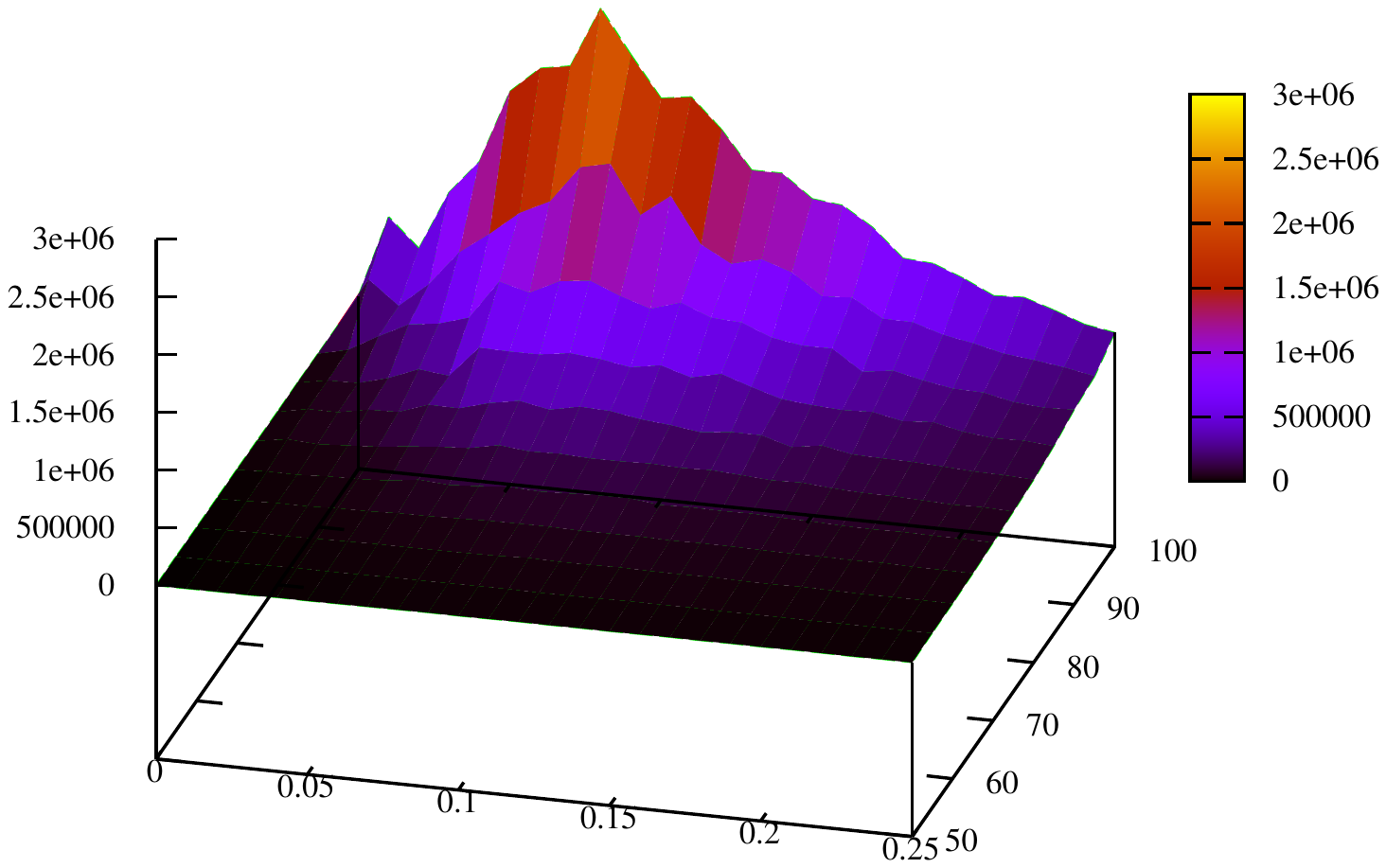}
\end{minipage}
\begin{minipage}[t]{0.49\textwidth}
\includegraphics[height=14.0cm]{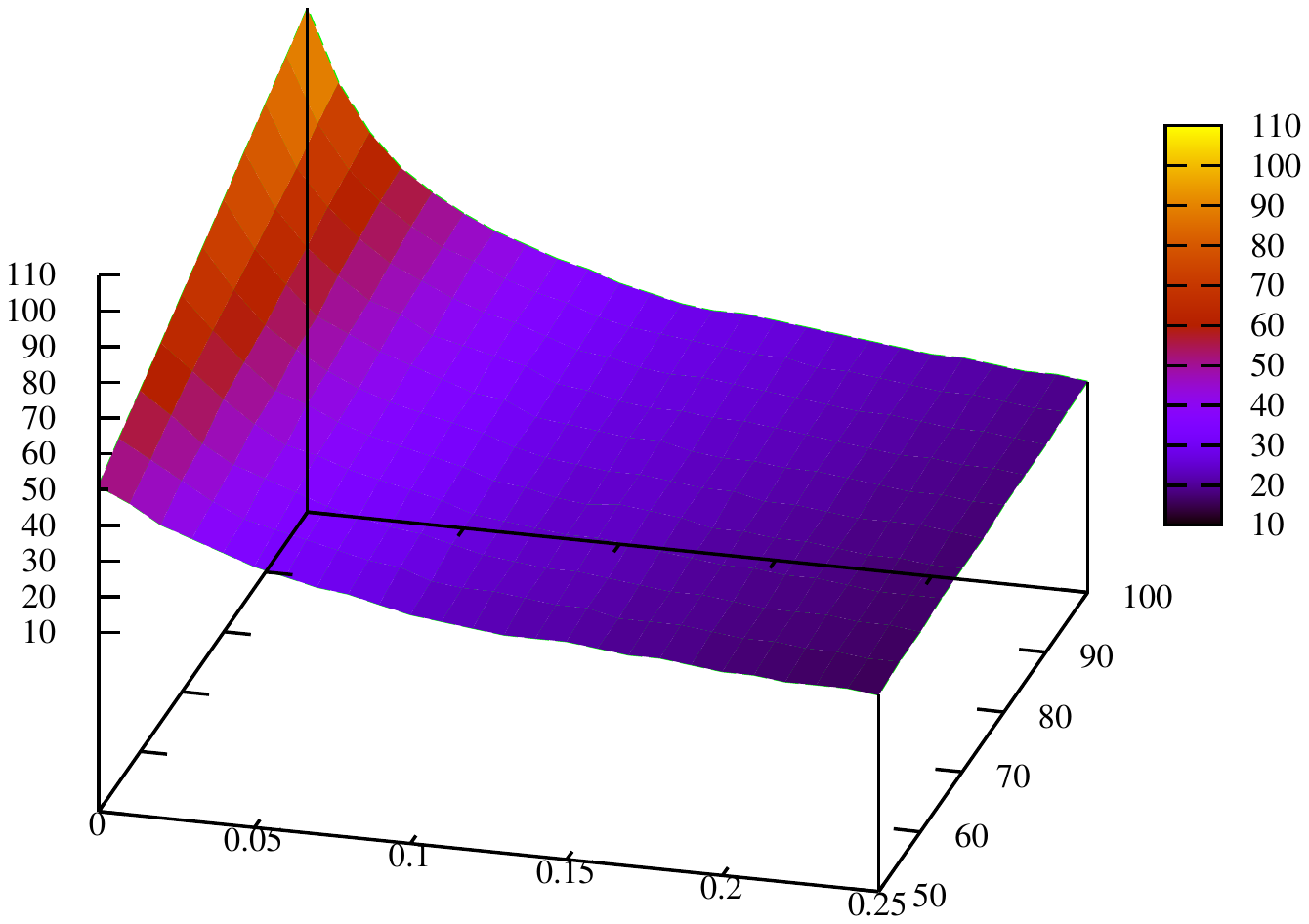}
\end{minipage}
\end{center}
\vspace{-1cm}
\caption{Small world graphs $SW(200,k,p)$. On the left search effort and on the right maximum clique size.}
\label{smallWorld}
\end{figure}

Small-world graphs $SW(n,k,p)$ were then generated using graphgen. This takes three parameters: $n$ the number of vertices, $k$ where each vertex
is connected to $k$ nearest neighbours to the right in a ring topology (i.e. vertices start with uniform degree $2k$), and $p$ is a 
rewiring probability. This corresponds to the
graphs in Figure 1 of \cite{smallWorld}. Small-world graphs were generated with $n = 1,000$, $50 \leq k \leq 100$ in steps of 5,
$0.0 \leq p \leq 0.25$ in steps of 0.01, 10 graphs at each point. BBMC1 was then applied to each instance to investigate
how difficulty of finding a maximum clique varies with respect to $k$ and $p$ and also how size of maximum clique varies,
i.e. this is an investigation of the problem.
The results are shown as three dimensional plots in Figure \ref{smallWorld}: on the left average search effort and on the right
average maximum clique size. Looking at the graph on the left: when $p = 0.0$ problems are easy, as $p$ increases and randomness
is introduced problems quickly get hard, but as $p$ continues to increase the graphs tend to become predominantly random 
and behave more like large sparse random graphs and get easier. We also see that as neigbourhood size $k$ increases problems get harder. We
can compare the $SW(1000,100,p)$ to the graphs $G(1000,0.2)$ in Table \ref{tableBig}: $G(1000,0.2)$ took on
average 39,478 nodes whereas $SW(1000,100,0.01)$ took 709,347 nodes, $SW(1000,100,0.08)$ took 2,702,199 nodes and 
$SW(1000,100,0.25)$ 354,430 nodes. Clearly small-world instances are relatively hard. Looking at the graph on the right (average maximum clique size)
we see that as rewiring probability $p$ increases maximum cliques size decreases and as $k$ increases so too does maximum clique size.

\subsection{Calibration of results}
To compare computational results across publications authors compile and run a standard C program, dfmax, against a 
set of benchmarks. These run times are then used as a conversion factor, 
and results are then taken from one publication, scaled accordingly, and then included in another publication. Recent examples of this are
\cite{prjo2002} including rescaled results from \cite{sewell98};
\cite{regin2003} including rescaled results from \cite{prjo2002}, \cite{wood97} and \cite{fahle};
\cite{tomita2007} including rescaled results from \cite{prjo2002} and \cite{sewell98};
\cite{segundo2011} including rescaled  results from \cite{Konc_Janezic_2007}; 
\cite{segundo2011b} including rescaled  results from \cite{segundo2011};
\cite{aaai2010} including rescaled results from \cite{tomita2007} and \cite{regin2003}.
Is this procedure safe?

To test this we take two additional machines, Fais and Daleview, and calibrate them with respect to our reference machine Cyprus. 
We then run experiments 
on each machine using the Java implementations of the algorithms implemented here against some of the DIMACS benchmarks. These results are
then rescaled. If the rescaling gives substantially different results from those on the reference machine
this would suggest that this technique is not safe.

Table \ref{rosettaStone} gives a ``Rosetta Stone'' for the three machines used in this study.
The standard program dfmax \footnote{Available from ftp://dimacs.rutgers.edu/pub/dsj/clique} was compiled using 
gcc and the -O2 compiler option on each machine and then run
on the benchmarks r* on each machine. Run times in seconds are tabulated for the five benchmark instances,
each machine's /proc/cpuinfo is given and a conversion factor relative to the reference machine Cyprus
is then computed in the same manner as that reported in
\cite{segundo2011} \emph{``... the first two graphs from the benchmark were removed (user time was considered too small) and the rest
of the times averaged ...''.}  Therefore when rescaling the run times from Fais we
multiply actual run time by 0.41 and for Daleview by 0.50.

\begin{table}
\begin{center}
\begin{tabular}{|l|c|c|c|c|c|c|c|c|c|c|} \hline 
machine & r100.5 & r200.5 & r300.5 & r400.5 & r500.5 & Intel(R) & GHz & cache & Java & scaling factor\\ \hline
Cyprus & 0.0 & 0.02 & 0.24 & 1.49 & 5.58 & Xeon(R) E5620 & 2.40 & 12,288KB & 1.6.0\_07 & 1 \\ \hline
Fais & 0.0 & 0.08 & 0.58 & 3.56 & 13.56 & XEON(TM) CPU & 2.40& 512KB & 1.5.0\_06 & 0.41 \\ \hline
Daleview & 0.0 & 0.09 & 0.53 & 3.00 & 10.95 & Atom(TM) N280 & 1.66 & 512KB & 1.6.0\_18 & 0.50 \\ \hline
\end{tabular}
\end{center}
\caption{Conversion factors using dfmax on three machines: Cyprus, Fais and Daleview}
\label{rosettaStone}
\end{table}

\begin{table}
\begin{center}
\begin{tabular}{|l|c c|c c|c c|c c|c c|c c|} \hline 
\multicolumn{1}{|c|} {} & \multicolumn{6}{|c|}{MCSa1} & \multicolumn{6}{|c|}{BBMC1}\\
\multicolumn{1}{|c|} {instance} & \multicolumn{2}{|c}{Fais} & \multicolumn{2}{c}{Daleview} & \multicolumn{2}{c|}{Cyprus} & \multicolumn{2}{|c}{Fais} & \multicolumn{2}{c}{Daleview} & \multicolumn{2}{c|}{Cyprus} \\ \hline
brock200-1 & 0.25 & (19,343) & 0.27 & (17,486) & 1.00 & (4,777) & 0.15 & (15,365) & 0.09 & (25,048) & 1.00 & (2,358) \\ 
brock200-4 & 0.40 & (1,870) & 0.43 & (1,765) & 1.00 & (755) & 0.20 & (1,592) & 0.13 & (2,464) & 1.00 & (321) \\ 
hamming10-2 & 0.18 & (1,885) & 0.14 & (2,299) & 1.00 & (333) & 0.25 & (608) & 0.21 & (710) & 1.00 & (151) \\ 
hamming8-4 & 0.24 & (1,885) & 0.28 & (1,647) & 1.00 & (455) & 0.23 & (1,625) & 0.19 & (1,925) & 1.00 & (367) \\ 
johnson16-2-4 & 0.35 & (2,327) & 0.38 & (2,173) & 1.00 & (823) & 0.26 & (1,896) & 0.14 & (3,560) & 1.00 & (495) \\ 
MANN-a27 & 0.21 & (32,281) & 0.22 & (31,874) & 1.00 & (6,912) & 0.14 & (12,335) & 0.10 & (16,491) & 1.00 & (1,676) \\ 
p-hat1000-1 & 0.25 & (8,431) & 0.28 & (7,413) & 1.00 & (2,108) & 0.14 & (8,359) & 0.12 & (9,389) & 1.00 & (1,169) \\ 
p-hat1500-1 & 0.19 & (77,759) & 0.22 & (66,113) & 1.00 & (14,421) & 0.11 & (90,417) & 0.10 & (92,210) & 1.00 & (9,516) \\ 
p-hat300-3 & 0.25 & (53,408) & 0.26 & (51,019) & 1.00 & (13,486) & 0.14 & (41,669) & 0.09 & (60,118) & 1.00 & (5,711) \\ 
p-hat500-2 & 0.27 & (13,400) & 0.30 & (12,091) & 1.00 & (3,659) & 0.14 & (10,177) & 0.11 & (13,410) & 1.00 & (1,428) \\ 
p-hat700-1 & 0.40 & (1,615) & 0.51 & (1,251) & 1.00 & (641) & 0.29 & (1,169) & 0.24 & (1,422) & 1.00 & (344) \\ 
san1000 & 0.11 & (94,107) & 0.12 & (89,330) & 1.00 & (10,460) & 0.10 & (57,868) & 0.11 & (54,816) & 1.00 & (5,927) \\ 
san200-0.9-1 & 0.29 & (4,918) & 0.31 & (4,705) & 1.00 & (1,444) & 0.18 & (4,201) & 0.11 & (6,588) & 1.00 & (748) \\ 
san200-0.9-2 & 0.22 & (23,510) & 0.25 & (20,867) & 1.00 & (5,240) & 0.15 & (14,572) & 0.09 & (23,592) & 1.00 & (2,218) \\ 
san400-0.7-1 & 0.25 & (10,230) & 0.27 & (9,607) & 1.00 & (2,573) & 0.15 & (8,314) & 0.12 & (10,206) & 1.00 & (1,260) \\ 
san400-0.7-2 & 0.23 & (84,247) & 0.27 & (72,926) & 1.00 & (19,565) & 0.13 & (71,360) & 0.11 & (87,325) & 1.00 & (9,219) \\ 
san400-0.7-3 & 0.24 & (45,552) & 0.27 & (40,792) & 1.00 & (10,839) & 0.13 & (39,840) & 0.11 & (46,818) & 1.00 & (5,162) \\ 
sanr200-0.7 & 0.31 & (5,043) & 0.33 & (4,676) & 1.00 & (1,548) & 0.19 & (4,079) & 0.12 & (6,652) & 1.00 & (795) \\ 
sanr400-0.5 & 0.28 & (9,898) & 0.31 & (8,754) & 1.00 & (2,745) & 0.16 & (9,177) & 0.12 & (12,658) & 1.00 & (1,484) \\ \hline
ratio (total)    & 0.21 & (491,709) & 0.23 & (446,788) & 1.00 & (102,784) & 0.13 & (394,623) & 0.11 & (475,402) & 1.00 & (50,349) \\ \hline
\end{tabular}
\end{center}
\caption{Calibration experiments using 3 machines, 2 algorithms and a subset of DIMACS benchmarks}
\label{calibration}
\end{table}

Table \ref{calibration} shows the results of the calibration experiments. Tabulated are DIMACS benchmark instances that
took more than 1 second and less than 2 minutes to solve using MCSa1 on our second slowest machine (Fais). Run times are tabulated in
milliseconds (in brackets) and the actual ratio of Cyprus-time over Fais-time (expected to be 0.41) is given as well as Cyprus-time over
Daleview-time (expected to be 0.50) for each data point. Two algorithms are used, MCSa1 and BBMC1. The last row of
Table \ref{calibration} gives the 
relative performance ratios computed using the sum of the run times in the table. Referring back to Table \ref{rosettaStone} we expect a 
Cyprus/Fais ratio of 0.41 but empirically get
0.21 when using MCSa1 and 0.13 when using BBMC1, and expect a Cyprus/Daleview ratio of 0.50 but empirically
get an average 0.23 with MCSa1 and 0.11 with BBMC1. The conversion factors in Table \ref{rosettaStone} consistently
over-estimate the speed of Fais and Daleview. For example, we would expect MCSa1 applied to brock200-1 on Fais to have a run time
of $19,343 \times 0.41 = 7,930$ milliseconds on Cyprus. In fact it takes 4,777 milliseconds. If we use the derived ratio
in the last row of Table \ref{calibration} we get $19,343 \times 0.21 = 4,062$ milliseconds, closer to actual performance on Cyprus. 
As another example consider san1000 using BBMC1 on Daleview. We would expect this to take $54,816 \times 0.50 = 27,408$
milliseconds on Cyprus. In fact it takes 5,927 milliseconds! If we use the conversion ratio from the last row of Table \ref{calibration}
we get a more accurate estimate $54,816 \times 0.11 = 6,030$ milliseconds.

\begin{table}
\begin{center}
\begin{tabular}{|l|c c|c c|c c|c c|c c|c c|} \hline 
\multicolumn{1}{|c|} {} & \multicolumn{6}{|c|}{Cliquer} & \multicolumn{6}{|c|}{dfmax}\\
\multicolumn{1}{|c|} {instance} & \multicolumn{2}{|c}{Fais} & \multicolumn{2}{c}{Daleview} & \multicolumn{2}{c|}{Cyprus} & \multicolumn{2}{|c}{Fais} & \multicolumn{2}{c}{Daleview} & \multicolumn{2}{c|}{Cyprus} \\ \hline
brock200-1 & 0.66 & (9,760) & 0.43 & (18,710) & 1.00 & (6,490) & 0.39 & (25,150) & 0.42 & (23,020) & 1.00 & (9,730) \\
brock200-4 & 0.64 & (690) & 0.47 & (1,190) & 1.00 & (440) & 0.41 & (1,510) & 0.46 & (1,360) & 1.00 & (620) \\
p-hat1000-1 &0.62 & (1,750) & 0.36 & (3,020) & 1.00 & (1,090) & 0.41 & (1,680) & 0.45 & (1,540) & 1.00 & (690) \\
p-hat700-1 & 0.67 & (150) & 0.37 & (270) & 1.00 & (100) & --- & --- & --- & --- & --- & --- \\
san1000 & 0.75 & (120) & 0.30 & (300) & 1.00 & (90) & --- & --- & --- & --- & --- & --- \\
san200-0.7-1 & 0.48 & (1,750) & 0.20 & (4,220) & 1.00 & (840) & --- & --- & --- & --- & --- & --- \\
san200-0.9-2 & 0.61 & (18,850) & 0.21 & (53,970) & 1.00 & (11,530) & --- & --- & --- & --- & --- & --- \\
san400-0.7-3 & 0.62 & (6,800) & 0.26 & (16,100) & 1.00 & (4,230) & --- & --- & --- & --- & --- & --- \\
sanr200-0.7 & 0.65 & (2,940) & 0.36 & (5,270) & 1.00 & (1,900) & 0.40 & (5,240) & 0.44 & (4,770) & 1.00 & (2,080) \\
sanr400-0.5 & 0.62 & (1,490) & 0.38 & (2,420) & 1.00 & (930) & 0.41 & (3,550) & 0.47 & (3,080) & 1.00 & (1,460) \\ \hline
ratio (total) & 0.62 & (44,300) & 0.26 & (105,470)& 1.00 & (27,640) & 0.39 & (37,130) & 0.43 & (33,770) & 1.00 & (14,580) \\ \hline
\end{tabular}
\end{center}
\caption{Calibration experiments for Cliquer and dfmax using 3 machines.}
\label{cliquerCalibration}
\end{table}

But maybe this is because we have used a C program (dfmax) to calibrate a Java program. Would we get a reliable calibration if a C 
program was used? \"{O}sterg\aa{}rd's Cliquer program was downloaded and compiled on our three machines and run against
DIMACS benchmarks, i.e. the experiments in Table \ref{calibration} were repeated using Cliquer and dfmax with a different, and easier, 
set of problems. The results are shown
in Table \ref{cliquerCalibration}\footnote{An entry --- was a run of dfmax that was terminated after 2 minutes.}. What we see is an actual 
scaling factor of 0.62 for Cliquer on Fais when dfmax predicts 0.41 
and for Cliquer on Daleview 0.26 when we expect 0.50; again we see that the rescaling procedure fails. The last three
columns show a dfmax calibration using problems other than the r* benchmarks and here we see an
error of about 5\% on Fais (expected 0.41, actual 0.39) and  about 16\% on Daleview (expected 0.50, actual 0.43). 
Therefore it appears that rescaling results using dfmax and the five r* benchmarks is not a safe procedure and can result in wrong
conclusions being drawn regarding the relative performance of algorithms.

\subsection{Relative algorithmic performance on different machines}
But is it even safe to draw conclusions on our algorithms when we base those conclusions on experiments
performed on a single machine? Previously, in Table \ref{tableMCSvBBMC} we compared MCSa against BBMC on our
reference machine Cyprus and concluded that BBMC was typically twice as fast as MCSa. Will that hold on
Fais and on Daleview? Table \ref{calibrationAlg} takes the data from Table \ref{calibration} and divides the
run time of MCSa by BBMC for each instance on our three machines. 
On Fais BBMC is rarely more than 50\% faster than MCSa and on Daleview BBMC is slower than MCSa more often than not! 
If experiments were performed
only on Daleview using only the DIMACS instances we might draw entirely different conclusions and claim that BBMC is slower than MCSa.
This change in relative algorithmic ordering has been observed on five different machines (four using the Java 1.6.0) 
using all of the algorithms.\footnote{The -server and -client options were also tried. The -server option sometimes
gave speedups of a factor of 2 sometimes a factor of 0.5, and this can also affect relative algorithmic performance.}

\begin{table}
\begin{center}
\begin{tabular}{|l|c|c|c|}  \hline
instance & ~~~Fais~~~ & Daleview & Cyprus \\ \hline
brock200-1 & 1.26  & 0.70  & 2.03  \\ 
brock200-4 & 1.17  & 0.72  & 2.35  \\ 
hamming10-2 & 3.10  & 3.24  & 2.21  \\ 
hamming8-4 & 1.16  & 0.86  & 1.24  \\ 
johnson16-2-4 & 1.23  & 0.61  & 1.66  \\ 
MANN-a27 & 2.62  & 1.93  & 4.12  \\ 
p-hat1000-1 & 1.01  & 0.79  & 1.80  \\ 
p-hat1500-1 & 0.86  & 0.72  & 1.52  \\ 
p-hat300-3 & 1.28  & 0.85  & 2.36  \\ 
p-hat500-2 & 1.32  & 0.90  & 2.56  \\ 
p-hat700-1 & 1.38  & 0.88  & 1.86  \\ 
san1000 & 1.63  & 1.63  & 1.76  \\ 
san200-0.9-1 & 1.17  & 0.71  & 1.93  \\ 
san200-0.9-2 & 1.61  & 0.88  & 2.36  \\ 
san400-0.7-1 & 1.23  & 0.94  & 2.04  \\ 
san400-0.7-2 & 1.18  & 0.84  & 2.12  \\ 
san400-0.7-3 & 1.14  & 0.87  & 2.10  \\ 
sanr200-0.7 & 1.24  & 0.70  & 1.95  \\ 
sanr400-0.5 & 1.08  & 0.69  & 1.85  \\ \hline
\end{tabular}
\end{center}
\caption{Calibration experiment part 2, does hardware affect relative algorithmic performance? 
Values greater than 1 imply BBMC is faster than MCSa, less than 1 MCSa is faster.}
\label{calibrationAlg}
\end{table}

\section{Conclusion}
\vspace{-1.5mm}
We have seen that small implementation details (in MC) can result in large changes in performance. Modern
programming languages with rich constructs and large libraries of utilities makes it easier for the programmer
to do this. We have also drifted away from the days when algorithms were presented along with their 
implementation code (examples here are \cite{bk73} and \cite{pardalosRodgers92}) to presenting algorithms only in
pseudo-code. Fortunately we are moving into a new era where code is being made publicly available (examples here are 
\"{O}sterg\aa{}rd's Cliquer and Konc and Jane\u{z}i\u{c}'s MaxCliqueDyn). Hopefully this will 
grow and allow Computer Scientist to be better able to perform reproducible empirical studies.

Tomita \cite{tomita2010} presented MCS as an improvement on MCR brought about via two modifications: (1) a static
colour ordering and (2) a colour repair step. Our study has shown that modification (1) improves performance
and (2) degrades performance, i.e. MCSa is better than MCSb. 

BBMC is algorithm MCSa with sets represented as bit strings, i.e. BitSet is used rather than ArrayList. Experiments
on the reference machine showed a speed up typically of a factor of 2. The three styles of ordering were investigated. 
The orderings were quickly disrupted by MCQ, but in the other algorithms minimum width ordering was best 
in random problems but in the DIMACS instances there was no clear winner. 

It was demonstrated in our basic algorithm MC (which does not use a colouring bound) that the order vertices were
selected can have an enormous effect on search effort (and this is well known). The best order was to select vertices of low degree first, 
i.e. the \emph{worst-out heuristic} in \cite{maxClique}. Incorporating this into MCSa as a tie breaker had negligible 
effect and the reason for this was because of the small size of colour classes in hard (dense) instances left 
little scope for tie-breaking.

New benchmark problems (i.e. problems rarely investigated by the maximum clique community) were investigated such as
BHOSLIB, k-regular and small-world graphs. Motivation for this study was partly to compare algorithms but also to
explore these problems to determine if and when they are hard.

Finally we demonstrated that the standard procedure for calibrating machines and rescaling results is unsafe, and that
running our own code on different machines can lead to different relative algorithmic performance. This is disturbing.
First, it suggests that to perform a fair and reliable empirical study we should not rescale other's results: we must either code
up the algorithms ourselves, as done here and also by Carmo and Z{\"u}ge \cite{carmoZuge},  or download and run code on our machines. And
secondly, we should run our experiments on different machines.\\

\noindent
All the code used in this study is available at http://www.dcs.gla.ac.uk/$\sim$pat/maxClique

\section*{Acknowledgments}
Pablo San Segundo patiently helped me to understand and implement BBMC. Jeremy Singer helped analyse Java code
and better understand how some of the Java utilities actually work. Ciaran McCreesh commented on an early version of this paper, 
installed and modified the SNAP graphgen program and helped with the analysis of the tie-breaking in colour classes results.
Quintin Cutts encouraged me to do this work. The School of Computing Science supported this work.

\section*{Appendix 1}
\label{sec:maxCliqueCode}
Listing \ref{codeMaxClique} shows how we read in a graph in DIMACS clq format
(lines 10 to 27, delivering an adjacency matrix $A$ and an integer array of degrees $degree$),
create one of our classes of styled algorithm (lines 32 to 44) and then to search for a maximum clique and
print it out (lines 46 to 52) along with run time statistics. An example of running from the command line is as follows:
\begin{verbatim}
      > java MaxClique BBMC1 brock200_1.clq 14400
\end{verbatim}
This will apply $BBMC$ with $style = 1$ to the first brock200 DIMACS instance allowing 14400 seconds of cpu time.

\begin{figure}
\lstset{caption={MaxClique},label=codeMaxClique}
\begin{lstlisting}
import java.util.*;
import java.io.*;

public class MaxClique {

    static int[] degree;   // degree of vertices
    static int[][] A;      // 0/1 adjacency matrix
    static int n;

    static void readDIMACS(String fname) throws IOException {
	String s = "";
	Scanner sc = new Scanner(new File(fname));
	while (sc.hasNext() && !s.equals("p")) s = sc.next();
	sc.next();
	n      = sc.nextInt();   
	int m  = sc.nextInt();   
	degree = new int[n];
	A      = new int[n][n];
	while (sc.hasNext()){
	    s     = sc.next(); // skip "edge"
	    int i = sc.nextInt() - 1;
	    int j = sc.nextInt() - 1; 
	    degree[i]++; degree[j]++;
	    A[i][j] = A[j][i] = 1;
	}
	sc.close();
    }

    public static void main(String[] args)  throws IOException {

	readDIMACS(args[1]);
	MC mc = null;
	if (args[0].equals("MC"))         mc = new MC(n,A,degree);
	else if (args[0].equals("MC0"))   mc = new MC0(n,A,degree);
	else if (args[0].equals("MCQ1"))  mc = new MCQ(n,A,degree,1);
	else if (args[0].equals("MCQ2"))  mc = new MCQ(n,A,degree,2);
	else if (args[0].equals("MCQ3"))  mc = new MCQ(n,A,degree,3);
	else if (args[0].equals("MCSa1")) mc = new MCSa(n,A,degree,1);
	else if (args[0].equals("MCSa2")) mc = new MCSa(n,A,degree,2);
	else if (args[0].equals("MCSa3")) mc = new MCSa(n,A,degree,3);
	else if (args[0].equals("MCSb1")) mc = new MCSb(n,A,degree,1);
	else if (args[0].equals("MCSb2")) mc = new MCSb(n,A,degree,2);
	else if (args[0].equals("MCSb3")) mc = new MCSb(n,A,degree,3);
	else if (args[0].equals("BBMC1")) mc = new BBMC(n,A,degree,1);
	else if (args[0].equals("BBMC2")) mc = new BBMC(n,A,degree,2);
	else if (args[0].equals("BBMC3")) mc = new BBMC(n,A,degree,3);
	else return;

	System.gc();
	if (args.length > 2) mc.timeLimit = 1000 * (long)Integer.parseInt(args[2]);
	long cpuTime = System.currentTimeMillis();
	mc.search();
	cpuTime = System.currentTimeMillis() - cpuTime;
	System.out.println(mc.maxSize +" "+ mc.nodes +" "+ cpuTime);
	for (int i=0;i<mc.n;i++) if (mc.solution[i] == 1) System.out.print(i+1 +" ");
	System.out.println();
    }
}
\end{lstlisting}
\end{figure}

\section*{Appendix 2}
\label{sec:randomGraph}
Listing \ref{randomGraph} is our code for generating Erd\'{o}s-R\"{e}nyi random graphs $G(n,p)$ where $n$ is the number of vertices and
each edge is included in the graph with probability $p$ independent from every other edge. It produces
a random graph in DIMACS format with vertices numbered 1 to $n$ inclusive. It can be run from the command line as follows to produce 
a clq file
\begin{verbatim}
      > java RandomGraph 100 0.9 > 100-90-00.clq
\end{verbatim}

\begin{figure}
\lstset{caption={RandomGraph},label=randomGraph}
\begin{lstlisting}
import java.util.*;

public class RandomGraph { 
  
    public static void main(String[] args) throws Exception {
	int n      = Integer.parseInt(args[0]);
	double p   = Double.parseDouble(args[1]);
	Random gen = new Random();
	System.out.println("p edge "+ n + " 0");
	for (int i=0;i<n-1;i++)
	    for (int j=i+1;j<n;j++)
		if (p >= gen.nextDouble())
		    System.out.println("e "+ (i+1) +" "+ (j+1));
    }
}
\end{lstlisting}
\end{figure}

\bibliographystyle{plain}

\end{document}